\documentclass[11pt,a4paper]{article}
\usepackage{jheppub}
\usepackage[dvipsnames]{xcolor}
\usepackage{amsfonts}
\usepackage{cancel}
\usepackage{physics}
\usepackage[normalem]{ulem}

\renewcommand{\>}{\rangle}
\newcommand{\<}{\langle}
\newcommand{\cO}{\mathcal{O}}
\newcommand{\bcO}{\overline{\mathcal{O}}}
\newcommand{\cW}{\mathcal{W}}
\newcommand{\cM}{\mathcal{M}}
\newcommand{\tcM}{\widetilde{\mathcal{M}}}
\newcommand{\p}{\partial}
\newcommand{\ii}{\mathrm{i}}

\newcommand{\delg}{\delta\gamma}

\newcommand{\nn}{\nonumber}
\usepackage{mathtools}
\usepackage[]{stackengine}

\newlength\correct
\newcommand{\sfrac}[2]{{\textstyle\frac{#1}{#2}}}

\newcommand{\beq}{\begin{equation}}
\newcommand{\eeq}{\end{equation}}
\newcommand{\beqq}{\begin{equation*}}
\newcommand{\eeqq}{\end{equation*}}
\newcommand{\la}[1]{\label{#1}}

\title{Exact Defect Correlation Functions in Chern--Simons Matter Theories}

\author[a]{Gwenaël Ferrando}
\author[b]{and Elior Urisman}

\affiliation[a]{Bethe Center for Theoretical Physics, Universität Bonn, Wegelerstr. 10, D-53115, Germany}
\affiliation[b]{School of Physics and Astronomy, Tel Aviv University, Ramat Aviv 69978, Israel}

\abstract{We study defect correlators in three-dimensional, large-$N$, conformal field theories with slightly broken higher-spin symmetry. Focusing on the quasi-fermionic theory, we bootstrap correlation functions of four defect-changing operators, i.e. of two collinear conformal line defects with boundaries, at arbitrary values of the coupling constant. We can compute an infinite number of such correlators and present a detailed analysis for three typical cases. In addition, we obtain some defect OPE coefficients that give us access to the relative conformal dimension between defect-changing operators at order $O(1/N)$. In the course of our analysis, we also derive explicit expressions for some Feynman integrals that might be of independent interest, including an extension of the usual star-triangle relation. Our bootstrap assumptions naturally apply to Chern--Simons matter theories, so our results provide a non-trivial example of explicit, non-perturbative defect four-point functions in a strongly coupled gauge theory.}

\begin{document}

\maketitle

\section{Introduction}

In principle, knowledge of the spectrum and structure constants of a conformal field theory (CFT) characterises it completely since they suffice to compute all correlation functions through repeated use of the operator product expansion (OPE). In practice, there are relatively few examples of interacting, solvable CFTs where the conformal data is either already known or expected to be computable, especially in dimension $d>2$. Moreover, even in these solvable theories, the resulting expressions for four- or higher-point functions are often unpractical as they involve infinite sums or integrals over conformal blocks, which are themselves fairly complicated functions. And properties of the correlation functions such as crossing are obscured. It thus makes sense to study higher-point correlation functions independently.

In two dimensions, there are many examples of exactly solvable CFTs: (generalised) minimal models, Liouville theory, and critical loop models, which all enjoy Virasoro symmetry, as well as Toda theories and (some) Wess--Zumino--Witten models, which have an extended chiral algebra. The conformal data of many of these theories can be computed by solving the conformal bootstrap equations, see \cite{Ribault:2024rvk} for a recent review. This is essentially possible because of the infinite-dimensional symmetry algebra---either Virasoro or a larger one---and the existence of degenerate fields whose OPE with any other field contains finitely many primaries, as first shown in \cite{Teschner:1995yf} for the Liouville theory.

Solvable, interacting CFTs in higher dimensions exist but they are still far from being completely solved.\footnote{Parisi--Sourlas supersymmetric CFTs that are uplifts of solvable 2D CFTs are slightly special: part of the conformal data of these four-dimensional theories simply coincide with that of the two-dimensional ones~\cite{Trevisani:2024djr}. Yet, it remains to understand whether the uplifts are entirely solvable.} In the intensely studied planar $\mathcal{N}=4$ SYM theory for instance, integrability has provided a solution for the spectrum but exact structure constants are still beyond our reach. Regarding higher-point functions, many perturbative results are known to relatively high loop orders but there are very few examples of exact results. We know a few four-point correlators in a degenerate version of $\mathcal{N}=4$ SYM, namely the planar fishnet theory \cite{Grabner:2017pgm,Gromov:2018hut}, and one that involves four half BPS operators with large R-charge in the planar $\mathcal{N}=4$ SYM theory \cite{Coronado:2018cxj,Kostov:2019stn,Belitsky:2019fan}.

Large-$N$, three-dimensional Chern--Simons matter (CS matter) theories are also among the rare solvable CFTs in higher dimension. They describe a Chern--Simons gauge field coupled to a fundamental boson (quasi-bosonic theory), a fundamental fermion (quasi-fermionic theory), or both for the $\mathcal{N}=2$ CS matter theory. At large $N$, these theories exhibit an approximate higher-spin symmetry---even at finite values of the 't Hooft coupling~$\lambda$ \cite{Aharony:2011jz,Giombi:2011kc}---that allows to obtain many exact results. The structure constants have thus been bootstrapped at leading order in~$N$~\cite{Maldacena:2012sf} and the spectrum of local operators is known through order $O(1/N)$~\cite{Giombi:2016zwa}. Some four-point functions have also been computed through order\footnote{We normalise the local operators so that their two-point functions are $O(N)$. This means that the four-point functions are $O(N^2)$ if they admit a disconnected contribution, and at most $O(N)$ otherwise.} $O(N)$ using various methods. The correlation function of four scalar operators for instance was bootstrapped in \cite{Turiaci:2018nua} combining the knowledge of the structure constants and the Lorentzian inversion formula \cite{Caron-Huot:2017vep}, in \cite{Aharony:2018npf} using the analytic large-$N$ conformal bootstrap, and in \cite{Li:2019twz} using the slightly broken higher-spin symmetry as in \cite{Maldacena:2012sf}. Correlators where one of the scalars is replaced by a spinning operator are also known \cite{Silva:2021ece}. Though all the previous results were derived in position space, several four-point correlators have been computed in momentum space in the collinear limit and for specific components of the higher-spin currents \cite{Bedhotiya:2015uga,Kalloor:2019xjb,Kukolj:2024yyo}. A recursion relation on momentum-space, collinear, $n$-point correlators of the scalar operator was also derived in \cite{Yacoby:2018yvy}. And some conjectures about the structure of general four-point functions were proposed in \cite{Jain:2022ajd}.

So far, we have only mentioned results involving local operators, but CFTs also possess extended, defect operators that can themselves be renormalisation group fixed points. Defects have been the topic of many recent works and there are still many open problems, see \cite{Andrei:2018die,Komargodski:2025jbu} and references therein. The simplest such defects are one-dimensional, line defects and in this work we will be interested in conformal line defects. For interacting CFTs with defects in dimension $d>2$, there are not many non-perturbative explicit results, some of the known ones compute expectation values of Wilson loops \cite{Witten:1988hf,Pestun:2007rz,Giombi:2018qox}, Wilson loops with two insertions \cite{Correa:2012at} and correlators of a local operator with a Wilson line \cite{Giombi:2018hsx}.

In the present work, we will derive some new exact results for correlators on the so-called fundamental line defects of large-$N$ CS matter theories. These can start and end on fundamental and anti-fundamental boundary (defect-changing) operators. Moreover, they can be fused to generate the other conformal line defects at large-$N$. Conformal line defects preserve a $\mathrm{SL}(2,\mathbb{R})\times \mathrm{U}(1)$ subgroup of the three-dimensional conformal symmetry of the bulk theory. The defect spectrum can thus be organised in multiplets characterised by the conformal dimension and the transverse spin of their primary operator.

Following \cite{Maldacena:2012sf} and as in \cite{Li:2019twz,Jain:2020puw,Jain:2022ajd,Kukolj:2024yyo,Ferrando:2025ufj}, we will employ a bootstrap approach relying on the existence of a slightly broken higher-spin symmetry in a large-$N$ limit. This approach is adapted to both quasi-bosonic and quasi-fermionic theories which differ by the conformal dimension of the scalar, single-trace operator $J_0$, respectively $1+O(1/N)$ and $2+O(1/N)$. For clarity of the exposition, we focus in this work on the quasi-fermionic theory.

For the fundamental line defect of 3D CFTs with slightly broken higher-spin symmetry, the planar boundary spectrum has been determined in \cite{Gabai:2022vri,Gabai:2022mya}. It depends on a single parameter $\Delta\in[\frac{1}{2};\frac{3}{2}]$. The fundamental $\cO_s$ and anti-fundamental $\bcO_s$ operators are labeled by $s\in\mathbb{Z}+\sfrac{1}{2}$. They have opposite, non-trivial, anomalous transverse spin independent of~$s$. The conformal dimension of $\bcO_{\frac{1}{2}}$ and $\cO_{-\frac{1}{2}}$ is $\Delta$, whereas all the other boundary operators have planar dimension in $\Delta + \mathbb{N}^*$ or in $2-\Delta + \mathbb{N}$. Since defect-changing operators must come in pairs, we also use the short-hand notation
\begin{equation}
    \cM^{(\bar s,s)}_{10} = \bcO_{\bar s}(x_1)\cW \cO_s(x_0)\, ,
\end{equation}
where $\cW$ is a finite-length, straight defect between points $x_0$ and $x_1$, which we will always take to lie on the third direction with $x_1^3>x_0^3$. We refer to $\cM^{(\bar s,s)}$ as a mesonic line operator.

Together with A. Sever, we showed in our previous paper \cite{Ferrando:2025ufj} that all the correlators $\<\cM^{(\bar s,s)} J_{\tilde{s}}\>$ of one mesonic line operator and one local operator can be fully bootstrapped---up to normalisation. Such correlators are only constrained by conformal symmetry up to $2\tilde{s}+1$ functions of one invariant cross ratio. Yet, we computed all of them for $J_0$ and $J_1$ with arbitrary $\bar s,s$, and all those for which $\bar s s>0$ but $\tilde{s}$ is arbitrary, which appeared to be remarkably simpler than when $\bar s s<0$. The $\<\cM^{(\bar s,s)} J_{\tilde{s}}\>$ correlators encode all the bulk-defect OPE coefficients, but we did not try to extract these.

\subsection{Summary of Results}

Our main results are the calculations of connected correlation functions of two aligned mesonic line operators---in other words, of four defect-changing operators---at first non-trivial order in the large-$N$ limit. Using the bulk-defect OPE coefficients, one could write such correlators as a sum of 1D conformal blocks. This is however not a trivial task as the defect spectrum is richer than the bulk spectrum: each local, bulk operator decomposes into infinitely many line-defect primaries.\footnote{To illustrate it consider the local operator $J_0$. From the 3D conformal group perspective, this is the only single-trace scalar primary. However, from the 1D viewpoint, it is part of an infinite tower of primaries. For instance, the next primary after $J_0$ is $[(2\Delta(J_0)+1)\p_+\p_- + \p^2_3]J_0$, where $\Delta(J_0) = 1$ or $2$.} We obtain these four-point functions instead by solving pseudo Ward--Takahashi identities coming from the slightly broken higher-spin symmetry. We treat in details three typical examples that are, in order of increasing complexity of the result,
\begin{equation}\label{list 4pt}
    \<\mathcal{M}_{32}^{\left(-\frac{1}{2},-\frac{1}{2}\right)}\mathcal{M}_{10}^{\left(\frac{1}{2},\frac{1}{2}\right)}\>\, ,\quad \<\mathcal{M}_{32}^{\left(\frac{3}{2},-\frac{1}{2}\right)}\mathcal{M}_{10}^{\left(-\frac{1}{2},-\frac{1}{2}\right)}\>\, ,\quad \<\mathcal{M}_{32}^{\left(\frac{1}{2},-\frac{1}{2}\right)}\mathcal{M}_{10}^{\left(\frac{1}{2},-\frac{1}{2}\right)}\>_c\, ,
\end{equation}
where the four points are on the third axis, with $x^3_3>x^3_2>x^3_1>x^3_0$. The subscript $c$ in the third example indicates that we determine the connected part of the four-point function, which is $O(1/N)$ in our normalisation. The full correlator also contains a disconnected part that is $O(1)$, but our bootstrap approach does not constrain it. In other words, the parameter~$\Delta$ can be taken to be the exact, non-planar dimension of $\bcO_{\frac{1}{2}}$ and $\cO_{-\frac{1}{2}}$.

Conformal covariance fixes each of the correlators \eqref{list 4pt} up to a function of one invariant cross ratio. We find explicit expressions for each of these functions in terms of a small number of one-loop Feynman-like integrals that are straightforward to evaluate numerically.

For instance, we obtain the following normalised defect four-point function:
\begin{equation}\label{equal equal}
    \frac{\langle\cM_{32}^{\left(-\frac{1}{2},-\frac{1}{2}\right)}\cM_{10}^{\left(\frac{1}{2},\frac{1}{2}\right)}\rangle}{\langle \cM^{\left(\frac{1}{2},-\frac{1}{2}\right)} \rangle \langle \cM^{\left(-\frac{1}{2},\frac{1}{2}\right)} \rangle} = \frac{\sin^2(\pi\Delta)[\mathcal{F}_{\text{ee}}(\chi) + \mathcal{\widetilde{F}}_{\text{ee}}(\chi)]}{N\pi^3(1-\Delta)|x_{30}|^{4-2\Delta} |x_{21}|^{2\Delta}} + O\left(\frac{1}{N^2}\right)\, ,
\end{equation}
where the cross ratio is $\chi = x^3_{32} x^3_{10}/x^3_{30} x^3_{21}>0$ , the function $\mathcal{F}_{\text{ee}}$ is defined by 
\begin{equation}
    2\mathcal{F}_{\text{ee}}'(\chi) = (1+\chi) \mathcal{G}_{\text{ee}}'(\chi) + (4-2\Delta) \mathcal{G}_{\text{ee}}(\chi)\, ,\quad \mathcal{F}_{\text{ee}}(0) = \pi^2 \frac{(1+4(\Delta-1)^2)}{(2\Delta-3) (2\Delta-1)}\cot(\pi\Delta)
\end{equation}
and the functions $\mathcal{\widetilde{F}}_{\text{ee}}$ and $\mathcal{G}_{\text{ee}}$ are given by the following integrals
\begin{align}
    \mathcal{\widetilde{F}}_{\text{ee}}(\chi) &= \chi\! \int \frac{(\Delta-1)\, r^2\, \dd^3 y}{|y+\chi \hat{e}_3|^{5-2\Delta} |y|^{2\Delta+1} |y-\hat{e}_3|^{2\Delta} }\, ,\label{Ftildeee}\\
    \mathcal{G}_{\text{ee}}(\chi) &= \chi \! \int \frac{(1-2\Delta)\, r^2\, \dd^3 y}{|y+\chi \hat{e}_3|^{5-2\Delta} |y|^{2\Delta+1} |y-\hat{e}_3|^{2\Delta+1}}\, ,\label{Gee}
\end{align}
where $\hat e_3 =(0,0,1)$ is the unit vector in the third direction. 

The integrals \eqref{Ftildeee} and \eqref{Gee} are not trivial but we can compute them explicitly\footnote{One of the authors (GF) expresses his thanks to Florian Loebbert for sharing this result.}, as well as the function $\mathcal{F}_{\text{ee}}$, in terms of hypergeometric functions by solving Yangian differential equations \cite{Chicherin:2017cns,Chicherin:2017frs,Loebbert:2024qbw,Loebbert:2019vcj}. For the other two examples listed in \eqref{list 4pt}, the relevant integrals are more complicated because they already contain hypergeometric functions in the integrand. We did not make any effort towards computing them, but we coincidentally prove a generalisation of the star-triangle relation involving hypergeometric functions. The simplest example of such a relation is\footnote{We also give a $d$-dimensional version of this relation in Appendix \ref{app:int}. But only the $d=3$ case is needed for this paper's calculations.}
\begin{multline}
    \int\! \frac{{}_2F_1\!\left[\genfrac{}{}{0pt}{1}{2\alpha,2\kappa}{1/2+\alpha+\kappa};\sfrac{1}{2} \pm \sfrac{x_{30}\cdot x_{31}}{2|x_{30}| |x_{31}|}\right] \dd^3x_3}{|x_{30}|^{2\alpha} |x_{31}|^{2\beta} |x_{32}|^{2\gamma}} = \pi^{\frac{3}{2}} \frac{\Gamma(3-2\beta)\, \Gamma(3/2-\gamma)}{2^{2\gamma-3} \, \Gamma(2\alpha)\, \Gamma(\gamma)}\\
    \times \frac{\Gamma(3/2-\kappa-\alpha)\, \Gamma(1/2+\alpha+\kappa)}{\Gamma(2+\kappa-\beta)\, \Gamma(\beta-\kappa)} \frac{{}_2F_1\!\left[\genfrac{}{}{0pt}{1}{3-2\beta,2\kappa}{2+\kappa-\beta};\sfrac{1}{2} \pm \sfrac{x_{20}\cdot x_{21}}{2|x_{20}| |x_{21}|}\right]}{|x_{10}|^{3-2\gamma} |x_{20}|^{3-2\beta} |x_{21}|^{3-2\alpha}}\, ,
\end{multline}
which holds when $\alpha+\beta+\gamma = 3$, and $\kappa$ is arbitrary. For $\kappa=0$, this reduces to the usual star-triangle relation.

Moreover, we compute some explicit $O(1/N)$ corrections to the conformal dimensions of boundary operators $\cO$ and $\bcO$ and of factorised operators $:\!\cO\times \bcO\!:$. For the boundary operators, we relate differences of their dimensions to defect structure constants involving the displacement operator, which we can also bootstrap. For the factorised operators, we consider the $x_2\to x_1$ limit of four-point functions $\<\cM_{32}^{(s,-s)}\cM_{10}^{(s',-s')}\>$ that contain a disconnected part. Having bootstrapped the connected contribution at $O(1/N)$, we can access at the same order the difference between the conformal dimension of factorised operators and the sum of their constituents' dimensions.

As mentioned before, our results apply to $\mathrm{SU}(N_c)$ Chern--Simons theories with fermionic matter, at large $N_c$. They give observables at $O(1/N_c)$ provided one identifies the mapping between our bootstrap parameters $(\Delta,N)$ and the parameters of the field theory $(\lambda,N_c)$.

The paper is organised as follows. The next subsection contains some interesting future directions. We present our setup in Section \ref{sec: frame}, namely our bootstrap assumptions and normalisation conventions. In Section \ref{sec:three}, as a warm-up, we bootstrap some three-point functions of operators on the line. A special structure constant is calculated that controls the relative conformal dimension between boundary operators at~$O(1/N)$. In Section \ref{sec:bootstrap2lines}, we bootstrap four-point functions made out of two collinear mesonic line operators. Several appendices contain further details. For convenience, we recall in Appendix~\ref{app:corr} some of the results of \cite{Ferrando:2025ufj}. We perform in Appendix~\ref{app:coeffs} some auxiliary computations needed to determine the coefficients appearing in our bootstrap equations. We derive a generalisation of the star-triangle relation and compute the Feynman integrals \eqref{Ftildeee} and \eqref{Gee} in Appendix~\ref{app:int}. We test our bootstrap results against one-loop computations in the fermionic CS matter theory in Appendix~\ref{app:perturbative}. Finally, Appendix~\ref{app:4pt} contains the computations of (infinitely) more correlators of four defect-changing operators.

\subsection{Future Directions}
\label{sec:fut}

In this work we are able to compute four-point functions at order $O(1/N)$. This allows us to extract some conformal data at the same order. It would be interesting to extend these results beyond $O(1/N)$. It is possible that Ward--Takahashi identities are not enough and that we would have to use additional techniques such as the conformal bootstrap. The analytical approaches of \cite{Mazac:2016qev} for one-dimensional CFTs and of \cite{Lemos:2017vnx} for defect CFTs could be particularly useful. It might also be fruitful to use numerical approaches that have recently been adapted to defect CFTs with defect-changing operators \cite{Lanzetta:2025xfw}.

It is natural to study more general correlation functions. Natural generalisations include five-point functions on the line, correlation functions of two collinear lines and one local operator in the bulk, or of three collinear lines. In other theories, these would be very difficult to compute but, in view of the results presented in this paper, it is possible than for theories with slightly broken higher spin symmetry such correlators could be studied non-perturbatively. One could also consider two parallel mesonic line operators.

When $\Delta = \frac{1}{2}$ or $\frac{3}{2}$, the theories we study are expected to coincide with the so-called critical boson vector model. For the four-point functions that we compute, this is immediate because the integrals we find, such as $\widetilde{\mathcal{F}}_{\text{ee}}$ given by \eqref{Ftildeee}, naturally reduce to the Feynman integrals contributing to the critical boson four-point functions at this order in $N$ \cite{Lang:1992pp}. Moving away from these values of $\Delta$, only the propagator powers change, suggesting the existence of an effective theory with certain Feynman rules that reproduce our integrals. It would be interesting to investigate this further and understand if one can indeed come up with Feynman rules that would directly reproduce the four-point functions and could ideally be applied to higher-point correlators.

Finally, it would be nice to reproduce our results in the holographic dual to CS matter theories. This would require to understand how to incorporate fundamental conformal lines in the dual picture. The results we present could help shed some light on these AdS theories.

\section{Theories with Slightly Broken Higher-Spin Symmetry} \label{sec: frame}

We briefly recall here our setup, conventions, as well as the bootstrap techniques we shall use. Part of this was already summarised in \cite{Ferrando:2025ufj} so we will be rather succinct.

\subsection{Setup}\label{sec:setup}

We assume that the spectrum of primary single-trace operators consists of operators $J_{\tilde s}$ for $\tilde{s}\in\mathbb{N}$, where $J_0$ is a scalar and the operators $J_{\tilde s}$ for $\tilde{s}\geqslant1$ are symmetric traceless tensors of rank $\tilde{s}$. The currents $J_1$ and $J_2$ are exactly conserved, whereas the higher-spin currents are only conserved up to corrections of order $O(1/N)$. The dimension of $J_{\tilde s}$ is thus $\tilde{s}+1+O(1/N)$. We study the so-called quasi-fermionic theory where $J_0$ is of dimension $2$ in the planar limit.

The theories we study also include a conformal line defect along a smooth path, that is to say a line operator\footnote{For brevity, we suppress the dependence of the line operator on the path. In this work, the line will always be straight, along the third direction.} $\cW$ that transforms covariantly under conformal transformations, with zero dimension. A straight conformal line preserves an $SL(2,{\mathbb R})\times U(1)$ symmetry. The sign of the transverse $U(1)$ spin is defined such that the spin of $x_\pm = (x^1\mp\ii x^2)/\sqrt 2$ is equal to $\pm 1$. 
A fundamental conformal line is a conformal line whose operators factorise~as
\begin{equation}\label{largeN2}
    \cO_{\text{line}} =\,\, :\!\cO\times\bcO\!: + O(1/N)\,,
\end{equation}
at large $N$. Here, $\cO$ and $\overline{\cO}$ are boundary fundamental and anti-fundamental operators. A fundamental conformal line can start or end on such boundary operators, which are thus defect-changing operators. The spectrum of boundary operators was bootstrapped in~\cite{Gabai:2023lax}. It is parameterised by $s\in\mathbb{Z}+\frac{1}{2}$, and the conformal dimension and transverse spin of the fundamental boundary operator $\cO_s$ are
\begin{equation}\label{spectrumR}
    (\Delta_s,\mathfrak{s}_s) = \left\{\begin{array}{lcl}\left(\Delta-s-\frac{1}{2},s-\Delta+1\right) +O(1/N) & \qquad & s<0 \\
    \left(s-\Delta+\frac{3}{2},s-\Delta+1\right) +O(1/N) & \qquad & s>0\end{array}\right. , \qquad s\in{\mathbb Z}+\frac{1}{2}\, .
\end{equation}
Similarly, the anti-fundamental boundary operator $\bcO_s$ has dimension $\Delta_{-s}$ and spin\linebreak $-\mathfrak{s}_{-s} = s+\Delta-1$. The signs of the transverse spins we have given assume that the line extends from the fundamental boundary operator towards the positive $\hat x^3$ direction and from the anti-fundamental boundary operator towards negative $\hat x^3$. We will always consider such configurations.

Among the various line-defect operators, the so-called displacement operator $\mathbb D$ plays a distinguished role. This is a protected operator of dimension $2$ and transverse spin (plus or minus) $1$ that generates deformations of the line. Namely, under a small deformation of the path, the line operator transforms as
\begin{equation}
    \delta \mathcal{W} =\int \mathrm{v}^\mu (\tau) \, \mathbb{D}_\mu(x(\tau))\,\cW\, |\dot x(\tau)|\dd\tau\,,
\end{equation}
where $\tau\mapsto x(\tau)$ is some parameterisation of the path, $\tau\mapsto x(\tau)+\mathrm{v}(\tau)$ is the deformed path, and $\mathbb D$ is inserted at $x(\tau)$ along the line. In our notation, the displacement operator reads
\begin{align}\label{displacement}
    \mathbb{D}_+ &= \eta_+\, \cO_{+\frac12}\times\overline \cO_{+\frac12} + \frac{\nu}{N\mathcal{N}_1} J_{1,+}+O(1/N^2)\,,\\
    \mathbb{D}_- &= \eta_-\, \cO_{-\frac12}\times\overline \cO_{-\frac12}-\frac{\nu}{N\mathcal{N}_1} J_{1,-}+O(1/N^2)\,,\nn
\end{align}
where $\eta_\pm$ and $\nu$ depend on our choice of normalisation of the operators and $\mathcal{N}_1$ is defined below in \eqref{normalization Js}. The second term of the displacement operator stands for a factorised (normal ordered) product of the line and $J_1$ at the point on the line. The subleading corrections cannot contain any new operator, they can only affect the coefficients in front of those already appearing.

Similarly, we can also displace the boundary operators
\begin{equation}\label{boundary var}
    \delta \cO = \mathrm{v}^\mu \delta_\mu \cO + \mathrm{dv}^\mu \delta_\mu^{(1)} \cO + \mathrm{ddv}^\mu \delta^{(2)}_\mu \cO + \dots\,,
\end{equation}
where $\mathrm{dv}= \mathrm{\dot{v}}/|\dot x|$ and $\mathrm{d^nv}$ are higher-order reparameterisation-invariant derivatives of $\mathrm{v}$ along the path, see \cite{Gabai:2023lax} for more details.

We shall denote a finite-length, straight, line operator stretched along the $\hat x^3$ axis between one fundamental and one anti-fundamental boundary operators by
\begin{equation}\label{def mesonic}
    \cM^{(\bar s,s)}_{\tau\sigma} = \overline\cO_{\bar s}(x_\tau)\,\cW\,\cO_s(x_\sigma)\, ,
\end{equation}
where $x_\tau=(0,0,x^3_\tau)$ and $x^3_\tau>x^3_\sigma$. Note that we could also introduce the infinite-length analogue of \eqref{def mesonic}, for which the boundary operators are connected through infinity along the third direction:
\begin{equation}
    \tcM^{(s,\bar s)}_{\tau\sigma} = \operatorname{Tr}[\cW\,\cO_s(x_\tau) \times \overline\cO_{\bar s}(x_\sigma)\,\cW]\, ,
\end{equation}
where $x^3_\tau>x^3_\sigma$ as before. These two line operators are not independent; they are related by a conformal transformation.

\subsection{Independent Parameters and Normalisation} \label{setup:norm}

First, we define the large parameter $N$ to be the expectation value of a straight, infinite conformal line:
\begin{equation}\label{def N}
    \<\operatorname{Tr}(\cW)\> = N\, .
\end{equation}
Then, as in \cite{Ferrando:2025ufj}, we use the following normalisation of the currents:
\begin{equation}\label{normalization Js}
    \<J_0(x) J_0(0)\> = \frac{N\mathcal{N}_0}{|x|^4}\quad \text{and}\quad \<J^{+\dots +}_{\tilde s}(x) J^{+\dots +}_{\tilde s}(0)\> = N \mathcal{N}_{\tilde s}\, \p^{2\tilde s}_{x^-}\frac1{|x|^2}\quad \text{for}\quad\tilde s\geqslant 1\,,
\end{equation}
where the normalisation constants $\mathcal{N}_{\tilde s}$ are of order $O(N^0)$. For the stress-tensor~$J_2$, we choose the canonical normalisation, which means that the momentum generator is $P^\mu = \int \dd^2 S_\nu J_2^{\mu\nu}$. Similarly, the current $J_1$ can be associated with a $U(1)$ symmetry and we choose a canonical normalisation for it, such that the boundary operators have charge~$\pm 1$.

Regarding the boundary operators, we relate the normalisation of all operators with $|s|>1/2$ to those of $\overline\cO_{\pm\frac12}$ and $\cO_{\pm\frac12}$ through
\begin{equation}\label{boundary tower}
    \overline\cO_{\bar s} = \begin{cases}
\delta^{\bar s - \frac12}_+\overline\cO_{\frac12} &\quad \bar s > 0\\
\delta^{-\bar s-\frac12}_-\overline\cO_{-\frac12} & \quad \bar s< 0
\end{cases}\, ,
\end{equation}
and similarly for $\cO_s$. Here, $\delta_\mu \cO$ stands for the {\it path derivative} of the boundary operator~$\cO$; it is defined in \eqref{boundary var}. In this normalisation it follows that \cite{Gabai:2022vri}
\begin{equation}\label{boundary eom}
    \delta_\mp \overline\cO_{\bar s \pm1} = -\frac12\delta_3^2\overline\cO_{\bar s} + O(1/N)\qquad\bar s\gtrless 0\,,
\end{equation}
and similarly for $\delta_{\pm}\cO_{s\mp 1}$. Here, the $O(1/N)$ terms contain both corrections to the leading order coefficient $-1/2$ and double-trace-like operators involving any of the bulk currents and the boundary operators. We refer to such equations as \textit{boundary equations of motions}.

In our conventions, we have
\begin{equation}\label{normalisation M}
    \<\cM_{10}^{(-s,s)}\> = \<\tcM_{10}^{(s,-s)}\>= \begin{cases}
\frac{c_+\,2^{\frac{1}{2}+s}\,\Gamma(2\Delta_s)}{\Gamma(2\Delta)\,  |x_1-x_0|^{2\Delta_s}} &\qquad s < 0\\
\frac{c_-\,2^{\frac{1}{2}-s}\,\Gamma(2\Delta_s)}{\Gamma(4-2\Delta)\,  |x_1-x_0|^{2\Delta_s}} & \qquad s> 0
\end{cases}\, ,
\end{equation}
for some normalisation-dependent coefficients $c_\pm$, and where we used conformal covariance to relate the expectation values of the finite- and infinite-length operators

Finally, let us stress that the various coefficients introduced until now are not independent: as shown in \cite{Gabai:2022mya,Gabai:2022vri}, the two-point function of the displacement operator is given by
\begin{equation}\label{2pt displacement}
    \eta_-\,\eta_+\, c_+\, c_- = -\frac1{2\pi}(2\Delta - 3)(2\Delta - 2)(2\Delta - 1)\sin(2\pi\Delta)\, .
\end{equation}

More generally, the CFTs we are studying depend on three parameters that we take to be $N$, $\Delta$, and $\nu$. The first one is defined in \eqref{def N}. The second parameter is defined as the scaling dimension of $\bcO_{\frac{1}{2}}$ and $\cO_{-\frac{1}{2}}$, namely $\Delta_{-\frac{1}{2}} = \Delta$ is taken to be independent of $N$. Lastly, $\nu$ is defined via \eqref{displacement} assuming that $J_1$ is canonically normalised. Equivalently,
\begin{equation}
    \<\operatorname{Tr}(\cW) J^\mu_1(x)\> =  \epsilon^{\mu3\rho}\frac{x_\rho}{r^3}(\ii\pi\nu N+ O(1))\, ,
\end{equation}
where $r^2= 2\,x^+x^-$ is the distance between $x$ and the line, which is oriented in the positive $\hat{x}^3$ direction.

\subsection{Bootstrapping Correlators}

The variations $\delta_\pm\cO_s$ and $\delta_\pm\bcO_s$ together with the explicit form of the displacement operator \eqref{displacement} are enough to write the (integrated) Ward--Takahashi identities coming from the conservation of the stress-tensor $J_2$. Indeed, they allow to express the action of the momentum generator on a mesonic line operator $\cM^{(\bar s,s)}$ in terms of other line operators, and thus to relate various correlators. In the simplest case, this can be used to arrive at~\eqref{normalisation M}. In the more complicated case of correlators of the form $\<\cM^{(\bar s,s)} J_{\tilde{s}}\>$, we argued in~\cite{Ferrando:2025ufj} that it is still enough to determine them up to a normalisation constant $d_{\tilde{s}}$ that only depends on the spin of the bulk operator. This is already a highly non-trivial statement as conformal covariance alone only determine these correlators up to $2\tilde{s}+1$ functions of a conformally invariant cross ratio.

The higher-spin currents can also be used to constrain the correlators of these CFTs. This was first realised in \cite{Maldacena:2012sf} where the planar structure constants of local operators were computed. The idea relies on the almost conservation of the higher-spin currents: because the twist of the bulk operators deviates from one only beyond the planar limit, their divergence is $O(1/N)$ and cannot contain any single-trace operators. In the case of $J_3$ for instance, one has \cite{Ferrando:2025ufj}
\begin{equation}\label{divergence J3}
    \p_\mu J_3^{\mu\nu\rho} = \frac{a_1}{N}\left[2J_0 \p^{(\nu}\! J_1^{\rho)} - 3 \p^{(\nu}\!J_0 J_1^{\rho)} + \eta^{\nu\rho} \p_{\mu} J_0 J_1^{\mu}\right] + \frac{2a_2}{N} \epsilon^{\alpha\beta(\nu} J_{1,\alpha} J_{2,\beta}^{\quad \rho)}\,,
\end{equation}
where $T^{(\nu\rho)} = (T^{\nu\rho}+T^{\rho\nu})/2$, and the coefficients $a_1$ and $a_2$ are $O(1)$ in the planar limit. Large-$N$ factorisation of the correlators then allows to write Ward--Takahashi identities for these slightly broken symmetries. We showed in \cite{Ferrando:2025ufj} that they can be used in the planar limit to relate all of the normalisation constants $d_{\tilde{s}}$, and to express the coefficients $a_1$ and $a_2$ in terms of the parameters $\Delta$, $\nu$, and the normalisation constants $\mathcal{N}_{\tilde{s}}$, $d_{\tilde{s}}$. These relations are recalled in Appendix \ref{app:relations}. We only note here that $a_2\propto \nu$.

\section{Three-Point Defect Correlators and Non-Planar Spectrum}
\label{sec:three}

In the limit where the bulk operator aligns with the mesonic line, the correlators $\<\cM^{(\bar s,s)} J_{\tilde{s}}\>$ give us access to some non-trivial defect three-point functions---one of the three defect operators being a factorised product of the line and a bulk operator. Three-point functions involving instead a defect operator of the form $:\!\cO_s\times\bcO_{\bar s}\!:$ could also be considered but they are essentially trivial in the planar limit owing to large-$N$ factorisation.

We show in this section that non-planar corrections to such correlators can be bootstrapped, provided one knows the precise form of the boundary equations of motion \eqref{boundary eom}, including non-planar contributions of the form $:\!J_{\tilde{s}}\cO_s\!:$. These, in turn, can be extracted from the $\<\cM^{(\bar s,s)} J_{\tilde{s}}\>$ correlators we have already computed \cite{Ferrando:2025ufj}. We then show that structure constants associated with the insertion of the displacement operator on a mesonic line measure the difference between conformal dimensions of boundary operators, and we compute the first few of these differences at order $O(1/N)$.

\subsection{Three-Point Functions on the Defect}

We examine here the constraints on defect three-point functions that can be deduced from the Ward--Takahashi identities coming from translation invariance.

As a warm-up, let us briefly consider three-point functions of the form $\<\cM^{(\bar s,s)} J_{\tilde{s}}\>$ where the local operator and the line are collinear. Of course, they are a particular case of the correlators we computed before \cite{Ferrando:2025ufj}. But recall that we only gave simple expressions for $\< \cM^{(\bar{s},s)} J^{+\dots +}_{\tilde{s}}\>$ when $\bar{s}$ and $s$ are both negative. As we explained, this is sufficient to determine all correlators $\<\cM^{(\bar{s},s)} J^{\mu_1\dots \mu_{\tilde{s}}}_{\tilde{s}}\>$ for the same $\bar{s}$ and $s$. However, correlators with $\bar{s}s<0$ are much more complicated. We just want to show here that it is not necessary to fully determine all of them if one is ultimately interested in defect correlation functions.

Indeed, we can easily derive relations between various structure constants. The only subtlety is the need to decompose the action of the transverse momentum generator on the third operator into (line) primary and descendants. For instance, for $s\leqslant{-1/2}$, the restriction of the Ward--Takahashi identity
\begin{equation}\label{3pt variation}
    \< \p_-J_2^{33}(x_2) \cM^{(1-s,s)}_{10} \> - \frac{1}{2}\p^2_{x^3_1} \< J_2^{33}(x_2) \cM^{(-s,s)}_{10} \> + \< J_2^{33}(x_2) \cM^{(1-s,s-1)}_{10} \> = O(\sfrac{1}{N})
\end{equation}
to $x^\pm_2=0$ can be verified provided one remembers that $J_2^{33}$ is a line primary but
\begin{equation}
    \p_-J_2^{33} = (\p_+J_2^{++} - \p_-J_2^{-+}) + \p_3J_2^{3+}
\end{equation}
is the sum of the primary $\p_+J_2^{++} - \p_-J_2^{-+}$ and a descendant. Solving \eqref{3pt variation} provides us with relations between planar structure constants for a given value of $\tilde{s}$, without having to determine the correlation functions $\<\cM^{(\bar{s},s)} J_{\tilde{s}}\>$ for an arbitrary position of the local operator. Then, as in \cite{Ferrando:2025ufj}, one could use the action of the pseudo-charge associated with $J_3$ to relate the structure constants for different $\tilde{s}$. This last step would only require the knowledge of the full bulk-defect-defect correlators with $J_0$, $J_1$, and $J_2$ if $\nu\neq 0$, since those are the only operators that enter the divergence of $J_3$, see \eqref{divergence J3}.

We now turn to three-point functions that could include factorised contributions in the planar limit, i.e.
\begin{equation}
    \<\bcO_{\bar{s}_2}(x_2) :\!\cO^{(j)}_{s_1}\!\times\!\bcO^{(k)}_{\bar{s}_1}\!\!:\!(x_1)\cW\cO_{s_0}(x_0) \> = \delta_{\bar{s}_2,- s_1}\, \delta_{\bar{s}_1, -s_0}\, (\p^j_{x^3_1} \cM^{(\bar{s}_2,s_1)}_{21}) (\p^k_{x^3_1} \cM^{(\bar{s}_1,s_0)}_{10}) + O(\sfrac{1}{N})\, ,
\end{equation}
where $x^3_2>x^3_1>x^3_0$, and we denoted by $\cO^{(j)}_{s} = \delta^j_3\cO_s$ the boundary descendants. In the planar limit, the spin and conformal dimension of $:\!\cO^{(j)}_{s}\times\bcO^{(\bar j)}_{\bar{s}}\!\!:$  are simply the sum of those of its constituents. This means that the spin is exactly $s+\bar s$, since the anomalous spins of the constituents cancel, whereas the conformal dimension is
\begin{equation}
    \Delta\left(:\!\cO^{(j)}_{s}\times\bcO^{(\bar j)}_{\bar{s}}\!\!:\right) = j+\bar{j} + \left\{\begin{array}{lcc}
    2\Delta+\bar{s}-s-1 & \quad & s<0\, ,\  \bar{s} >0 \\
    4-2\Delta-\bar{s}+s-1 & \quad & s>0\, ,\  \bar{s} <0\\
    |s+\bar{s}|+1 & \quad & s\bar{s} > 0
    \end{array}
    \right\} + O(\sfrac{1}{N}) \, .
\end{equation}
Thus, line operators with $s$ and $\bar{s}$ of different signs can only mix among themselves when~$\Delta$ is generic. However, for $s\bar{s} > 0$, these line operators have integer twist and are thus expected to mix with the operators considered above. The displacement operator \eqref{displacement} is the simplest example of such a mixing and we now show how to compute the following three-point functions:
\begin{equation}\label{3pt displ}
    \langle \bcO_{1-s}(x_2) \mathbb{D}_-(x_1) \cW \cO_{s}(x_0)\rangle = \frac{C_s}{N |x_{21}|^{3+\delg_{s}} |x_{10}|^{1-\delg_{s}} |x_{20}|^{\Delta_{s-1} + \Delta_s - 2}}
\end{equation}
where $s<0$, the structure constant $C_s = O(1)$, and $\delg_s = \Delta_{s-1} - \Delta_s - 1 = O(1/N)$. At leading order in the large-$N$ limit, $\delg_s$ could be ignored, but we chose to keep it here because we will see in the next subsection that it is simply related to $C_s$.

We begin with the simplest case, $\bar s =1/2$, and the identity
\begin{multline}\label{eqthree}
    \langle\delta_- \bcO_{\frac{5}{2}}(x_2) \mathbb{D}_-(x_1)\cW \cO_{-\frac{1}{2}}(x_0)\rangle + \langle \bcO_{\frac{5}{2}}(x_2) \delta_- \mathbb{D}_-(x_1)\cW \cO_{-\frac{1}{2}}(x_0) \rangle\\
    + \langle \bcO_{\frac{5}{2}}(x_2) \mathbb{D}_{-}(x_1)\cW \cO_{-\frac{3}{2}}(x_0)\rangle = O(\sfrac{1}{N^2})\, ,
\end{multline}
where $\delta_- \mathbb{D}_{-}$ is an operator with conformal dimension $3$ and transverse spin $2$. It is easy to determine it by solving some additional Ward--Takahashi identities but we will only use the fact that, given the spectrum of line operators, $\delta_- \mathbb{D}_{-}$ must be a primary.

The relevant terms of the boundary equation of motion are
\begin{equation} \label{bdvar}
  \delta_- \bcO_{\frac{5}{2}} = -\frac{1}{2}\delta^2_3 \bcO_{\frac{3}{2}} + \frac{1}{N} \left(\xi_1\,\p_+ J_0 + \xi_2\,\p_3 J_1^- + \xi_3\, J^-_1 \delta_3 + \xi_4\,\p_+J^3_1 + \xi_5\, J_{2,+3}\right)\bcO_{\frac{1}{2}} + \dots\,,
\end{equation}
where we ignored the potential $O(1/N)$ correction to the first term in the right-hand side because it would produce a $O(1/N^2)$ term in \eqref{eqthree}. The double-trace terms, however, are relevant because they produce factorised contributions $\<\cM\>\<\cM J\>=O(1)$. Their coefficients $\xi_i$ can be computed using the results of \cite{Ferrando:2025ufj}, and in the case of $\xi_5$, the correlators with $J_2$ given in Appendix \ref{app:J2corr}. For completeness, we give their values here:
\begin{align}\label{dtxi}
   \xi_1& = -\frac{d_0\,\eta_-}{2\,\mathcal{N}_{0}} (2\Delta+1)\,, \quad \xi_2 =\frac{d_1\,\eta_- - 4\,\Delta\,\nu}{8\,\mathcal{N}_1}\,, \quad \xi_3 =-\frac{\nu}{2\,\mathcal{N}_1} (2\Delta+1)\, ,\\
   \xi_4 &= \frac{8\Delta(\Delta+1)\nu - d_1\,\eta_-}{16\,\mathcal{N}_1}\, , \qquad \xi_5 = 0\, .
\end{align}

Using these values and the correlators between a mesonic line and $J_0$ or $J_1$ recalled in Appendix \ref{app:corr}, equation \eqref{eqthree} is enough to determine the structure constants $C_{-\frac{1}{2}}$, $C_{-\frac{3}{2}}$, and~$\< \bcO_{\frac{5}{2}}\delta_- \mathbb{D}_-\cW \cO_{-\frac{1}{2}}\>$. Setting $\nu=0$ for simplicity, we find
\begin{equation}\label{struc}
    C_{-\frac{1}{2}} = \frac{c_+\,\eta^2_-\,d^2_0\, (3-2\Delta)(2\Delta+1)}{3 \mathcal{N}_{0}}\, ,\quad C_{-\frac{3}{2}} = \frac{c_+\,\eta^2_-\,d^2_1}{96\, \mathcal{N}_1}\left(9-\frac{16(\Delta-\frac{3}{2})_4}{\cos^2(\pi\Delta)}\right),
\end{equation}
where we omitted the third structure constant as we do not use it further.

The natural next step is to solve $\<\delta_-(\bcO_{\frac{7}{2}}\mathbb{D}_{-}\cW\cO_{-\frac{3}{2}})\>=0$. This already requires us to first fix eight double-trace coefficients. Setting aside the explicit details we simply state that after taking everything into account, we consistently reproduce the value $C_{-\frac{3}{2}}$ and find in addition 
\begin{equation}
    C_{-\frac{5}{2}} = \frac{c_+\, \eta^2_-\, d^2_1}{192\,\mathcal{N}_1}\Bigg(45 + 144\Delta(\Delta+1) - 32(\Delta-\sfrac{3}{2})_5\frac{2\Delta+1}{\cos^2(\pi\Delta)}\Bigg)\,.
\end{equation}

It should be clear by now that computing the structure constants $C_s$ for $s<0$ by solving $\<\delta_-(\bcO_{2-s}\mathbb{D}_{-}\cW\cO_s)\>=0$ quickly becomes challenging. We did not try to push it further. We simply remark that, introducing the structure constants $C_s = N\<\bcO_{-1-s}\mathbb{D}_+\cW\cO_s\>$ for $s>0$, we have
\begin{equation}\label{Cs rel}
    C_s = C_{-s}|_{\substack{\Delta\to2-\Delta\\ c_+\to c_-}}\,.
\end{equation}
This can be traced back to a symmetry that the spectrum \eqref{spectrumR} possesses under the simultaneous mapping $\Delta\mapsto2-\Delta$ and $s\mapsto-s$, which sends $\Delta_s\to\Delta_{-s}$ and $\mathfrak{s}_s\to-\mathfrak{s}_{-s}$.

\subsection{Non-Planar Dimensions of Line Boundary Operators}
\label{dimspace}

We study in this section the first non-planar corrections to the conformal dimensions of the line operators $\cO_s$ and $\bcO_{\bar s}$. We will see that the integer spacing between operators in a given tower only holds in the planar limit, i.e. $|\Delta_s - \Delta_{s'}| = |s -s'| + O(1/N)$ when $s$ and~$s'$ are of the same sign.

\begin{figure}
\centering{}\includegraphics[scale=0.9]{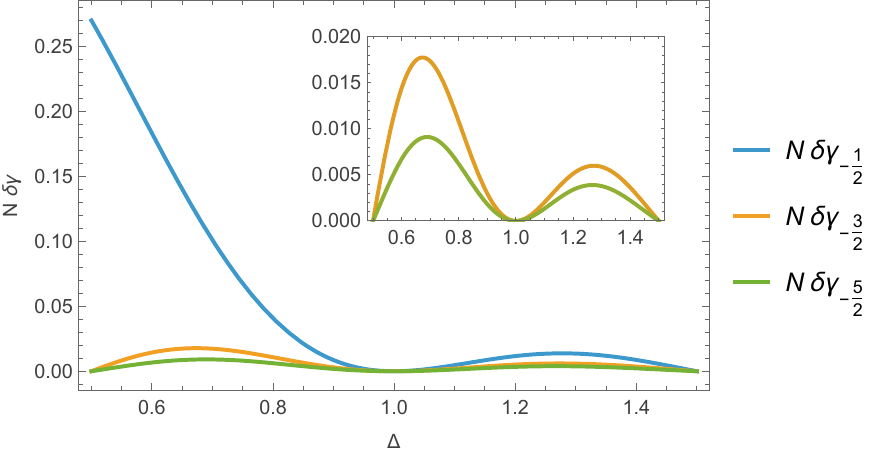}\caption{A plot of $N \delta\gamma_{s}$ as a function of $\Delta\in[\frac{1}{2};\frac{3}{2}]$, at leading order in the large-$N$ limit and for $-s\in\{\frac{1}{2}, \frac{3}{2}, \frac{5}{2}\}$.}
\label{fig:diff dimensions}
\end{figure}

Let us revisit the first-order variation of a single mesonic line to include the first non-planar correction to the anomalous dimensions. For example, the expectation value of a constant minus variation of $\cM^{(1-s,s)}_{10}$ for $s<0$ reads
\begin{equation}\label{first-order revisited}
    \langle (\delta_- \bcO_{1-s})\cW \cO_{-\bar s}\rangle + \langle \bcO_{1-s} \cW (\delta_- \cO_{s})\rangle + |x_{10}| \int_{\tilde{\epsilon}}^{1-\tilde{\epsilon}} \langle \bcO_{1-s} \mathbb{D}_-(x_s) \cW \cO_{s}\rangle\, \dd s = 0\, ,
\end{equation}
where $\epsilon = \tilde{\epsilon}|x_{10}|$ is the regulator. The integral term contains the line three-point function~\eqref{3pt displ}. Since it is of order $O(1/N)$, we used to neglect it entirely. But we have to be more careful because $\delg_{s}$ is of the same order. This is a particularly simple example where the integral can be computed exactly, and expanding the result as $N\to +\infty$, we obtain
\begin{equation}
    \frac{C_{s}}{N |x_{10}|^{\Delta_{s-1} + \Delta_{s} +1}} \int_{\tilde{\epsilon}}^{1-\tilde{\epsilon}}\!\!\! \frac{\dd s}{(1-s)^{3+\delg_{s}} s^{1-\delg_{s}}} = \frac{C_{s} (\tilde{\epsilon}^{-\delg_{s}} - \tilde{\epsilon}^{\delg_{s}} + O(N^{-1}))}{N \delg_{s} |x_{10}|^{\Delta_{s -1} + \Delta_{s} +1}}\, ,
\end{equation}
where we consider $\tilde{\epsilon}\ll \mathrm{e}^{-N}$, i.e. we take $\tilde{\epsilon}\to 0$ before $N\to +\infty$. Regarding the variation of the boundary operators, they should be altered to take into account the fact that $\delg_{s}\neq 0$. We actually have
\begin{equation}
    \delta_- \bcO_{1-s} = -\frac{\epsilon^{-\delg_{s}}}{2}\delta_3^2 \bcO_{-s} + O(\sfrac{1}{N})\, , \qquad 
    \delta_- \cO_{s} = \epsilon^{\delg_{s}} \cO_{s-1} + O(\sfrac{1}{N})\, .
\end{equation}
Combining everything, we deduce two relations from the planar limit of equation \eqref{first-order revisited}, namely
\begin{equation}\label{dimspaceeq}
    \frac{C_s}{N\delg_s} = c_{s-1} = \Delta_{s}(2\Delta_s + 1) c_s + O(\sfrac{1}{N})\, ,
\end{equation}
where $c_s = |x_{10}|^{2\Delta_s} \<\cM^{(-s,s)}_{10}\>$. The second equality in \eqref{dimspaceeq} is the one we could get by ignoring the integral term in \eqref{first-order revisited}. Solving it reproduces \eqref{normalisation M} with $c_+ = c_{-\frac{1}{2}}$. The first relation, however, is new. Using the structure constants computed in the previous subsection, supplemented with relations \eqref{deta/N} and \eqref{old rel norm}, we thus get the first $\delg_s$; they are plotted in Figure \ref{fig:diff dimensions}.

We observe that as we decrease $s$ the relative conformal dimension decreases as well, suggesting that $\delta \gamma_{s} \xrightarrow[s \to -\infty]{} 0$. In general $\delta \gamma_{s}\geqslant 0$, and it vanishes only for $\Delta = 1$ (free fermion) or $\Delta = \frac{1}{2}$ or $\frac{3}{2}$ (critical boson). The only exception being $\delg_{-\frac{1}{2}}(\Delta=\frac{1}{2})= \frac{8}{3 \pi ^2N}$. One should be able to reproduce this via a Feynman diagram computation in the critical boson theory, but we have not tried.\footnote{Still, we point out that, in the $O(N)$ critical boson model, the conformal dimension of a single component of the vector field, which is the analogue of $\cO_{-\frac{1}{2}}$, is $\frac{1}{2} + \frac{4}{3 \pi ^2N} + O(\frac{1}{N^2})$ \cite{Brezin:1972se}. One would have to check that the presence of the line modifies this so as to reproduce our result, without affecting any of the derivative operators.}

Recall from Section \ref{sec:setup} that there is second tower of operators in the spectrum, related to the one we considered by $\Delta\to 2-\Delta$ and $s\to-s$, see \eqref{spectrumR}. Performing the analogous analysis for their dimensional spacing we find, following the similar result for the structure constants \eqref{Cs rel}, that $\delg_s(\Delta) = \delg_{-s}(2-\Delta)$.

\section{Correlators of Two Collinear Mesonic Lines}
\label{sec:bootstrap2lines}

Our aim in this section is to present the method of bootstrapping correlation functions of the form $\<\cM^{(s_3,s_2)}_{32} \cM^{(s_1,s_0)}_{10}\>$ in the large-$N$ limit when all the points are collinear---along the third direction. Conservation of the transverse spin imposes $\sum_{i=0}^3s_i = 0$. Generally, these correlators are of order $O(1/N)$, the only exception being if $s_3=-s_2$ and $s_1=-s_0$. We thus have
\begin{equation}\label{decomposition disconnected}
    \<\cM^{(s_3,s_2)}_{32} \cM^{(s_1,s_0)}_{10}\> = \delta_{s_3,-s_2}\delta_{s_1,-s_0}\<\cM^{(s_3,s_2)}_{32}\> \<\cM^{(s_1,s_0)}_{10}\>+\<\cM^{(s_3,s_2)}_{32} \cM^{(s_1,s_0)}_{10}\>_c\, .
\end{equation}
We will determine the second term of the right-hand side, i.e. the connected part of the correlator. When it occurs, the first term is finite at large $N$, so it would also contain a sub-leading contribution. Indeed, both the dimension of boundary operators and the normalisation of the expectation values of single mesonic lines could receive corrections of order $O(1/N)$. Our bootstrap approach however is insensitive to such terms and will naturally yield the connected part of the correlator.

Using the OPE to decompose one of the mesonic lines into local operators, we could write a conformal block expansion for these four-point functions. Local operators $J_s$ of arbitrary spin $s$ would be exchanged. This is fundamentally different from the previous section, and our previous paper \cite{Ferrando:2025ufj}, where the observables only involved interaction between a mesonic line and at most one local operator. This also explains why the translation Ward--Takahashi identities are not sufficient to determine the four-point functions, as they do not encode any information about interaction between local operators.

\begin{figure}
\centering
\includegraphics[scale=3.5]{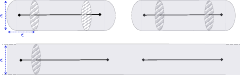}
\caption{Two acceptable choices for the excluded integration region in \eqref{Ward id}. For technical reasons, we favour the bottom one. The right-hand side of \eqref{Ward id} contains integrations of $J_3$ over the boundary of such regions. We split such terms into a finite cylinder that surrounds the segment away from its boundaries and two surfaces that surround the boundaries, as can be seen in the top figure.}
\label{fig:int domain}
\end{figure}

We solve instead the pseudo Ward--Takahashi identities coming from the slight non-conservation of $J_3$. Namely, we start from
\begin{multline}\label{Ward id}
    - \left[\int_{r>\epsilon}\< \mathcal{M}_{32}^{(s_3,s_2)}\mathcal{M}_{10}^{(s_1,s_0)}\,\p_{\mu}J_{3}^{\mu33}(x) \> \dd^3x \right]_{O(\epsilon^0)}\\
    = \<[Q^{(3)}_{33},\cM_{32}^{(s_3,s_2)}] \,\cM_{10}^{(s_1,s_0)}\> + \<\cM_{32}^{(s_3,s_2)}[Q^{(3)}_{33},\cM_{10}^{(s_1,s_0)}]\,\>\,, 
\end{multline}
where the domain of integration is $\mathbb{R}^3$ minus an infinite cylinder or radius $\epsilon$ around the third direction. In principle, it is sufficient to exclude small regions surrounding the line insertions, see Figure \ref{fig:int domain}. However, for technical reasons that will be explained later, it is more convenient to exclude an infinite cylinder. The subscript $O(\epsilon^0)$ means that we only keep the finite terms in the $\epsilon\to0$ limit. The right-hand side was obtained by applying the divergence theorem to the left-hand side. It thus involves the action of the pseudo-charge $Q_{33}^{(3)}$ on the mesonic lines and is defined by integrating $J_3$ on a small neighbourhood of the line. We decompose this action into an action on the line and one on the boundary operators. The former produces the {\it tilt} operator ${\mathbb D}^{(2)}$ and takes the form
\begin{equation}\label{lineact}
    [Q^{(3)}_{33},\cW] = \left[\epsilon\int_{S^1}\dd\hat n \int\dd\tau|\dot x_\tau|\,\hat n_\mu J_{3}^{\mu33}(x_\tau+\epsilon\hat n)\, \cW\right]_{O(\epsilon^0)}\!\!\equiv \int\!\dd\tau\,|\dot x_\tau|\, \mathbb{D}^{(2)}_{33}(x_\tau)\,\cW\,,
\end{equation}
where $\hat n$ is the unit vector parameterising an $S^1$ in the transverse space to the line. The action of $Q^{(3)}_{33}$ on the boundary operators takes the form
\begin{equation}\label{localact}
    [Q^{(3)}_{33},\cO(0)] = \left[\epsilon^2 \int_{S^2}n_\mu J_{3}^{\mu33}(\epsilon n)\, \cO(0) \dd^2 n\right]_{O(\epsilon^0)}\,,  
\end{equation}
where $S^2$ is the unit two sphere.

We now consider the action of $Q_{33}^{(3)}$ on boundary operators and the line to leading order. This charge has dimension 2 and spin 0 under $SL(2,\mathbb{R})\times U(1)$. Thus, the action on the boundary operators must be of the form
\begin{equation}\label{QonBoundary}
    [Q^{(3)}_{33},\cO_s] = q\, \delta^2_3 \cO_s + O(\sfrac{1}{N})\, ,\quad [Q^{(3)}_{33},\overline{\cO}_{\bar s}] = -q\, \delta^2_3 \overline{\cO}_{\bar s} + O(\sfrac{1}{N})\,,
\end{equation}
where we refer to Appendix \ref{app:Qaction} for the proof that the coefficients are all the same (up to a sign). The sub-leading terms include possible double-trace terms as was the case for the boundary equations of motion. In Appendix \ref{app:coeffs}, we present details on how to fix these terms that involve $J_0$ or $J_1$. 

In addition to the action of the charge on the boundary operators, we could expect a non-trivial action of the charge on the fundamental line, see \eqref{lineact}. The tilt operator $\mathbb{D}^{(2)}_{33}$ would be a linear combination of
\begin{equation}
    \p_3 J_0\, ,\quad \p_3 J_1^3 = -(\p_+ J_1^+ + \p_- J_1^-)\, ,\quad \p_+ J_1^+ - \p_- J_1^-\, , \quad \text{and}\quad J_2^{33} = -2J_2^{+-}\, .
\end{equation}
However, we ignore the first two options because they can always be absorbed in the~$O(1/N)$ terms of the boundary equations \eqref{QonBoundary}. And we show in Appendix \ref{app:D33zero} that the other two operators do not contribute either, so that $\mathbb{D}^{(2)}_{33} = O(1/N^2)$ can be ignored at the order we are working at.

Lastly, we recall that the divergence of $J_3$ reads
\begin{equation}\label{divJ3}
    \p_\mu J_3^{\mu33} = \frac{a_1}{N}\left(2J_0\p_3J_1^3 - 3 J_1^3 \p_3 J_0 + J_1^{\mu} \p_{\mu} J_0\right) + 2\ii\frac{a_2}{N} \left(J_1^+ J_2^{3-} - J_1^- J_2^{3+}\right)\, .
\end{equation}
For the rest of this section, for simplicity, we set $a_2 = 0$.

In the next subsections we solve \eqref{Ward id} for some low values of the spin---more general cases are treated in Appendix~\ref{app:4pt}. As was observed in the case of bulk-defect-defect correlators in \cite{Ferrando:2025ufj}, mesonic line operators whose boundary operators have spins of opposite sign are harder to deal with than their equal-sign counterparts. We thus split our four-point functions into classes which we dub equal-equal, unequal-equal and unequal-unequal, depending on the relative sign of the spins. We will compute explicitly one typical example for each of these classes.

Before we proceed to the examples, let us comment on equation \eqref{Ward id}. As we will see, it takes the form of a linear, second-order, partial differential equation on the four-point function, with an explicit source term. Using conformal symmetry to fix the form of the four-point function up to one function of one cross ratio, the homogeneous equation admits a solution only in those cases where the identity can be exchanged, i.e. when $s_3 = -s_2$ or $s_3 = -s_0$. This freedom will be fixed using the known OPE to constrain the behaviour of the correlator when $x_3\to x_2$ or $x_3\to x_0$.

\subsection{Equal-Equal Case}\label{equalequal}
\label{sec:equal-equal}

We start with the simplest case of $\<\mathcal{M}_{32}^{\left(-\frac{1}{2},-\frac{1}{2}\right)}\mathcal{M}_{10}^{\left(\frac{1}{2},\frac{1}{2}\right)}\>$ to illustrate the bootstrap method. Since both lines have non-zero transverse spin, there is no disconnected contribution to the correlator, see \eqref{decomposition disconnected}, and we do not need to consider the double-trace terms in the boundary equations \eqref{QonBoundary}. This second point can be seen as follows. For instance, acting with the charge on the boundary operator at $x_3$, the contributions of the double-trace terms take the form 
\begin{equation}
  \frac{1}{N}\langle :\!J \overline\cO_{-\frac{1}{2}}(x_3)\!:\cW\, \cO_{-\frac{1}{2}}(x_2)\,\mathcal{M}_{10}^{\left(\frac{1}{2},\frac{1}{2}\right)}\rangle=  \frac{1}{N}\<\cM_{32}^{(-\frac{1}{2},-\frac{1}{2})}\>\<J(x_3) \cM_{10}^{(\frac{1}{2},\frac{1}{2})}\>\, + O(\frac{1}{N^2}).
\end{equation}
where $J$ is some (derivative of a) single-trace operator. Using the fact that $ \<\cM_{32}^{(-\frac{1}{2},-\frac{1}{2})}\>=0$, we see that the double-trace terms are indeed not relevant for equal-equal configurations.

Hence, the identity \eqref{Ward id} reads simply
\begin{align}\label{Qrhs}
    &q\left(\p_{x_2^3}^2 - \p_{x_3^3}^2 + \p_{x_0^3}^2 - \p_{x_1^3}^2\right) \langle\mathcal{M}_{32}^{\left(-\frac{1}{2},-\frac{1}{2}\right)}\mathcal{M}_{10}^{\left(\frac{1}{2},\frac{1}{2}\right)}\rangle\\
    &= \frac{5a_{1}}{N} \int\! \Big[\langle\mathcal{M}_{32}^{\left(-\frac{1}{2},-\frac{1}{2}\right)} \p_3J_{0}(y)\rangle\langle\mathcal{M}_{10}^{\left(\frac{1}{2},\frac{1}{2}\right)}J_{1}^{3}(y)\rangle + \langle\mathcal{M}_{32}^{\left(-\frac{1}{2},-\frac{1}{2}\right)}J_{1}^{3}(y)\rangle \langle\mathcal{M}_{10}^{\left(\frac{1}{2},\frac{1}{2}\right)} \p_3 J_{0}(y) \rangle \Big] \, \dd^3 y \, .\nonumber
\end{align}

We now want to rewrite the right-hand side in a form similar to the left-hand side. Using the explicit correlators given in Appendix \ref{app:corr}, elementary manipulations first bring the right-hand side to the form 
\begin{multline}\label{RHS equal-equal}
    \int\! \Big[\langle\mathcal{M}_{32}^{\left(-\frac{1}{2},-\frac{1}{2}\right)} \p_3J_{0}(y)\rangle\langle\mathcal{M}_{10}^{\left(\frac{1}{2},\frac{1}{2}\right)}J_{1}^{3}(y)\rangle + \langle\mathcal{M}_{32}^{\left(-\frac{1}{2},-\frac{1}{2}\right)}J_{1}^{3}(y)\rangle \langle\mathcal{M}_{10}^{\left(\frac{1}{2},\frac{1}{2}\right)} \p_3 J_{0}(y) \rangle \Big]\, \dd^3 y\\
    = \frac{d_0 d_1 \eta_-}{2\eta_+} (\p_{x_2^3}+\p_{x_3^3}) \bigg[ \p_{x_0^3} \int \frac{(3-2\Delta)(2\Delta-1)^2\, |x_{32}|\, r^2\, \dd^3 y}{|y-x_3|^{5-2\Delta} (|y-x_2||y-x_1|)^{2\Delta+1} |y-x_0|^{3-2\Delta}}\\
    \hspace{9.1em} + \p_{x_3^3} \int \frac{(3-2\Delta)(2\Delta-1)^2\, |x_{10}|\, r^2\, \dd^3 y}{|y-x_3|^{3-2\Delta} (|y-x_2||y-x_1|)^{2\Delta+1} |y-x_0|^{5-2\Delta}} \bigg]\\
    \hspace{-1.4em} + \frac{d_0\tilde{d}_1 \eta_-}{2\eta_+}\bigg[ (\p^2_{x_0^3}-\p^2_{x_1^3}) \int \frac{(3-2\Delta)(2\Delta-1)(\Delta-1)\, |x_{32}|\, r^2\, \dd^3 y}{|y-x_3|^{5-2\Delta} |y-x_2|^{2\Delta+1} |y-x_1|^{2\Delta} |y-x_0|^{4-2\Delta}}\\
    +(\p^2_{x_2^3}-\p^2_{x_3^3}) \int \frac{(3-2\Delta)(2\Delta-1)(\Delta-1)\, |x_{10}|\, r^2\, \dd^3 y}{|y-x_3|^{4-2\Delta} |y-x_2|^{2\Delta} |y-x_1|^{2\Delta+1} |y-x_0|^{5-2\Delta}}\bigg]\, ,
\end{multline}
where $r^2= (y^1)^2+(y^2)^2$. Even though we work at $\nu = a_2=0$, which implies $\tilde{d}_1 = d_1$, we have chosen to keep using both to make it easier for the reader to follow the computations. We now notice that the integrals in the right-hand side can be simplified by making a conformal change of variables, see Appendix \ref{app:conf int} for details. In particular, this shows that the last two integrals are equal. We arrive at
\begin{multline}
    \int \Big[\langle\mathcal{M}_{32}^{\left(-\frac{1}{2},-\frac{1}{2}\right)} \p_3J_{0}(y)\rangle\langle\mathcal{M}_{10}^{\left(\frac{1}{2},\frac{1}{2}\right)}J_{1}^{3}(y)\rangle + \langle\mathcal{M}_{32}^{\left(-\frac{1}{2},-\frac{1}{2}\right)}J_{1}^{3}(y)\rangle \langle\mathcal{M}_{10}^{\left(\frac{1}{2},\frac{1}{2}\right)} \p_3 J_{0}(y) \rangle \Big] \dd^3 y\\
    = \frac{d_0 d_1 \eta_-}{2\eta_+} (3-2\Delta)(2\Delta-1) (\p_{x_1^3}+\p_{x_0^3}) \bigg[ \p_{x_0^3}  \frac{|x_{20}| \mathcal{G}_{\text{ee}}(\chi)}{|x_{30}|^{4-2\Delta} |x_{21}|^{2\Delta+1}} + \p_{x_3^3}  \frac{|x_{31}| \mathcal{G}_{\text{ee}}(\chi)}{|x_{30}|^{4-2\Delta} |x_{21}|^{2\Delta+1}}\bigg]\\
    + \frac{d_0\tilde{d}_1 \eta_-}{2\eta_+} (3-2\Delta)(2\Delta-1) (\p^2_{x_2^3}-\p^2_{x_3^3}+\p^2_{x_0^3}-\p^2_{x_1^3}) \frac{ \mathcal{\widetilde{F}}_{\text{ee}} (\chi)}{|x_{30}|^{4-2\Delta} |x_{21}|^{2\Delta}}\, ,
\end{multline}
where the cross ratio is $\chi = x^3_{32} x^3_{10}/x^3_{30} x^3_{21}>0$ and the integrals $\mathcal{\widetilde{F}}_{\text{ee}}$ and $\mathcal{G}_{\text{ee}}$ were defined in \eqref{Ftildeee} and \eqref{Gee} respectively. Let $\mathcal{F}_{\text{ee}}$ be such that 
\begin{equation}\label{ode equal-equal}
    2\mathcal{F}_{\text{ee}}'(\chi) = (1+\chi) \mathcal{G}_{\text{ee}}'(\chi) + (4-2\Delta) \mathcal{G}_{\text{ee}}(\chi)\, ,
\end{equation}
then
\begin{equation}
    \p_{x_0^3}  \frac{|x_{20}| \mathcal{G}_{\text{ee}}(\chi)}{|x_{30}|^{4-2\Delta} |x_{21}|^{2\Delta+1}} + \p_{x_3^3}  \frac{|x_{31}| \mathcal{G}_{\text{ee}}(\chi)}{|x_{30}|^{4-2\Delta} |x_{21}|^{2\Delta+1}} = (\p_{x_3^3}-\p_{x_2^3}+\p_{x_0^3}-\p_{x_1^3}) \frac{\mathcal{F}_{\text{ee}}(\chi)}{|x_{30}|^{4-2\Delta} |x_{21}|^{2\Delta}}\, .
\end{equation}
Hence, we may write
\begin{multline}
    \int \Big[\langle\mathcal{M}_{32}^{\left(-\frac{1}{2},-\frac{1}{2}\right)} \p_3J_{0}(y)\rangle\langle\mathcal{M}_{10}^{\left(\frac{1}{2},\frac{1}{2}\right)}J_{1}^{3}(y)\rangle + \langle\mathcal{M}_{32}^{\left(-\frac{1}{2},-\frac{1}{2}\right)}J_{1}^{3}(y)\rangle \langle\mathcal{M}_{10}^{\left(\frac{1}{2},\frac{1}{2}\right)} \p_3 J_{0}(y) \rangle \Big] \dd^3 y\\
    = \frac{d_0 \eta_-}{2\eta_+} (3-2\Delta)(2\Delta-1) \left(\p^2_{x_2^3}-\p^2_{x_3^3}+\p^2_{x_0^3}-\p^2_{x_1^3}\right) \frac{d_1\mathcal{F}_{\text{ee}}(\chi) + \tilde{d}_1 \mathcal{\widetilde{F}}_{\text{ee}}(\chi)}{|x_{30}|^{4-2\Delta} |x_{21}|^{2\Delta}}\, .
\end{multline}
Putting everything together and setting $d_1=\tilde d_1$, we have thus shown that
\begin{equation}\label{equal equal bis}
    \langle\mathcal{M}_{32}^{\left(-\frac{1}{2},-\frac{1}{2}\right)}\mathcal{M}_{10}^{\left(\frac{1}{2},\frac{1}{2}\right)}\rangle = \frac{c_+\, c_-\, \eta^2_-\, d^2_1\, [\mathcal{F}_{\text{ee}}(\chi) + \mathcal{\widetilde{F}}_{\text{ee}}(\chi)]}{32\pi\,N\,\mathcal{N}_1 (1-\Delta) \cos^2(\pi\Delta) |x_{30}|^{4-2\Delta} |x_{21}|^{2\Delta}}\, ,
\end{equation}
where we have used \eqref{2pt displacement} and a relation from Appendix \ref{app:relations} to rewrite the constant prefactor.

Note that this result is only determined up to one integration constant because $\mathcal{F}_{\text{ee}}$ is only required to satisfy \eqref{ode equal-equal}. This can be fixed by considering the limit $x_3\to x_2$. Using the OPE of the mesonic line, we find
\begin{equation}
    \langle\mathcal{M}_{32}^{\left(-\frac{1}{2},-\frac{1}{2}\right)}\mathcal{M}_{10}^{\left(\frac{1}{2},\frac{1}{2}\right)}\rangle \sim \frac{\langle\mathcal{M}^{\left(-\frac{1}{2},-\frac{1}{2}\right)} J_1^-\rangle}{\langle J^+_1 J^-_1\rangle} \langle J_1^+(x_2) \mathcal{M}_{10}^{\left(\frac{1}{2},\frac{1}{2}\right)}\rangle = \frac{\eta_-\,d^2_1}{16N\,\mathcal{N}_1\,\eta_+ |x_{20}|^{4-2\Delta} |x_{21}|^{2\Delta}}\, .
\end{equation}
Comparing it with the result \eqref{equal equal bis}, this implies that
\begin{equation}
   \mathcal{F}_{\text{ee}}(0)+ \mathcal{\widetilde{F}}_{\text{ee}}(0) = \frac{2\pi(1-\Delta)\cos^2(\pi\Delta)}{\eta_-\,\eta_+\, c_+\, c_-} = \frac{\pi^2 \cot(\pi\Delta)}{(2\Delta-3) (2\Delta-1)}\, ,
\end{equation}
where we used \eqref{2pt displacement}. Given the explicit integral definition of $\mathcal{\widetilde{F}}_{\text{ee}}$, see \eqref{Ftildeee}, one computes $\mathcal{\widetilde{F}}_{\text{ee}}(0) = 4\pi^2(\Delta-1)^2 \cot(\pi\Delta)/(3-2\Delta) (2\Delta-1)$ with the help of relation \eqref{twopropstar}, hence
\begin{equation}\label{f(0)}
    \mathcal{F}_{\text{ee}}(0) = \pi^2 \frac{1+4(\Delta-1)^2}{(2\Delta-3) (2\Delta-1)}\cot(\pi\Delta)\, .
\end{equation}

\begin{figure}
\centering{}\includegraphics[scale=1.05]{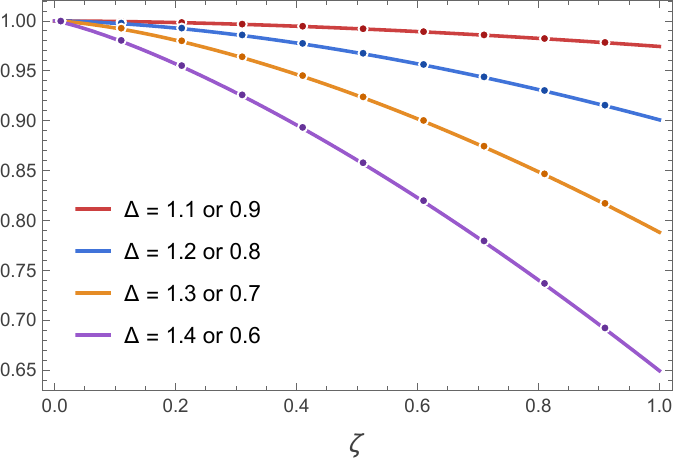}\caption{Plot of the cross-ratio function in \eqref{equal equal}, including the prefactor $\sin^2(\pi \Delta)/\pi^3(1-\Delta)$, for various values of the parameter $\Delta$, evaluated at $\chi = (1-\zeta)/\zeta$, where $\zeta\in[0;1]$. The dots are obtained using only the integral definitions of the functions, and the lines are the analytical expressions in terms of hypergeometric functions. The result is invariant under $\Delta\rightarrow2-\Delta$.}
\la{fplusftilde}
\end{figure}

We explain in Appendix \ref{app:3pt int} how to compute the integrals defining the functions $\mathcal{G}_{\text{ee}}$ and $\mathcal{\widetilde{F}}_{\text{ee}}$. We first find that 
\begin{equation}\label{g(chi)}
    \mathcal{G}_{\text{ee}}(\chi) = \frac{4\pi^2(\Delta-1)\cot(\pi\Delta)}{(2\Delta-3)(1+\chi)^{2\Delta+1}} {}_2F_{1}\!\left[\genfrac{}{}{0pt}{1}{2\Delta-1,2\Delta}{2};\sfrac{\chi}{1+\chi}\right]\,,
\end{equation}
which we then use to obtain
\begin{equation}\label{Fee}
    \mathcal{F}_{\text{ee}}(\chi) = \frac{\pi^3(1-\Delta)}{\sin^2(\pi\Delta)}+\frac{2\pi^2(\Delta-1)}{\tan(\pi\Delta)}\left[\frac{{}_2F_{1}\!\left[\genfrac{}{}{0pt}{1}{3-2\Delta,2\Delta}{2};-\chi\right]}{2\Delta-3}  + \int_{0}^{\frac{1}{1+\chi}} \!\!\!{}_2F_{1}\!\left[\genfrac{}{}{0pt}{1}{2\Delta-1,2\Delta}{2};1-t\right] \frac{\dd t}{t^{1-2\Delta}}\right]\
\end{equation}
by solving \eqref{ode equal-equal} with the boundary condition \eqref{f(0)}. The explicit expression for $\mathcal{\widetilde{F}}_{\text{ee}}$ is somehow more cumbersome so we do not display it here. All the details are given in Appendix \ref{app:3pt int}.

We end this subsection with an additional computation to explain how to go from~\eqref{equal equal bis} to \eqref{equal equal}. In the limit $x_2\rightarrow x_1$, the OPE between the middle two operators is dominated by the exchange of the identity operator:
\begin{equation}\label{newnorm}
    \mathcal{O}_{-\frac{1}{2}}(x_2) \times \overline{\mathcal{O}}_{\frac{1}{2}}(x_1) \sim \frac{c_+}{N |x_{21}|^{2\Delta}}\, ,
\end{equation}
as can be deduced from our definition of $N$ in \eqref{def N} and the expectation value of a single line operator, given in \eqref{normalisation M}. Inserting the OPE in the correlator gives
\begin{equation}
    \langle\mathcal{M}_{32}^{\left(-\frac{1}{2},-\frac{1}{2}\right)}\mathcal{M}_{10}^{\left(\frac{1}{2},\frac{1}{2}\right)}\rangle \sim \frac{c_+\, c_-}{N\, |x_{30}|^{4-2\Delta} |x_{21}|^{2\Delta}}\, .
\end{equation}
On the other hand, we have
\begin{equation}
    \lim_{\chi\to +\infty} [\mathcal{F}_{\text{ee}}(\chi) + \mathcal{\widetilde{F}}_{\text{ee}}(\chi)] = \frac{\pi^3(1-\Delta)}{\sin^2(\pi\Delta)}\, ,
\end{equation}
as can be seen from \eqref{Ftildeee} and \eqref{Fee}. Our explicit result \eqref{equal equal} thus becomes
\begin{equation}
    \langle\mathcal{M}_{32}^{\left(-\frac{1}{2},-\frac{1}{2}\right)}\mathcal{M}_{10}^{\left(\frac{1}{2},\frac{1}{2}\right)}\rangle \sim \frac{\pi^2\,c_+\, c_-\, \eta^2_-\,d^2_1}{8\,N\,\mathcal{N}_1 \sin^2(2\pi\Delta) |x_{30}|^{4-2\Delta} |x_{21}|^{2\Delta}}\, .
\end{equation}
This means that
\begin{equation}\label{deta/N}
    \frac{\eta^2_-\, d^2_1}{\mathcal{N}_1} = \frac{8}{\pi^2} \sin^2(2\pi\Delta)\, ,
\end{equation}
which was the quantity we needed to arrive at \eqref{equal equal}. In our previous paper \cite{Ferrando:2025ufj} we showed that the left-hand side of \eqref{deta/N} was invariant under the replacement $(d_1,\mathcal{N}_1)\to(d_s,\mathcal{N}_s)$ for $s>0$, and there was a slight modification for $s=0$, see \eqref{old rel norm}. But we could not fix the value of this invariant because our computations were not sensitive to the definition of~$N$. Here, however, it enters the OPE expansion \eqref{newnorm}.

We plot the results in Figure \ref{fplusftilde} and show some numerical checks. Finally, we perform a one-loop check of \eqref{equal equal bis} in the fermionic CS matter theory in Appendix~\ref{app:equalpert}. The generalisation to other equal-equal correlators---built out of boundary operators with higher spin---follows similar lines, and we present it in Appendix~\ref{app: equal-equal-gen}.

\subsection{Unequal-Equal Case}

We now consider an intermediate configuration where three of the four spins have the same sign. As before, we choose to focus on a specific example, namely $\<\cM^{\left(\frac{3}{2},-\frac{1}{2}\right)}_{32} \cM^{\left(-\frac{1}{2},-\frac{1}{2}\right)}_{10}\>$. Using the action of the pseudo-charge, we arrive at the equation
\begin{multline}\label{Quneqeq}
    q\left(\p_{x_2^3}^2 - \p_{x_3^3}^2 + \p_{x_0^3}^2 - \p_{x_1^3}^2\right) \langle\cM_{32}^{\left(\frac{3}{2},-\frac{1}{2}\right)} \cM_{10}^{\left(-\frac{1}{2},-\frac{1}{2}\right)}\rangle + \frac{\bar{\rho}^{(2)}_{\frac{3}{2},\frac{1}{2}}}{N} \<\cM_{32}^{\left(\frac{1}{2},-\frac{1}{2}\right)}\> \<\p_3J^-_1(x_3) \cM_{10}^{\left(-\frac{1}{2},-\frac{1}{2}\right)}\> \\
    = \frac{5 a_1}{N}\!\! \int \!\left[\langle\mathcal{M}_{32}^{\left(\frac{3}{2},-\frac{1}{2}\right)} \p_3 J_0(y)\rangle \langle\mathcal{M}_{10}^{\left(-\frac{1}{2},-\frac{1}{2}\right)} J_1^3(y)\rangle +\langle\mathcal{M}_{32}^{\left(\frac{3}{2},-\frac{1}{2}\right)} J_1^3(y)\rangle \langle\mathcal{M}_{10}^{\left(-\frac{1}{2},-\frac{1}{2}\right)} \p_3 J_0(y)\rangle \right]\! \dd^3 y\, .
\end{multline}
The last term of the first line comes from a double-trace term in the action of the pseudo-charge on $\bcO_{\frac{3}{2}}$. We compute it independently in Appendix \ref{app:coeffs} where we also show that all other admissible counter-terms do not occur---at least when $\nu=a_2=0$.

As in the previous section, we are able to manipulate the right-hand side to write it in the same form as the left-hand side. The manipulations are quite cumbersome, see Appendix \ref{app:unequal-equal} for details, but the end result can be written relatively compactly as
\begin{multline}\label{unequal equal}
    \<\cM^{\left(\frac{3}{2},-\frac{1}{2}\right)}_{32} \cM^{\left(-\frac{1}{2},-\frac{1}{2}\right)}_{10}\> = \frac{(2\Delta+1) c_+ \sin^3(\pi\Delta) \cos(\pi\Delta)}{\pi^4\, N\, \eta_- |x_{32}|^{2\Delta-1} |x_{31}|^{4-2\Delta} |x_{30}|^{2\Delta-1} |x_{20}|} \Bigg[ \frac{2\pi^2 \cot(\pi\Delta)}{(4\Delta^2-1)\zeta}\\
    - \widetilde{\mathcal{F}}_{\text{ue}}(\zeta) + \frac{(2\Delta-1)(2\Delta-3)}{4(\Delta-1)} \left( \frac{\mathcal{F}^{(1)}_{\text{ue}}(\zeta)}{2\Delta-1} + \frac{\mathcal{F}^{(2)}_{\text{ue}}(\zeta)}{(2\Delta-3)\zeta} + \frac{(1-\zeta)^2}{\zeta} \mathcal{F}^{(3)}_{\text{ue}}(\zeta)\right)\Bigg]\, ,
\end{multline}
where the cross ratio is $\zeta=x_{30}^3 x_{21}^3/x_{31}^3 x_{20}^3 \in ]0;1[$, and the auxiliary functions are defined as the following integrals:
\begin{align}
    \widetilde{\mathcal{F}}_{\text{ue}}(\zeta) &= \int \frac{F_1(\sfrac{y^3}{|y|})\, r^2\, \dd^3 y}{|y|^2 |y-\zeta \hat{e}_3|^{4-2\Delta} |y-\hat{e}_3|^{2\Delta}}\, ,\\
    \mathcal{F}^{(1)}_{\text{ue}}(\zeta) &= \int \frac{F_1(\sfrac{y^3}{|y|})\, r^2\, \dd^3 y}{|y|^2 |y-\zeta\hat{e}_3|^{5-2\Delta} |y-\hat{e}_3|^{2\Delta-1}}\, ,\\
    \mathcal{F}^{(2)}_{\text{ue}}(\zeta) &= \int \frac{F_1(\sfrac{y^3}{|y|})\, r^2\, \dd^3 y}{|y|^2 |y-\zeta\hat{e}_3|^{3-2\Delta} |y-\hat{e}_3|^{2\Delta+1}}\, ,\\
    \mathcal{F}^{(3)}_{\text{ue}}(\zeta) &= \int \frac{F_2(\sfrac{y^3}{|y|})\, r^2\, \dd^3 y}{|y| |y-\zeta\hat{e}_3|^{5-2\Delta} |y-\hat{e}_3|^{2\Delta+1}}\, ,
\end{align}
where the hypergeometric functions $F_i$ are defined in \eqref{F123}.

At first glance, the result \eqref{unequal equal} may appear different from the previous case: we do not need to solve some differential equation like in \eqref{ode equal-equal} (or as in  \eqref{ode unequal-unequal} for the next example), but we need to know four integrals and not two. However, as we explain in Appendix~\ref{app:unequal-equal}, the last three of these integrals satisfy a system of two coupled, first-order differential equations, see \eqref{rel1} and \eqref{rel2}. This means that we could effectively define two of them in terms of the third one using these differential equations. Moreover, we compute in Appendix~\ref{app: equal-equal-gen} the general four-point functions in the equal-equal configuration and it appears there that the current situation is actually generic: in most cases, the four-point function is a linear combination of four or five integrals, without any derivatives. Hence, this suggests that the solution to \eqref{ode equal-equal} (or to \eqref{ode unequal-unequal} below) might itself be expressible in terms of similar integrals. We did not investigate this possibility. We also did not try to compute more general unequal-equal configurations.

\begin{figure}
\centering{}\includegraphics[scale=0.95]{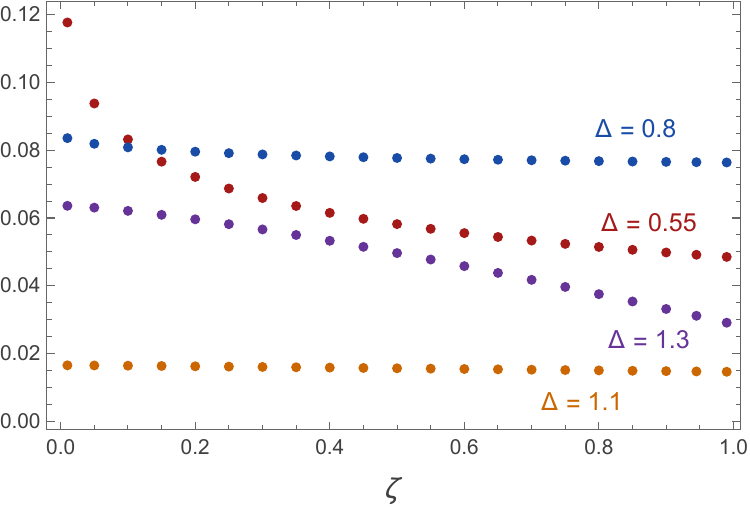}\caption{Numerical evaluation of the cross-ratio function in \eqref{unequal equal}, including the prefactor\\ $(2\Delta+1) \sin^3(\pi\Delta) \cos(\pi\Delta)/\pi^4$, for various values of $\Delta$.}
\label{eqplot}
\end{figure}

We have plotted in Figure \ref{eqplot} the cross-ratio functions of \eqref{unequal equal}---including the prefactor $(2\Delta+1) \sin^3(\pi\Delta) \cos(\pi\Delta)$---for various values of $\Delta$.

We perform a few tests of our result. First, in the limit $\zeta\to 0^+$, the four-point function reduces to one of the three-point functions with the displacement operator considered in Section \ref{sec:three}, see \eqref{3pt displ}. We checked that both results are consistent, i.e. that we reproduce~$C_{-\frac{1}{2}}$. Then, we perform a one-loop check of \eqref{unequal equal} in the fermionic CS matter theory in Appendix~\ref{app:structper}.

\subsection{Unequal-Unequal Case}
\label{sec:unequal-unequal}

As in the previous examples, let us focus on a example with boundary operators of lowest spins: we want to compute the connected correlator $\langle\mathcal{M}_{32}^{\left(\frac{1}{2},-\frac{1}{2}\right)}\mathcal{M}_{10}^{\left(\frac{1}{2},-\frac{1}{2}\right)}\rangle_c$.

We begin by stating the relevant variation equation:
\begin{align}\label{variation uu}
    &q\left(\p_{x_2^3}^2 - \p_{x_3^3}^2 + \p_{x_0^3}^2 - \p_{x_1^3}^2\right) \langle\cM_{32}^{\left(\frac{1}{2},-\frac{1}{2}\right)} \cM_{10}^{\left(\frac{1}{2},-\frac{1}{2}\right)}\rangle_c\\
    &= \frac{5 a_1}{N}\!\! \int \!\left[\langle\cM_{32}^{\left(\frac{1}{2},-\frac{1}{2}\right)} \p_3 J_0(y)\rangle \langle\cM_{10}^{\left(\frac{1}{2},-\frac{1}{2}\right)} J_1^3(y)\rangle +\langle\cM_{32}^{\left(\frac{1}{2},-\frac{1}{2}\right)} J_1^3(y)\rangle \langle\cM_{10}^{\left(\frac{1}{2},-\frac{1}{2}\right)} \p_3 J_0(y)\rangle \right]\! \dd^3 y\, .\nonumber
\end{align}
Note the absence of factorised terms coming from double-trace operators in the left-hand side. This is in part due to the vanishing of their coefficients when $\nu=0$, see Appendix~\ref{app:J1dt}, and in part because they are canceled by additional contributions to the right-hand side, see Appendix~\ref{app:J0dt} for details. We also stress that the variation equation applies only to the connected part of the four-point function, which explains why we did not keep terms like $\<\cM_{32}^{\left(\frac{1}{2},-\frac{1}{2}\right)}\> \int \<\p_\mu J_3^{\mu 33}(y)\cM_{10}^{\left(\frac{1}{2},-\frac{1}{2}\right)}\> \dd^3 y$ in the right-hand side.

As in the previous examples, we need to write the right-hand side in a form similar to the left-hand side. Using the explicit expressions for the correlators collected in Appendix~\ref{app:corr}, we first obtain 
\begin{multline}\label{RHS uneq-uneq}
    \int \!\left[\langle\mathcal{M}_{32}^{\left(\frac{1}{2},-\frac{1}{2}\right)} \p_3 J_0(y)\rangle \langle\mathcal{M}_{10}^{\left(\frac{1}{2},-\frac{1}{2}\right)} J_1^3(y)\rangle +\langle\mathcal{M}_{32}^{\left(\frac{1}{2},-\frac{1}{2}\right)} J_1^3(y)\rangle \langle\mathcal{M}_{10}^{\left(\frac{1}{2},-\frac{1}{2}\right)} \p_3 J_0(y)\rangle \right]\! \dd^3 y\\
    = \frac{d_0\, \eta^2_- c^2_+}{8(\Delta - 1)} \bigg[(\p_{x_0^3}^2 - \p_{x_1^3}^2) \int \frac{F_1(u_{32})\big[d_1 F_2(u_{10})/(1-\Delta) + 2 \tilde{d}_1 F_3(u_{10})/(2\Delta-1)\big]}{|x_{10}|^{2\Delta-1} |x_{32}|^{2\Delta-2} |y-x_3|^2 |y-x_2|^2 |y-x_1| |y-x_0|} \dd^3y\\
    \qquad\qquad \quad+ (\p_{x_2^3}^2 - \p_{x_3^3}^2) \int \frac{F_1(u_{10})\big[d_1 F_2(u_{32})/(1-\Delta) + 2 \tilde{d}_1 F_3(u_{32})/(2\Delta-1)\big]}{|x_{10}|^{2\Delta-2} |x_{32}|^{2\Delta-1} |y-x_3| |y-x_2| |y-x_1|^2 |y-x_0|^2} \dd^3y\\
    + \frac{2d_1}{1-\Delta} (\p_{x_0^3} + \p_{x_1^3})\bigg(\int \frac{F_1(u_{10}) F_2(u_{32})}{|x_{10}|^{2\Delta-2} |x_{32}|^{2\Delta} |y-x_3| |y-x_2| |y-x_1|^2 |y-x_0|^2} \dd^3y\\
    - \int \frac{F_1(u_{32}) F_2(u_{10})}{|x_{10}|^{2\Delta} |x_{32}|^{2\Delta-2} |y-x_3|^2 |y-x_2|^2 |y-x_1| |y-x_0|} \dd^3y\bigg)\bigg] + O(d_1 - \tilde{d}_1)\, ,
\end{multline}
where we use $u_{ij} = (y-x_i)\cdot(y-x_j)/|y-x_i||y-x_j|$. We now make appropriate conformal changes of variables in the integrals to arrive at
\begin{multline}
    \int \!\left[\langle\mathcal{M}_{32}^{\left(\frac{1}{2},-\frac{1}{2}\right)} \p_3 J_0(y)\rangle \langle\mathcal{M}_{10}^{\left(\frac{1}{2},-\frac{1}{2}\right)} J_1^3(y)\rangle +\langle\mathcal{M}_{32}^{\left(\frac{1}{2},-\frac{1}{2}\right)} J_1^3(y)\rangle \langle\mathcal{M}_{10}^{\left(\frac{1}{2},-\frac{1}{2}\right)} \p_3 J_0(y)\rangle \right]\! \dd^3 y\\
    = \frac{d_0\, \eta^2_- c^2_+}{8(\Delta - 1)}\bigg[(\p_{x_2^3}^2 - \p_{x_3^3}^2+\p_{x_0^3}^2 - \p_{x_1^3}^2) \frac{d_1 \mathcal{G}(\chi)/(1-\Delta) + 2 \tilde{d}_1 \widetilde{\mathcal{F}}(\chi)/(2\Delta-1)}{|x_{32}|^{2\Delta} |x_{10}|^{2\Delta}}\\
    + \frac{2d_1}{1-\Delta} (\p_{x_0^3} + \p_{x_1^3})\frac{(|x_{10}|-|x_{32}|)\mathcal{G}(\chi)}{|x_{32}|^{2\Delta+1} |x_{10}|^{2\Delta+1}}\bigg]\, ,
\end{multline}
where the cross ratio is $\chi = x^3_{32} x^3_{10}/x^3_{30} x^3_{21}>0$, and we introduced the functions\footnote{Observe that we have the property $\mathcal{G}_{\text{uu}}(\chi) =-\mathcal{G}_{\text{uu}} \left(\frac{-\chi}{1+\chi}\right)$ and similarly for $\widetilde{\mathcal{F}}_{\text{uu}}$.}
\begin{equation}\label{Guu}
    \mathcal{G}_{\text{uu}}(\chi) = \chi \int \frac{F_1\Big(\sfrac{(1-y^3)}{|\hat e_3-y|}\Big) F_2\Big(\sfrac{(y+\chi \hat{e}_3)\cdot y}{|y+\chi \hat{e}_3| |y| }\Big)}{|y+\chi \hat{e}_3| |y| |y-\hat{e}_3|^{2} } \dd^3 y
\end{equation}
and
\begin{equation}\label{Ftildeuu}
    \widetilde{\mathcal{F}}_{\text{uu}}(\chi) = \chi \int \frac{F_1\Big(\sfrac{(1-y^3)}{|\hat e_3-y|}\Big) F_3\Big(\sfrac{(y+\chi \hat{e}_3)\cdot y}{|y+\chi \hat{e}_3| |y| }\Big)}{|y+\chi \hat{e}_3| |y| |y-\hat{e}_3|^{2} } \dd^3 y\, .
\end{equation}
 We are thus finally able to write
\begin{multline}
    \int \!\left[\langle\mathcal{M}_{32}^{\left(\frac{1}{2},-\frac{1}{2}\right)} \p_3 J_0(y)\rangle \langle\mathcal{M}_{10}^{\left(\frac{1}{2},-\frac{1}{2}\right)} J_1^3(y)\rangle +\langle\mathcal{M}_{32}^{\left(\frac{1}{2},-\frac{1}{2}\right)} J_1^3(y)\rangle \langle\mathcal{M}_{10}^{\left(\frac{1}{2},-\frac{1}{2}\right)} \p_3 J_0(y)\rangle \right]\! \dd^3 y\\
    = \frac{d_0\, \eta^2_- c^2_+}{8(\Delta - 1)} (\p_{x_2^3}^2 - \p_{x_3^3}^2+\p_{x_0^3}^2 - \p_{x_1^3}^2) \frac{d_1 (\mathcal{F}_{\text{uu}}(\chi)+\mathcal{G}_{\text{uu}}(\chi))/(1-\Delta) + 2 \tilde{d}_1 \widetilde{\mathcal{F}}_{\text{uu}}(\chi)/(2\Delta-1)}{|x_{32}|^{2\Delta} |x_{10}|^{2\Delta}}\, ,
\end{multline}
where we introduced $\mathcal{F}_{\text{uu}}$ such that
\begin{equation}\label{ode unequal-unequal}
    \chi\mathcal{F}_{\text{uu}}'(\chi) - 2\Delta\mathcal{F}_{\text{uu}}(\chi) = \mathcal{G}_{\text{uu}}(\chi)\, .
\end{equation}
Putting everything together, reinstating $\tilde{d}_1 = d_1$, using \eqref{a1a2} and \eqref{deta/N} to simplify the constant prefactor, we thus get the normalized four-point
\begin{equation}\label{unequal-unequal}
    \frac{\langle\cM_{32}^{\left(\frac{1}{2},-\frac{1}{2}\right)} \cM_{10}^{\left(\frac{1}{2},-\frac{1}{2}\right)}\rangle_c }{\<\cM_{32}^{(\frac{1}{2},-\frac{1}{2})}\> \<\cM_{10}^{(\frac{1}{2},-\frac{1}{2})}\>}= \frac{\sin^3(\pi\Delta) \cos(\pi\Delta)}{2\pi^4(1-\Delta) N} \left(\frac{\mathcal{F}_{\text{uu}}(\chi)+\mathcal{G}_{\text{uu}}(\chi)}{1-\Delta} + \frac{2\widetilde{\mathcal{F}}_{\text{uu}}(\chi)}{2\Delta-1} \right)\, .
\end{equation}

\begin{figure}
\centering{}\includegraphics[scale=0.75]{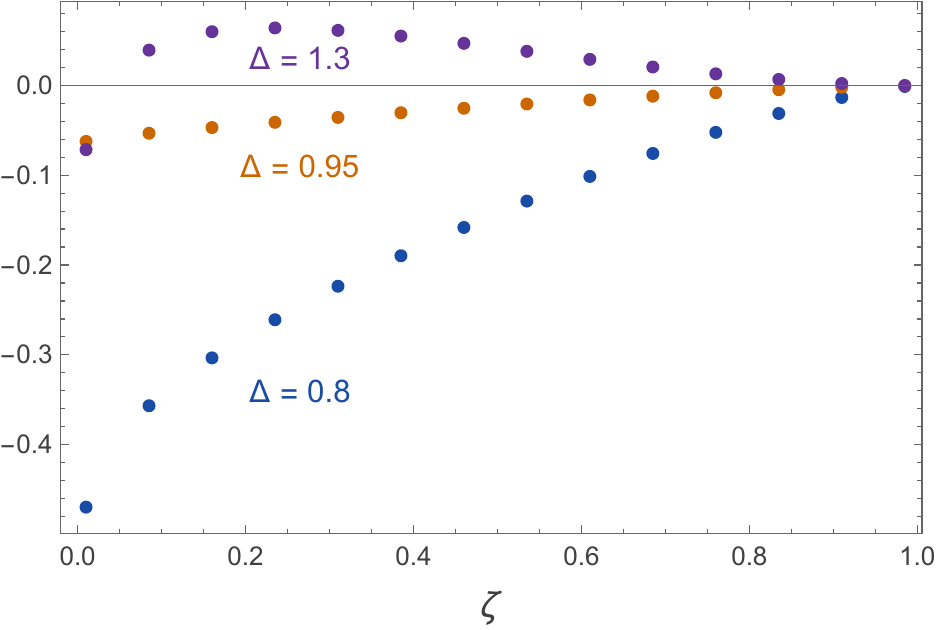}\caption{The plot presents numerical evaluations of the normalised connected four-point in $N\times$\eqref{unequal-unequal} for various $\Delta$ with $\chi=(1-\zeta)/\zeta$, where $\zeta\in[0;1]$. We have subtracted the identity contribution $b\chi^{2\Delta}$ from $\mathcal{F}_{\text{uu}}(\chi)$ as it overwhelmed the other contributions. For the computation of $\mathcal{F}_{\text{uu}}$ we have used an interpolation function of $\mathcal{G}_{\text{uu}}$.}
\label{fig:unequalplot1}
\end{figure}

Note that the function $\mathcal{F}_{\text{uu}}$ is only determined up to one integration constant, which we now determine. The homogeneous solution to \eqref{ode unequal-unequal} is $\chi^{2\Delta}$ so we cannot use the $\chi\to 0$ limit to probe it. We focus instead on the opposite limit corresponding to the segments touching, i.e. $x_2\rightarrow x_1$ or $\chi\rightarrow+\infty$. On the one hand, from the expression we just derived, we find 
\begin{equation}\label{unequalsol}
    \langle\mathcal{M}_{32}^{\left(\frac{1}{2},-\frac{1}{2}\right)}\mathcal{M}_{10}^{\left(\frac{1}{2},-\frac{1}{2}\right)}\rangle \sim \frac{c_+^2\, \sin^3(\pi\Delta) \cos(\pi\Delta)\, b\,\chi^{2\Delta}}{2\pi^4(1-\Delta)^2N\, |x_{32}|^{2\Delta} |x_{10}|^{2\Delta}}\, ,
\end{equation}
where, in the limit $\chi\rightarrow\infty$, we have $\mathcal{F}_{\text{uu}}(\chi) \sim b \chi^{2\Delta}$. We have used the fact that $\mathcal{G}_{\text{uu}}$ and $\widetilde{\mathcal{F}}_{\text{uu}}$ diverge only logarithmically, see details below, and the absence of divergence in the disconnected part of \eqref{decomposition disconnected} to neglect all of them . On the other hand, the OPE is dominated by the exchange of the identity operator, see \eqref{newnorm}, so we should get
\begin{equation}\label{unequalOPE}
    \langle\mathcal{M}_{32}^{\left(\frac{1}{2},-\frac{1}{2}\right)}\mathcal{M}_{10}^{\left(\frac{1}{2},-\frac{1}{2}\right)}\rangle \sim \frac{c^2_+\, \chi^{2\Delta}}{ N|x_{32}|^{2\Delta}|x_{10}|^{2\Delta}} \,.
\end{equation}
Comparing \eqref{unequalsol} with \eqref{unequalOPE}, we arrive at
\begin{equation}\label{boundary unequal}
    b =\frac{2 \pi ^4 (\Delta -1)^2}{\sin ^3(\pi  \Delta ) \cos (\pi  \Delta )}\, . 
\end{equation}
We plot some numerical evaluations of the cross-ratio functions in Figure \ref{fig:unequalplot1}.

A non-trivial check of our result is to study its behaviour when $\chi\to 0$. From the definitions \eqref{Guu} and \eqref{Ftildeuu}, we see that $\mathcal{G}_{\mathrm{uu}}(\chi)\sim\widetilde{\mathcal{F}}_{\mathrm{uu}}(\chi) = O(\chi)$. This implies that the combination that appears inside the brackets in \eqref{unequal-unequal} is only $O(\chi^2)$, which is exactly what we expect since the lightest exchanged operators in this channel are $J_0$ and $J_1^3$, which have dimension $2$. And we have used formula \eqref{2F1 integral} to verify that our result for the four-point function reproduces correctly the OPE data at this order in $\chi$.

As for the previous cases, we perform a one-loop check of \eqref{unequal-unequal} in the fermionic CS matter theory, see Appendix~\ref{app: pert unequal unequal}.

\paragraph{Large-$\chi$ behaviour}
As alluded to above, the functions $\mathcal{G}_{\text{uu}}$ and $\widetilde{\mathcal{F}}_{\text{uu}}$ diverge logarithmically as $\chi\to+\infty$. We compute 
\begin{align}\label{ln Guu}
    \mathcal{G}_{\text{uu}}(\chi) &\sim 2\pi\left[\int_{-1}^1 F_1(u) F_2(-u) \dd u \right]\ln\chi= \frac{8\pi^2(\Delta-1)^2}{1-2\Delta} \cot(\pi\Delta) \ln\chi\, ,\\
    \widetilde{\mathcal{F}}_{\text{uu}}(\chi) &\sim 2\pi\left[\int_{-1}^1 F_1(u) F_3(-u) \dd u \right]\ln\chi=8\pi^2\frac{(\Delta-1)(2\Delta-1)}{1-4\Delta} \cot(2\pi\Delta) \ln\chi\, .
\end{align}
The function $\mathcal{F}$ also contains a power law divergent term:
\begin{equation}
    \mathcal{F}_{\text{uu}}(\chi) = b\,\chi^{2\Delta} + \frac{4\pi^2(\Delta-1)^2}{\Delta(2\Delta-1)} \cot(\pi\Delta) \ln\chi + O(1)\, ,
\end{equation}
where the coefficient of the logarithm can be deduced from \eqref{ln Guu} and the differential equation \eqref{ode unequal-unequal}. These logarithmic divergences can be interpreted using the conformal block expansion of the (full) four-point function:
\begin{equation}\label{uu block expansion}
    \langle\mathcal{M}_{32}^{\left(\frac{1}{2},-\frac{1}{2}\right)}\mathcal{M}_{10}^{\left(\frac{1}{2},-\frac{1}{2}\right)}\rangle = \frac{c_+^2\, \chi^{2\Delta}}{|x_{32}|^{2\Delta} |x_{10}|^{2\Delta}} \left[\frac{1}{N} + \sum_{\cO_{\mathrm{line}}} \frac{C^2_{\cO_{\mathrm{line}}}}{(1+\chi)^{\Delta_{\mathrm{line}}}} {}_2F_{1}\!\left[\genfrac{}{}{0pt}{1}{\Delta_{\mathrm{line}},\Delta_{\mathrm{line}}}{2\Delta_{\mathrm{line}}};\sfrac{1}{1+\chi}\right] \right]\, ,
\end{equation}
where the explicit $1/N$ corresponds to the identity exchange and the sum runs over all the other primary operators that appear in the OPE between the middle two operators $\cO_{-\frac{1}{2}}(x_2) \times \bcO_{\frac{1}{2}}(x_1)$. The normalised structure constants $C_{\cO_{\mathrm{line}}}$ do not all scale the same way in the planar limit. In particular, the only ones that are of order $O(1)$ are associated with the exchange of operators built out of $:\!\!\cO^{(j)}_{-\frac{1}{2}}\times \bcO^{(k)}_{\frac{1}{2}}\!\!\!:$ where $\cO^{(j)}_{-\frac{1}{2}} = \delta^j_3\cO_{-\frac{1}{2}}$ is a descendant. All the other operators appearing in the OPE come with a structure constant $C_{\cO_{\mathrm{line}}} = O(1/N)$, so they do not contribute to the four-point function at the order we are probing.

Focusing on the operators that contribute, it is easy to see that primary operators are indexed by an integer $k\in\mathbb{N}$ and have dimension $2\Delta+k+O(1/N)$. In the planar limit, the explicit form of these operators and the structure constants can be found easily using factorisation of the two- and three-point functions. We have
\begin{equation}
    \cO_{\mathrm{line},k} = \sum_{j=0}^k \binom{k}{j}\frac{(-1)^j (2\Delta)_k}{(2\Delta)_j (2\Delta)_{k-j}} :\!\cO^{(k-j)}_{-\frac{1}{2}} \times \bcO^{(j)}_{\frac{1}{2}}\!\!: + O(\sfrac{1}{N})\, ,
\end{equation}
and
\begin{equation}
    C^2_{\cO_{\mathrm{line},k}} = \frac{(2\Delta)^2_k}{k! (4\Delta + k-1)_k} + O(\sfrac{1}{N})\, .
\end{equation}
This then correctly reproduces the planar, factorised part of the correlator, cf. \eqref{decomposition disconnected}, since
\begin{equation}
    \sum_{k=0}^{+\infty} \frac{(2\Delta)^2_k (1+\chi)^{-2\Delta-k}}{k! (4\Delta + k-1)_k} {}_2F_{1}\!\left[\genfrac{}{}{0pt}{1}{2\Delta+k,2\Delta+k}{4\Delta+2k};\sfrac{1}{1+\chi}\right] = \chi^{-2\Delta}\, .
\end{equation}
Next, we turn to the first non-planar contributions. It is clear that the logarithmic behaviour observed in the large $\chi$ limit comes from the expansion of $(1+\chi)^{-\Delta_{\mathrm{line}}}$ in \eqref{uu block expansion} as $N,\chi\to+\infty$. Putting everything together, we obtain the relative conformal dimension of the lightest line operator, namely
\begin{equation}\label{relano1}
    \Delta(:\!\cO_{-\frac{1}{2}} \times \bcO_{\frac{1}{2}}\!\!:) - 2\Delta = \frac{(1-4\Delta+\cos(2\pi\Delta))}{N\,\pi^2\Delta(1-4\Delta)}\sin^2(\pi\Delta)+O(\sfrac{1}{N^2})\, ,
\end{equation}
where we recall that $\Delta = \Delta_{-\frac{1}{2}}$ is the exact (non-planar) dimension of each of the boundary operators constituents. We can generalise the result and obtain\footnote{To be precise, we can only say that we are computing the conformal dimension of the primary that behaves as $:\!\cO_{s} \times \bcO_{-s}\!\!:$ in the large-$N$ limit. Beyond that, mixing with operators of the form $:\!\cO^{(j)}_{s'} \times \bcO^{(\bar j)}_{-s''}\!\!:$ for $s',s''<0$ and $j+\bar j - s' - s'' = -2s$ would have to be resolved.}
\begin{equation}\label{relanol}
    \Delta(:\!\cO_{s} \times \bcO_{-s}\!\!:) - 2\Delta_{s} = \frac{(1-4\Delta_s+\cos(2\pi\Delta))}{N\,\pi^2\Delta_s(1-4\Delta_s)}\sin^2(\pi\Delta)+O(\sfrac{1}{N^2})\, ,
\end{equation}
which is simply obtained from the preceding equation by the replacement $\Delta\rightarrow \Delta_s$, see details in
Appendix~\ref{app:unequal-unequal}. We also present there another example of an unequal-unequal configuration, where the double-trace terms of the boundary equations combine to give an explicit contribution to the four-point function.

\section*{Acknowledgments}

We thank A. Sever and F. Loebbert for interesting discussions and collaborations on related topics. We are grateful to O. Aharony for comments on the manuscript. We also thank F. Loebbert for showing us how to derive the differential equation \eqref{eqdiff J0}. The work of GF is funded by the Deutsche Forschungsgemeinschaft (DFG, German Research Foundation)---Projektnummer 508889767. EU is supported by the Israel Science Foundation, grant number 1099/24.

\begin{appendix}

\section{Useful Correlators}
\label{app:corr}

In this section we recall some of the correlators of a mesonic line and a local operator that we bootstrapped in \cite{Ferrando:2025ufj}, presenting them in a way adapted to the current paper's calculations. We also give some new results for correlators involving the stress-tensor.

Since some of the expressions are long, we make use of the short-hand notation 
\begin{equation}\label{F123}
       F_1(u) = {}_2F_{1}\!\left[\genfrac{}{}{0pt}{1}{2,1-2\Delta}{2-\Delta};\sfrac{1-u}{2}\right],\quad F_{2}(u) = {}_2F_{1}\!\left[\genfrac{}{}{0pt}{1}{1,2-2\Delta}{2-\Delta};\sfrac{1-u}{2}\right],\quad F_{3}(u) = {}_2F_1\!\left[\genfrac{}{}{0pt}{1}{1,1-2\Delta}{3/2-\Delta};\sfrac{1-u}{2}\right],
\end{equation}
and more generally, for $s>0$, replacing $\Delta$ by $\Delta_s = \Delta -s-1/2$, we introduce  
\begin{equation} \label{2f1gen1}
    F_{1,s}(u) = {}_2F_{1}\!\left[\genfrac{}{}{0pt}{1}{2,2+2s-2\Delta}{5/2+s-\Delta};\sfrac{1-u}{2}\right],\quad F_{2,s}(u) = {}_2F_{1}\!\left[\genfrac{}{}{0pt}{1}{1,3+2s-2\Delta}{5/2+s-\Delta};\sfrac{1-u}{2}\right],
\end{equation}
\begin{equation}\label{2f1gen2}
    F_{3,s}(u) = F_{2,s-\frac{1}{2}}(u) = {}_2F_1\!\left[\genfrac{}{}{0pt}{1}{1,2+2s-2\Delta}{2+s-\Delta};\sfrac{1-u}{2}\right],
\end{equation}
where $u = (x_{21}\cdot x_{20})/|x_{21}||x_{20}|$. We also remark that
\begin{equation}
F_{i,s}(1) = F_{1,s}(-1) = 1\,, \qquad F_{2,s}(-1) = F_{3,s}(-1) = -1\, ,
\end{equation}
and that
\begin{equation}\label{F2eqdiff}
    (1-u^2)F'_{2,s}(u) = (2\Delta-2s-3)[1-u\,F_{2,s}(u)]\, ,
\end{equation}
\begin{equation}\label{F2eqdiff2}
    F_{1,s}(u) = 3+2s-2\Delta + (2\Delta-2s-2)u\, F_{2,s}(u)\, .
\end{equation}

\subsection{$J_0$ Correlators}
\label{app:J0corr}

\subsubsection{Unequal Sign}

When $\bar s + s\geqslant 0$, and $s<0$
\begin{equation}
    \langle \mathcal{M}_{10}^{(\bar{s},s)} J_0(x_2) \rangle = \frac{2^{\bar{s}-\frac{3}{2}}}{\Delta - 1} (\Delta+\sfrac{1}{2})_{\bar s-\frac{1}{2}} (\Delta-1)_{-s-\frac{1}{2}} \frac{d_0\, \eta_-\, c_+ (x_2^-)^{\bar{s}+s} F_{1,s}(u)}{|x_{10}|^{2\Delta-2s-3} |x_{20}|^{2} |x_{21}|^{2(1+\bar{s}+s)}}\,.
\end{equation}
Note that in the case $\bar s + s = 0$, the limits $u\to 1^-$ and $u\to -1^+$ both give
\begin{equation}
    \lim_{r\to 0}\langle \mathcal{M}_{10}^{(\bar s,-\bar s)} J_0(x_2) \rangle = \frac{2^{\bar{s}-\frac{3}{2}}}{\Delta - 1} (\Delta+\sfrac{1}{2})_{\bar s-\frac{1}{2}} (\Delta-1)_{\bar{s}-\frac{1}{2}} \frac{d_0\, \eta_-\, c_+ }{|x_{10}|^{2\Delta+2\bar{s}-3} |x_{20}|^{2} |x_{21}|^{2}}\,.
\end{equation}

\subsubsection{Equal Sign}

\begin{align}
    \langle \mathcal{M}_{10}^{\left(\bar s<0,s<0\right)} J_0(x_2)\rangle &= d_0\, 2^{1-\bar s - s} \left(\sfrac{3}{2}-\Delta\right)_{\frac{1}{2}-\bar s} \left(\Delta-\sfrac{1}{2}\right)_{\frac{1}{2}-s} \frac{(x^+_2)^{-\bar s-s}|x_{10}|}{|x_{20}|^{2\Delta-2 s} |x_{21}|^{4-2\Delta-2\bar s}}\\
    &= \frac{d_0}{x_2^+} \p^{\frac{1}{2}-\bar s}_{x^-_1} \p^{\frac{1}{2}-s}_{x^-_0} \frac{|x_{10}|}{|x_{20}|^{2\Delta-1} |x_{21}|^{3-2\Delta}}\, ,
\end{align}

\begin{align}
    \langle \mathcal{M}_{10}^{\left(\bar s>0,s>0\right)} J_0(x_2)\rangle &= \frac{d_0\, \eta_-}{\eta_+} 2^{1+\bar s + s} \left(\sfrac{3}{2}-\Delta\right)_{\frac{1}{2}+s} \left(\Delta-\sfrac{1}{2}\right)_{\frac{1}{2}+\bar s} \frac{(x^-_2)^{\bar s+s}|x_{10}|}{|x_{20}|^{4-2\Delta+2 s} |x_{21}|^{2\Delta+2\bar s}}\\
        &= \frac{d_0\, \eta_-}{\eta_+\,x_2^-} \p^{\frac{1}{2}+\bar s}_{x^+_1} \p^{\frac{1}{2}+s}_{x^+_0} \frac{|x_{10}|}{|x_{20}|^{3-2\Delta} |x_{21}|^{2\Delta-1}}\, .
\end{align}

\subsection{$J_1$ Correlators}
\label{app:J1corr}

\subsubsection{Equal Sign}

The correlators with $J_1^+$ are
\begin{multline}
    \langle\mathcal{M}_{10}^{(\bar{s}<0,s<0)} J_1^+ (x_2) \rangle = \p_{x_1^-}^{-\bar{s}-\frac{1}{2}} \p_{x_0^-}^{-s-\frac{1}{2}} \bigg[\tilde{d}_1 \left(\p_{x_0^-} -\p_{x_1^-}\right) \frac{(\Delta-1) x_2^+}{|x_{20}|^{2\Delta} |x_{21}|^{4-2\Delta}}\\
    + d_1\p_{x_0^-}\p_{x_1^-} \frac{1}{|x_{20}|^{2\Delta-1} |x_{21}|^{3-2\Delta}}\bigg]\, ,
\end{multline}
\begin{multline}
    \langle\mathcal{M}_{10}^{(\bar{s}>0,s>0)} J_1^+(x_2) \rangle = \frac{\eta_-}{2\,\eta_+} \p_{x_1^+}^{\bar{s}-\frac{1}{2}} \p_{x_0^+}^{s-\frac{1}{2}} \bigg[\tilde{d}_1 \! \left(\p_{x_1^3}\frac{(\Delta-1) x_{21}^{3}}{|x_{20}|^{4-2\Delta} |x_{21}|^{2\Delta}}-\p_{x_0^3}\frac{(\Delta-1) x_{20}^3}{|x_{20}|^{4-2\Delta} |x_{21}|^{2\Delta}}\right)\\
    + d_1\p_{x_0^3} \p_{x_1^3}\frac{1}{|x_{20}|^{3-2\Delta} |x_{21}|^{2\Delta-1}}\bigg]\, .
\end{multline}
When $\bar s = s = 1/2$ and $x_2$ goes to the line, this becomes
\begin{equation}
    \lim_{\substack{r\to 0 \\ u\to \pm 1}}\langle\mathcal{M}_{10}^{(\frac{1}{2},\frac{1}{2})} J_1^+(x_2) \rangle = \frac{\eta_-\left[4\,\tilde{d}_1(\Delta-1)^2 \pm d_1 (3-2\Delta)(2\Delta-1)\right]}{2\,\eta_+\,|x_{20}|^{4-2\Delta} |x_{21}|^{2\Delta}}\, .
\end{equation}

The correlators with $J_1^-$ are
\begin{multline}
    \langle\mathcal{M}_{10}^{(\bar{s}<0,s<0)} J_1^- (x_2)\rangle = -\frac{1}{2}\p_{x_{1}^{-}}^{-\bar{s}-\frac{1}{2}}\p_{x_{0}^{-}}^{-s-\frac{1}{2}}\bigg[\tilde{d}_{1}\!\left(\!\p_{x_0^3}\frac{(\Delta-1) x_{20}^3}{|x_{20}|^{2\Delta} |x_{21}|^{4-2\Delta}}-\p_{x_1^3}\frac{(\Delta-1) x_{21}^3}{|x_{20}|^{2\Delta} |x_{21}|^{4-2\Delta}}\right)\\
    + d_{1}\p_{x_0^3}\p_{x_1^3}\frac{1}{|x_{20}|^{2\Delta-1} |x_{21}|^{3-2\Delta}}\bigg]\, ,
\end{multline}
which becomes, when $\bar s = s = -1/2$, and $x_2$ goes to the line,
\begin{equation}
    \lim_{\substack{r\to 0 \\ u\to \pm 1}}\langle\mathcal{M}_{10}^{(-\frac{1}{2},-\frac{1}{2})} J_1^-(x_2) \rangle = -\frac{4\,\tilde{d}_1(\Delta-1)^2 \pm d_1 (3-2\Delta)(2\Delta-1)}{2\,|x_{20}|^{2\Delta} |x_{21}|^{4-2\Delta} }\, ,
\end{equation}
and
\begin{multline}
    \langle\mathcal{M}_{10}^{(\bar{s}>0,s>0)} J_1^-(x_2) \rangle = \frac{\eta_-}{\eta_+} \p_{x_1^+}^{\bar{s}-\frac{1}{2}} \p_{x_0^+}^{s-\frac{1}{2}} \bigg[\tilde{d}_1 \left(\p_{x_{0}^{+}}-\p_{x_{1}^{+}}\right) \frac{(\Delta-1) x_2^-}{|x_{20}|^{4-2\Delta} |x_{21}|^{2\Delta}}\\
    - d_1 \p_{x_0^+}\p_{x_1^+}\frac{1}{|x_{20}|^{3-2\Delta} |x_{21}|^{2\Delta-1}}\bigg]\, .
\end{multline}

For the third component of $J_1$, the correlators are
\begin{multline}
    \langle \mathcal{M}_{10}^{\left(\bar s<0,s<0\right)} J_1^3(x_2)\rangle = \p^{-\bar s-\frac{1}{2}}_{x^-_1} \p^{-s-\frac{1}{2}}_{x^-_0}\bigg[\tilde{d}_1 \left(\!\p_{x_0^3} - \p_{x_1^3}\!\right) \frac{(\Delta-1) x_2^+}{|x_{20}|^{2\Delta} |x_{21}|^{4-2\Delta}}\\
    + \frac{d_1}{2}(\p_{x_0^3}\p_{x_1^-} + \p_{x_1^3}\p_{x_0^-}) \frac{1}{|x_{20}|^{2\Delta-1} |x_{21}|^{3-2\Delta}} \bigg]\, ,
\end{multline}
\begin{multline} \label{ssbarpJ13}
    \langle \mathcal{M}_{10}^{\left(\bar s>0,s>0\right)} J_1^3(x_2)\rangle = \frac{\eta_-}{\eta_+} \p^{\bar s-\frac{1}{2}}_{x^+_1} \p^{s-\frac{1}{2}}_{x^+_0}\bigg[\tilde{d}_1 \left(\!\p_{x_0^3} - \p_{x_1^3}\!\right) \frac{(\Delta-1) x_2^-}{|x_{20}|^{4-2\Delta} |x_{21}|^{2\Delta}}\\
    - \frac{d_1}{2}(\p_{x_0^3}\p_{x_1^+} + \p_{x_1^3}\p_{x_0^+}) \frac{1}{|x_{20}|^{3-2\Delta} |x_{21}|^{2\Delta-1}} \bigg]\, .
\end{multline}

\subsubsection{Unequal Sign}

For $\bar s + s\geqslant 0$ and $s<0$, we have
\begin{multline}\label{J1uneq}
    \frac{\langle\cM_{10}^{(\bar s,s)} J_1^3(x_2)\rangle}{\eta_-\, c_+\, C_{\bar s,s} (x_2^-)^{\bar s+s}} = - \frac{d_1 (\Delta_s+\sfrac{1}{2})_{\bar s+s}}{2(\Delta_s-1) |x_{10}| |x_{21}|^{2(\bar s+s)}} \left(\!\p_{x_0^3} - \p_{x_1^3}\!\right) \frac{F_{2,s}(u)}{|x_{10}|^{2(\Delta-s)-3} |x_{20}| |x_{21}|}\\
    + \frac{\tilde{d}_1 (\Delta_s)_{\bar s+s}}{2\Delta_s-1} \left(\!\p_{x_0^3} - \p_{x_1^3}\!\right) \frac{F_{3,s}(u)}{|x_{10}|^{2(\Delta-s-1)} |x_{20}| |x_{21}|^{1+2(\bar s+s)}}\\
    + \frac{2(d_1 (\Delta_s+\sfrac{1}{2})_{\bar s+s} - \tilde{d}_1 (\Delta_s)_{\bar s+s})}{|x_{10}|^{2(\Delta-s)-1} |x_{21}|^{2(1+\bar s+s)}} + \delta_{\bar s,-s} \frac{d_1 - \tilde{d}_1}{|x_{10}|^{2(\Delta - s)-1}} \left(\frac{1}{|x_{20}|^2} - \frac{1}{|x_{21}|^2}\right)\, ,
\end{multline}
where we recall that $\Delta_s = \Delta - s -1/2$ for $s<0$, and we introduced
\begin{equation}
    C_{\bar s,s} = 2^{\bar s+2s-\frac{1}{2}} (2\Delta)_{-1-2s}\, .
\end{equation}
When $\bar s + s = 0$ and $x_2$ goes to the line, this correlator reduces to 
\begin{equation}
    \lim_{r\to 0} \langle\cM_{10}^{(- s,s)} J_1^3(x_2)\rangle = \frac{2^{s-\frac{1}{2}}\, \eta_-\, c_+\,(1-2\Delta_s)_{-1-2s}}{|x_{10}|^{2(\Delta-s)-3} |x_{20}|^2 |x_{21}|^2} \left[\frac{d_1 (2\Delta_s-1)}{2(\Delta_s-1)} + \frac{2\,\tilde{d}_1 \Delta_s}{1-2\Delta_s}\right]\, .
\end{equation}
Finally, we also recall the Ward identity
\begin{equation}\label{contact J1}
    \p_{x_2^\mu}\langle\cM_{10}^{(\bar s,s)} J_1^\mu(x_2)\rangle = \delta_{\bar s,-s}(\delta^{(3)}(x_{21}) - \delta^{(3)}(x_{20})) \langle\cM_{10}^{(\bar s,s)}\rangle\, ,
\end{equation}
which reflects our choice of a canonical normalisation for $J_1$.

\subsection{$J_2$ Correlators}
\label{app:J2corr}

Following \cite{Ferrando:2025ufj}, let us define the following three conformal structures:
\begin{equation}\label{conf structures}
    Q^\mu_1=\frac{x^\mu_{21}}{|x_{21}|^2} - \frac{x^\mu_{20}}{|x_{20}|^2}\,,\quad\quad Q^\mu_2= \frac{r^\mu}{r^2} - \frac{x^\mu_{20}}{|x_{20}|^2} - \frac{x^\mu_{21}}{|x_{21}|^2}\,,\quad\quad Q^\mu_3 = \epsilon^{\mu\nu 3} \frac{x_{2,\nu}}{r^2}\,,
\end{equation}
where $r^\mu = (x^1_2,x^2_2,0)$. We also let 
\begin{equation}
    \mathcal{Q}_{ij}^{\mu\nu}\equiv\frac{1}{2}\left(Q_{i}^{\mu}Q_{j}^{\nu}+Q_{i}^{\nu}Q_{j}^{\mu}\right)-\frac{1}{3}\delta^{\mu\nu}Q_{i}^{\rho}Q_{j\rho}\, .
\end{equation}
Conformal covariance then implies that the correlators with the stress-tensor are of the form
\begin{multline}\label{j2corr}
    \<\cM_{10}^{(\bar{s},s)}J^{\mu \nu}_{2}(x_2)\>  = \frac{\left(\!\frac{x_{2}^{\pm}|x_{10}|}{|x_{20}||x_{21}|}\!\right)^{|\bar{s}+s|}}{|x_{10}|^{\Delta_{-\bar{s}}+\Delta_{s}-1}|x_{20}|^{1+\Delta_{s}-\Delta_{-\bar{s}}}|x_{21}|^{1+\Delta_{-\bar{s}}-\Delta_{s}}}\\
    \times \left[h_{\bar{s},s,2}^{(0)}(u) \mathcal{Q}_{22}^{\mu\nu}\! + h_{\bar{s},s,2}^{(1)}(u) \mathcal{Q}_{12}^{\mu\nu}\! + h_{\bar{s},s,2}^{(2)}(u) \mathcal{Q}_{11}^{\mu\nu}\! + k_{\bar{s},s,2}^{(0)}(u) \mathcal{Q}_{23}^{\mu\nu}\! + k_{\bar{s},s,2}^{(1)}(u) \mathcal{Q}_{13}^{\mu\nu} \right]\, ,
\end{multline}
where $\pm = -\operatorname{sign}(\bar s +s)$, and the five functions of $u$ are not constrained by conformal symmetry. The completely explicit form of these correlators was only written in \cite{Ferrando:2025ufj} for $\bar s$ and $s$ negative and $\mu=\nu=+$, for which we found a simple, compact expression. This is sufficient to uniquely determine the five functions, so it effectively gives all correlators with $\bar s$ and $s$ negative. In Appendix \ref{app:coeffs}, we obtain the cross-ratio functions in the case $\bar s + s>0$ and $s<0$ as a by-product of our investigation of the double-trace terms generated by the action of the pseudo-charge on the boundary operators. We present these functions here but keep the computations in Appendix \ref{app:coeffs} for ease of reading. They are:
\begin{align}\label{J2h0}
    \frac{h_{\bar{s},s,2}^{(0)}(u)}{\eta_-\,c_+\,d_2\,C_{\bar s,s}} &= \frac{2}{u}\left[(3\Delta_{-\bar s}-\sfrac{1}{2})(\Delta_s)_{\bar s+s} - 2(\Delta_s+\Delta_{-\bar s}-1) (\Delta_s+\sfrac{1}{2})_{\bar s + s}\right]\\
    &+ 2(2-\Delta_s-\Delta_{-\bar s})\frac{ (\Delta_s-\sfrac{1}{2})_{\bar s + s + 1}(1-u^2)}{(\Delta_s-1)u^2} F_{2,s}(u) - \frac{(\Delta_s)_{\bar s+s} F_{3,s}(u)}{u^2(2\Delta_s - 1)}\nonumber\\
    &\times \left[(\Delta_s+\Delta_{-\bar s}-1) ((\Delta_s+\Delta_{-\bar s})(u^2-2) + 3) + (1-2
    u^2) (s+\bar{s})^2\right]\nonumber\, ,\\
    \label{J2h1}
    \frac{h_{\bar{s},s,2}^{(1)}(u)}{\eta_-\,c_+\,d_2\,C_{\bar s,s}} &= \frac{2}{u}\left[ (3\Delta_s+\Delta_{-\bar s}) (\Delta_s+\sfrac{1}{2})_{\bar s + s} - (\Delta_s+3\Delta_{-\bar s})(\Delta_s)_{\bar s+s} \right]\\
    &+ 2(\bar s+s)\bigg[ \frac{(\Delta_s-\sfrac{1}{2})_{\bar s + s + 1}}{\Delta_s-1} \, F_{2,s}(u) - (\Delta_s+\Delta_{-\bar s}+1) \frac{(\Delta_s)_{\bar s+s}}{2\Delta_s-1} F_{3,s}(u)\bigg]\, , \nonumber\\
    \label{J2h2}
    \frac{h_{\bar{s},s,2}^{(2)}(u)}{\eta_-\,c_+\,d_2\,C_{\bar s,s}} &= \frac{u}{1-u^2}\left[(\Delta_s)_{\bar s+s} - 2(\Delta_s+\Delta_{-\bar s}-1) (\Delta_s+\sfrac{1}{2})_{\bar s + s} \right]\\
    &+ 2(1-2\Delta_s-2\Delta_{-\bar s}) \frac{(\Delta_s-\sfrac{1}{2})_{\bar s + s + 1}}{\Delta_s-1} \, F_{2,s}(u) - \frac{(\Delta_s)_{\bar s+s} F_{3,s}(u)}{(2\Delta_s-1)(1-u^2)} \nonumber\\
    &\times \left[3(2\Delta_s+2\Delta_{-\bar s}-1)+(u^2-2)((\Delta_s+\Delta_{-\bar s})(\Delta_s+\Delta_{-\bar s}+1)+4\Delta_s \Delta_{-\bar s})\right]\, ,\nonumber\\
    \frac{k_{\bar{s},s,2}^{(0)}(u)}{\eta_-\,c_+\,d_2\,C_{\bar s,s}} &= \frac{2\ii}{u}[(\Delta_{s}+3\Delta_{-\bar{s}}-2)(\Delta_{s})_{\bar{s}+s} - (3\Delta_{s}+\Delta_{-\bar{s}}-2)(\Delta_{s}+\sfrac{1}{2})_{\bar{s}+s}]\\
    &+2\ii(s+\bar{s}) \bigg[-u(\Delta_{s}+\sfrac{1}{2})_{\bar{s}+s} + \frac{(\Delta_{s}-\frac{1}{2})_{\bar{s}+s+1}}{1-\Delta_{s}}(1-u^{2})F_{2,s}(u) \nonumber\\
    &+(2-\Delta_{-\bar{s}}-\Delta_{s})\frac{(\Delta_{s})_{\bar{s}+s}}{1-2\Delta_{s}}F_{3,s}(u)\bigg]\,,\nonumber\\
    \frac{k_{\bar{s},s,2}^{(1)}(u)}{\eta_-\,c_+\,d_2\,C_{\bar s,s}} &=2\ii  \bigg[(\Delta_{-\bar{s}}+\Delta_{s}+1)(\Delta_{s}+\sfrac{1}{2})_{\bar{s}+s}u\\
    &+ ((2-u^{2})(\Delta_{s}+\Delta_{-\bar{s}}+1)-3)\frac{(\Delta_{s}-\frac{1}{2})_{\bar{s}+s+1}}{\Delta_{s}-1}F_{2,s}(u) \nonumber\\
    &- (4\Delta_{-\bar{s}}+\frac{\bar s+ s+2 }{2\Delta_{s}-1}(\bar{s}+s)) (\Delta_{s})_{\bar{s}+s} F_{3,s}(u)\bigg]\,.\nonumber
\end{align}

\subsection{Relations Between Constants} \label{app:relations}

For the reader's convenience, we recall here some useful relations between the various constants derived in \cite{Ferrando:2025ufj}:
\begin{equation}\label{old rel norm}
   16\cos^2(\pi\Delta)\frac{d^2_0}{\mathcal{N}_0} = \frac{1}{\mathcal{N}_1}\left[d^2_1 \cos^2(\pi\Delta)  + \tilde{d}^2_1 \sin^2(\pi\Delta) \right] = \frac{d^2_s}{\mathcal{N}_s}\, ,\quad s\geqslant 2\,,
\end{equation}
\begin{equation} \label{e01}
     e_{0,1}  = \frac{q\, \mathcal{N}_0 \, d_1}{8\,\mathcal{N}_1 \,d_0}\,,\quad e_{1,2}= \frac{q}{2\,d_2} \left(d_1 \cos^2(\pi\Delta) + \tilde{d}_1 \sin^2(\pi\Delta)\right)\,,\quad e_{2,3}=\frac{q\, d_2}{2\,d_3}\,.
\end{equation}
The parameter $\nu$ is also encoded in $\eta_-(\tilde{d}_1 - d_1) = 2\nu$. The coefficients appearing in the divergence of $J_3$ are
\begin{equation}\label{a1a2}
    a_1 = -\frac{q\,\tilde{d}_1\tan(\pi\Delta)}{40\pi^2\mathcal{N}_1\, d_0}\, ,\qquad a_2 = \ii\frac{q\,\nu\sin(2\pi\Delta)}{8\pi^2\mathcal{N}_1\, d_2\,\eta_-}\, .
\end{equation}

If the currents $J_1$ and $J_2$ are canonically normalised, then $e_{2,3} = 2$ and
\begin{equation}
    16\,\mathcal{N}_{2}=\frac{\mathcal{N}_{1}}{1+4\pi^4\nu^2}\, ,\quad \eta_- d_1 = \frac{\sin(2\pi\Delta)}{2\pi^{2}}-2\nu\sin^2(\pi\Delta)\, , \quad \eta_{-}d_2 = -\frac{\sin(2\pi\Delta)}{8\pi^2}\, .
\end{equation}
Furthermore, we showed in Section \ref{sec:equal-equal} that if $N$ is defined as the expectation value of the infinite line defect, see \eqref{def N}, then \eqref{deta/N} holds, which means that
\begin{equation}
    \mathcal{N}_1 = 2^{-5} \pi^{-2} + O(\sfrac{1}{N}) \quad\text{if}\quad\nu=0\, .
\end{equation}

\section{Action of the Pseudo-Charge}
\label{app:coeffs}

We present in this appendix details on how to fix the $O(1/N)$ double-trace coefficients appearing in \eqref{QonBoundary}. These double-trace terms can \emph{a priori} involve any bulk operator $J_s$. Below we compute the coefficients associated with $s=0$ or $1$. 

We begin by resolving the action of the pseudo-charge $Q^{(3)}_{33}$ on local bulk operators. We then use it to bootstrap the double-trace coefficients. The procedure also yields correlation functions with $J_2$ that were not computed in \cite{Ferrando:2025ufj}, see the results in Appendix \ref{app:J2corr}. We then finish by showing that $\mathbb{D}_{33}\cW=O(1/N^2)$, which is an important input in Section \ref{sec:bootstrap2lines}.

\subsection{Local Operators}

In this section we fix the action of the pseudo-charge $Q^{(3)}_{\mu\nu}$ on some of the local operators. The idea is to write all terms with the correct dimension, spin and index structure and then constrain their coefficients using the action of the pseudo-charge on two-point functions of local operators. Below we showcase the results.

For the scalar operator $J_0$ we have
\begin{equation}\label{QJ0gebra}
    [Q^{(3)}_{\mu\nu},J_0] = \ii\, e_{0,1}\,  \epsilon_{\alpha\beta(\mu}\p_{\nu)}\p^\alpha J_1^\beta\, ,
\end{equation}
and for $J^\mu_1$ we have
\begin{equation} \label{actionQJ1}
    [Q^{(3)}_{\mu\nu},J_{1,\rho}] = e_{1,2}\, \p_{(\mu} J_{2,\nu)\rho} + 2\ii\, e_{1,0}\,  \epsilon_{\alpha\rho(\mu} \p_{\nu)}\p^\alpha J_0 + O(\sfrac{1}{N})\, ,
\end{equation}
where the explicit factor of $2\ii$ is such that $e_{1,0}\, \mathcal{N}_0 = e_{0,1}\, \mathcal{N}_1 $. We also verified that necessarily
\begin{multline}\label{q3j2}
    [Q^{(3)}_{\mu\nu},J_{2,\rho_1\rho_2}] = e_{2,3}\, \p_{(\mu} J_{3,\nu)\rho_1\rho_2} + e_{2,1}\Big[2\p_\mu\p_\nu \p_{(\rho_1}J_{1,\rho_2)} - \p_{\rho_1}\p_{\rho_2} \p_{(\mu}J_{1,\nu)} \\
    + \eta_{\rho_1\rho_2} \square \p_{(\mu}J_{1,\nu)}- 2\eta_{(\rho_1(\mu} \square \p_{\nu)} J_{1,\rho_2)}\Big]+  O(\sfrac{1}{N})\, .
\end{multline}

\subsection{Leading Order of the Boundary Equation}
\label{app:Qaction}

Let us show that the action of the pseudo-charge on the boundary operators is given by~\eqref{QonBoundary} at leading order in the large-$N$ limit. We have to show that the constant $q$ is independent of the spin of the boundary operator. We consider the action of $Q_{33}^{(3)}$ on $\mathcal{M}_{10}^{(\bar{s},s)}J_{0}(x_{2})$
with $\bar{s},s>0$ and arrive at the following equation
\begin{multline}\label{qeq}
\left(\bar q_{\bar{s}}\p_{x_{1}^{3}}^{2}+q_{s}\p_{x_{0}^{3}}^{2}\right) \langle\mathcal{M}_{10}^{(\bar{s},s)} J_0(x_{2}) \rangle + e_{0,1}\langle\mathcal{M}_{10}^{(\bar{s},s)} \p_3(\p_+J_1^+ - \p_-J_1^-)(x_2) \rangle\\
= -\int \langle\mathcal{M}_{10}^{(\bar{s},s)} J_0(x_2) \p^\mu J_{3,\mu33}(x_3)\rangle \dd^3 x_3\,,
\end{multline}
where we used \eqref{QJ0gebra}. Notice that because we chose $\bar s, s>0$, double-traces contributions at the boundary or insertions on the line cannot contribute in the planar limit. They would indeed produce factorised terms that include some $\langle\mathcal{M}_{10}^{(\bar{s}',s')}\rangle = 0$ since $\bar{s}'+s'\neq0$. Inserting then the divergence of $J_3$ from \eqref{divJ3} and the explicit expression for the correlators, see Appendix \ref{app:corr}, we can compute the right-hand side using the generalisation \eqref{2F1 integral} of the star-triangle relation. The equation is then verified provided
\begin{equation}
   q_s = -\bar q_{\bar s}= -\frac{40 \pi^2 \mathcal{N}_1\,a_1\, d_0}{\tilde{d}_1 \tan(\pi \Delta)}\,,
\end{equation}
where we used the known expression for $e_{0,1}$ from \eqref{e01}. In addition, comparing with notation from \cite{Ferrando:2025ufj}, see \eqref{a1a2}, it is consistent to write $q_s=q$. 

The coefficients associated to operators of negative spin are then straightforward to obtain by requiring that $Q_{33}^{(3)}\<\cM^{(\bar s>0, s<0)}_{10}\>=0$. We thus proved \eqref{QonBoundary}.

\subsection{Bootstrapping $:\!J_0\cO:$ Coefficients}
\label{app:J0dt}

The purpose of this subsection is to probe the double-trace terms in \eqref{QonBoundary} that involve the operator $J_0$. For $\bar s>0$, the relevant $1/N$ part of the action on the anti-fundamental boundary operator takes the form 
\begin{equation}\label{QonBoundary2}
    \left[Q^{(3)}_{33},\bcO_{\bar s}(x)\right] = -q\, \delta^2_{3}\overline\cO_{\bar s} + \sum_{k=1/2}^{\bar s}\frac{\overline\omega_{\bar s,k}}{N} :\!\p^{\bar s -k}_+ J_0\, \overline\cO_k\!: +\,  O(\sfrac{1}{N})\,,
\end{equation}
where $O(1/N)$ now denotes contributions involving $J_{\tilde s>0}$, which do not affect the computations in this subsection. A similar equation can be written for the action on the fundamental boundary operator.

We consider the action of $Q_{33}^{(3)}$ on $\cM^{(\bar s, s)}_{10}J_0(x_2)$ for $\bar s + s\geqslant 0$ and $s<0$. This results in the following equation
\begin{align}
    q\left(\p_{x_0^3}^2 -\p_{x_1^3}^2\right) &\langle\mathcal{M}_{10}^{(\bar{s},s)} J_0(x_2)\rangle + \frac{\overline{\omega}_{\bar{s},-s}}{N} \langle\mathcal{M}_{10}^{(-s,s)}\rangle \langle\p_+^{\bar{s}+s} J_0(x_1) J_0(x_2)\rangle \nonumber\\
    &+ \delta_{\bar{s},-s} \frac{\omega_{s,s}}{N} \langle\mathcal{M}_{10}^{(\bar s,s)}\rangle \langle J_0(x_0) J_0(x_2)\rangle + e_{0,1} \langle\mathcal{M}_{10}^{\left(\bar{s},s\right)}\p_3(\p_+J_1^+-\p_-J_1^-)(x_2)\rangle \nonumber\\
    &= - 5 \frac{a_1}{N} \p_{x_2^3} \int \langle\mathcal{M}_{10}^{(\bar{s},s)} J_1^3(x_3)\rangle \langle J_0(x_3) J_0(x_2)\rangle\, \dd^3 x_3\nonumber\\
    &+ \delta_{\bar s,-s} \frac{a_1}{N}\langle\mathcal{M}_{10}^{(\bar s,s)}\rangle(\langle J_0(x_1) J_0(x_2)\rangle - \langle J_0(x_0) J_0(x_2)\rangle)\, .
\end{align}
Several comments are in order. First, due to our choice of spins, the double-trace terms in the action on the fundamental boundary operator can only contribute if $\bar s+s =0$. Second, we did not include an insertion of $\p_3 J_0$ on the line as this can be absorbed in a redefinition of the double-trace coefficients. Third, the last line occurs because, when $\bar s+s =0$, the correlator $\langle\mathcal{M}^{(\bar{s},s)} \p_\mu J_1^\mu\rangle$ contains contact terms, see \eqref{contact J1}. Indeed the $J_1^\mu \p_\mu J_0$ term from the divergence of $J_3$, which usually does not give any contribution, produces here
\begin{align}\label{boundary RHS}
    -\frac{a_1}{N} \bigg[\int_{\mathcal{D}_\epsilon} \langle\mathcal{M}_{10}^{(\bar{s},s)} J_1^\mu(x_3)\rangle &\langle \p_\mu J_0(x_3) J_0(x_2)\rangle\, \dd^3 x_3\bigg]_{O(\epsilon^0)} \nonumber\\
    &= \frac{a_1}{N} \lim_{\epsilon\to 0}\int_{r=\epsilon} \langle\mathcal{M}_{10}^{(\bar{s},s)} J_1^\mu(x_3)\rangle \frac{r_\mu}{r} \langle J_0(x_3) J_0(x_2)\rangle\, \dd^2 x_3 \nonumber\\
    &= \delta_{\bar s,-s} \frac{a_1}{N}\langle\mathcal{M}_{10}^{(\bar s,s)}\rangle(\langle J_0(x_1) J_0(x_2)\rangle - \langle J_0(x_0) J_0(x_2)\rangle)\, .
\end{align}
The initial integration region is $\mathcal{D}_\epsilon = \{r>\epsilon\}\setminus \mathcal{B}_\epsilon(x_2)$. This example also explains why we chose the integration region to exclude the whole cylinder $\{r<\epsilon\}$ and not just some neighbourhood of the mesonic line, see Figure \ref{fig:int domain}. The boundary terms are indeed simpler to write down since we do not have to worry about integrating by parts for integrals over~$x_3^3$.

Plugging in correlators from Appendix \ref{app:corr} and using the integral relation \eqref{2F1 integral}, we find
\begin{equation}
    \omega_{s,s} = - \overline{\omega}_{-s, -s} = -a_1-2\nu\frac{e_{0,1}}{\mathcal{N}_0}\, ,
\end{equation}
and
\begin{equation}
    \omega_{s,k} = 0 \quad \text{if} \quad k \neq s\, ,\quad \overline{\omega}_{\bar s,k} = 0 \quad \text{if} \quad k \neq \bar s\, .
\end{equation}

A computation similar to \eqref{boundary RHS}, would also give rise to factorised contributions to the right-hand side of \eqref{variation uu}. However, the left-hand side would also receive a contribution from the double-trace terms with coefficients $\overline{\omega}_{\frac{1}{2},\frac{1}{2}}$ and $\omega_{-\frac{1}{2},-\frac{1}{2}}$. And it is clear that the $a_1$ part of these coefficients would simply reproduce the factorised contribution in the right-hand side. Practically, this means that when $\nu=0$, we can just ignore both the $\omega$ coefficients and the $J_1^\mu \p_\mu J_0$ term from the divergence of $J_3$.

\subsection{Bootstrapping $:\!J_1\cO:$ Coefficients}
\label{app:J1dt}

We compute here the coefficients of the double-trace terms involving the conserved current~$J_1$ in the action of the pseudo-charge $Q_{33}^{(3)}$ on the boundary operators. The relevant $1/N$ part of \eqref{QonBoundary} takes the form
\begin{multline}
\left[Q_{33}^{(3)},\overline{\mathcal{O}}_{\bar{s}}\right] = \sum_{k=1/2}^{\bar s}\frac{\bar\rho_{\bar{s},k}^{(1)}}{N} :\!\left(\p_{+}^{\bar{s}-k}J_{1}^{3}\right)\bcO_{k}:\\
+\sum_{k=1/2}^{\bar{s} -1}\bigg[\frac{\bar\rho_{\bar{s},k}^{(2)}}{N} :\!\left(\p_{+}^{\bar{s}-k-1}\p_{3}J_{1}^{-}\right)\bcO_{k}: + \frac{\bar\rho_{\bar{s},k}^{(3)}}{N} :\!\left(\p_{+}^{\bar{s}-k-1}J_{1}^{-}\right)\delta_{3}\bcO_{k}:\bigg] + \dots\,,
\end{multline}
with a similar equation for the fundamental boundary operator. The computations of this subsection are done for $\nu=a_2=0$.

We consider mesonic lines with $\bar s+ s\geqslant0$ and $s <0$. For simplicity, we first start with the action of $Q_{33}^{(3)}$ on $\<\cM^{(\bar s, s)}_{10}J^3_1(x_2)\>$ with $\bar s=-s$ and find the equation
\begin{multline}\label{Q3J1eq}
    q\left(\p_{x_{0}^{3}}^{2}-\p_{x_{1}^{3}}^{2}\right)\langle\cM_{10}^{(-s,s)}J^3_1 (x_2) \rangle + \frac{\langle\cM_{10}^{(-s,s)}\rangle}{N} \left(\bar{\rho}^{(1)}_{-s,-s}\langle J^3_1(x_1) J^3_1(x_2)\rangle + \rho_{s,s}^{(1)} \langle J^3_1(x_0) J^3_1(x_1)\rangle\right)\\
    + e_{1,2}\, \p_{x^3_2} \langle\cM_{10}^{(-s,s)} J^{33}_2(x_2)\rangle = 5 \frac{a_1}{N}\p_{x_2^3} \int \langle\cM_{10}^{(-s,s)} J_0(x_3)\rangle \langle J_1^3(x_3) J_1^3(x_2)\rangle \dd^3 x_3\,.
\end{multline}
Notice that we did not include terms associated to the action of the pseudo-charge on the line: any contribution from $\p_3 J^3_1=-\p_+ J^+_1-\p_- J^-_1$ can be absorbed in a redefinition of the coefficients $\rho^{(1)}$, whereas $\p_+ J^+_1-\p_- J^-_1$ is not probed at this order in the planar limit since $\langle(\p_+ J^+_1-\p_- J^-_1)(x) J^3_1(y)\rangle = 0$ up to contact terms. The correlators involving $J_0$ and $J^3_1$ are given in Appendix~\ref{app:J0corr} and Appendix~\ref{app:J1corr} respectively. The integral in the right-hand side can be computed using \eqref{2F1 integral}. Restricting first \eqref{Q3J1eq} to $x^\pm_2 = 0$ and using the fact that the kinematical dependence of the three-point function $\langle\cM_{10}^{(-s,s)} J^{33}_2(x_2)\rangle$ is fixed by conformal symmetry is enough to show that
\begin{equation}
   \rho^{(1)}_{s,s} = \bar\rho^{(1)}_{-s,-s} = 0\, .
\end{equation}

Next, we consider the situation $\bar s +s >0$ and arrive at the equation
\begin{multline}\label{Q3J1uneq}
    q \left(\p_{x^3_0}^2-\p_{x^3_1}^2\right) \langle\cM_{10}^{(\bar{s},s)} J^3_1(x_2)\rangle + \frac{\bar{\rho}_{\bar{s},-s}^{(1)}}{N} \langle\cM_{10}^{(-s,s)} \rangle \langle \p_+^{\bar{s}+s}J^3_1(x_1) J^3_1(x_2)\rangle\\
    + \frac{\bar{\rho}_{\bar{s},-s}^{(2)}}{N} \langle\cM_{10}^{(-s,s)}\rangle \langle \p_+^{\bar{s}+s-1} \p_3 J^-_1(x_1) J^3_1(x_2)\rangle + \frac{\bar{\rho}_{\bar{s},-s}^{(3)}}{N} (\p_{x^3_1}\langle \cM_{10}^{(-s,s)}\rangle) \langle\p_+^{\bar{s}+s-1} J^-_1(x_1) J^3_1(x_2)\rangle\\
    + e_{1,2}\,\p_{x^3_2} \langle\cM_{10}^{(\bar{s},s)} J^{33}_2(x_2)\rangle = 5 \frac{a_1}{N}\p_{x_2^3} \int \langle\cM_{10}^{(\bar{s},s)} J_0(x_3)\rangle \langle J_1^3(x_3) J_1^3(x_2)\rangle \dd^3 x_3\,.
\end{multline}
We find that this implies $\bar\rho_{\bar s, -s}^{(1)}=\bar\rho_{\bar s, -s}^{(3)}=0$ and
\begin{equation}
    \bar{\rho}^{(2)}_{\bar s,-s} = \frac{\eta_-\, q}{\mathcal{N}_1\, \Gamma(\bar{s}+s+2)} \left(\tilde{d}_1(\Delta_{-\bar s}+\Delta_s-1)(\Delta_s)_{\bar s+s} - 2d_1 (\Delta_s-\sfrac{1}{2})_{\bar s + s + 1} \right)\, .
\end{equation}
We also use Equation \eqref{Q3J1uneq} to determine some of the cross-ratio functions appearing in the correlators of a mesonic line with $J_2$, namely the three functions $h^{(i)}_{\bar s, s,2}$, see Appendix~\ref{app:J2corr}. Even though only a derivative of the correlator with $J^{33}_2$ enters our equation, there is no ambiguity in reconstructing the correlator because the equation
\begin{equation}
    \p_{x^3_2} \langle\cM_{10}^{(\bar s,s)} J^{33}_2(x_2)\rangle = 0
\end{equation}
admits no solution that is both consistent with \eqref{j2corr} and regular as $u\to 1^-$. Repeating the same exercise with $J^3_1$ replaced by $J^+_1$ allows to compute the functions $k^{(i)}_{\bar s ,s, 2}$, we also give the resulting functions in Appendix~\ref{app:J2corr}. This fully determines the correlators with~$J_2$.

Finally, we remark that even though all the calculations in this appendix are done assuming $\nu=0$, the results for the $J_2$ correlators obtained from them and presented in Appendix~\ref{app:J2corr} are expected not to depend on $\nu$. They could indeed be computed independently using only the translation Ward identities. The coefficients $\bar{\rho}^{(i)}_{\bar s,s}$, however, must depend on $\nu$, and we have not tried to compute them.

\subsection{Line Defect} \label{app:D33zero}

We show here that the pseudo-charge acts trivially on the line defect through order $O(1/N)$. As explained before, it could \emph{a priori} generate insertions of the form
\begin{equation}
    \mathbb{D}_{33} = \frac{\Omega_1}{N} (\p_+ J^+_1- \p_- J^-_1) + \frac{\Omega_2}{N} J^{33}_2 + O(\sfrac{1}{N^2})\, .
\end{equation}

Consider first the action of the pseudo-charge on the one-point function of $J^+_1$ in the presence of an infinite fundamental line stretched along the $x^3$ axis. This reads
\begin{multline}
    \Omega_{1} \int_{-\infty}^{+\infty}\!\!\! \<\left(\p_+ J_{1}^{+}-\p_- J_{1}^{-}\right)(y\hat{e}_3) J_{1}^{+}(x) \>\, \dd y + \p_{x^3} \langle \mathrm{Tr}(\mathcal{W}) (e_{1,2}\,J_{2}^{3+} - 2\,e_{1,0}\,\p_-J_{0})(x)\rangle\\
    = \frac{5\,a_{1}}{N}\p_{x^{3}}\int \langle\mathrm{Tr}(\mathcal{W}) J_{0}(y) \rangle\langle J^3_1(y) J^+_1(x)\rangle\, \dd^3 y + O(1)\,.
\end{multline}
Now notice that the right-hand side and the second term of the left-hand side vanish because the correlators are independent of $x^3$ and we have to take a derivative with respect to it. This means that the only other term must also vanish, i.e. that $\Omega_1=o(1)$.

We can repeat the same exercise for the integrated operator $J^{33}_2$ by considering the action of the pseudo-charge on the one point function of $J^{33}_2$, this results in the equation 
\begin{align}
    \Omega_{2} &\int_{-\infty}^{+\infty} \langle J_{2}^{33}(y\hat{e}_{3}) J_{2}^{33}(x) \rangle\, \dd y + \p_{x^3} \langle\text{tr}(\mathcal{W})(e_{2,3}J_{3}^{333} - 2e_{2,1} \p_{x^{+}}\p_{x^{-}}J_{1}^{3}) (x)\rangle\\
    &=2\ii \frac{a_2}{N} \int \Big[\<\mathrm{Tr}(\mathcal{W}) J_1^-(y) \> \< J_2^{3+}(y) J_2^{33}(x)\> - \<\mathrm{Tr}(\mathcal{W}) J_1^+(y) \> \< J_2^{3-}(y) J_2^{33}(x)\>\Big] \dd^3 y + O(1)\,. \nonumber
\end{align}
In the left-hand side, the integral in the first term is finite and the second term is zero due to the derivative in the $x^3$ direction. In the right-hand side, the correlators are all fixed by conformal symmetry and it is easy to check that the integral gives vanishes since the integrand is odd in $y^3-x^3$. Hence $\Omega_2=o(1)$ as well.

\section{Some Feynman Integrals}
\label{app:int}

\subsection{A New Integral Relation}

We begin by recalling the star-triangle relation
\begin{equation}\label{star-triangle}
    \int\frac{\dd^3 x_3}{|x_{30}|^{2\alpha} |x_{31}|^{2\beta} |x_{32}|^{2\gamma}} = \pi^{\frac{3}{2}} \frac{\Gamma(3/2-\alpha)\, \Gamma(3/2-\beta)\, \Gamma(3/2-\gamma)}{\Gamma(\alpha)\, \Gamma(\beta)\, \Gamma(\gamma)\, |x_{10}|^{3-2\gamma} |x_{20}|^{3-2\beta} |x_{21}|^{3-2\alpha}}\, ,
\end{equation}
where $\alpha+\beta+\gamma = 3$. We also use the following useful generalisation of the chain relation:
\begin{equation}\label{twopropstar}
    \int\frac{r^{2S}\, \dd^3 x_3}{|x_{30}|^{2\alpha} |x_{31}|^{2\beta}} = \pi^{\frac{3}{2}} S! \frac{\Gamma(3/2+S-\alpha)\, \Gamma(3/2+S-\beta)\, \Gamma(\alpha+\beta-3/2-S)}{\Gamma(\alpha)\, \Gamma(\beta)\, \Gamma(3+2S-\alpha-\beta)\, |x_{10}|^{2(\alpha+\beta-S)-3} }\, ,
\end{equation}
where $\alpha$ and $\beta$ are arbitrary, $S\in\mathbb{N}$, and we denote $r^2 = |x_3|^2 - (x_3\cdot x_{10})^2/|x_{10}|^2$ .

The following two integral relations hold:
\begin{multline}\label{2F1 integral}
    \int\! \frac{\left(x_{3}^{-}\right)^S {}_2F_1\!\left[\genfrac{}{}{0pt}{1}{2\alpha,1-2\kappa}{1+\alpha-\kappa};\sfrac{1}{2} \pm \sfrac{x_{30}\cdot x_{31}}{2|x_{30}| |x_{31}|}\right] \dd^3x_3}{|x_{30}|^{2\alpha} |x_{31}|^{2(\beta+S)} |x_{32}|^{2\gamma}} = \pi^{\frac{3}{2}} \frac{\Gamma(3-2\beta)\, \Gamma(3/2-\gamma)}{2^{2\gamma-3} \, \Gamma(2\alpha)\, \Gamma(\gamma)}\\
    \times \frac{\Gamma(\kappa+S+1-\alpha)\, \Gamma(1+\alpha-\kappa)}{\Gamma(5/2-\kappa-\beta)\, \Gamma(\kappa+S+\beta-1/2)} \frac{\left(x_{2}^{-}\right)^S {}_2F_1\!\left[\genfrac{}{}{0pt}{1}{3-2\beta,1-2\kappa}{5/2-\beta-\kappa};\sfrac{1}{2} \pm \sfrac{x_{20}\cdot x_{21}}{2|x_{20}| |x_{21}|}\right]}{|x_{10}|^{3-2\gamma} |x_{20}|^{3-2\beta} |x_{21}|^{3+2(S-\alpha)}}\, ,
\end{multline}
where $S\in\mathbb{N}$, $\alpha,\beta,\gamma,\kappa$ are complex numbers, and we require $\alpha+\beta+\gamma = 3$. These relations only differ by the sign $\pm$ in the argument of the hypergeometric functions. They are equivalent because there is a conformal transformation that maps one relation to the other: if we first perform a translation and a dilation to arrive at $x_0 = -x_1$, and $|x_0|^2=1$, then the inversion $x\mapsto-x/|x|^2$ effectively changes the sign of $\pm$. When $S=0$, the points $x_0,x_1,$ and $x_2$ are arbitrary. When $S>0$, we have assumed that $x^\pm_0 = x^\pm_1 = 0$ for simplicity. To be more general, we should replace $x^-_2$ and $x^-_3$ by $\zeta\cdot x_2$ and $\zeta\cdot x_3$ where $\zeta\in\mathbb{C}^3$ is such that $\zeta\cdot\zeta =\zeta\cdot x_{10}= 0$. Another interesting integral relation is
\begin{multline}\label{2F1r integral}
    \int\! \frac{r^{2S} {}_2F_1\!\left[\genfrac{}{}{0pt}{1}{2\alpha,1-2\kappa}{1+\alpha-\kappa};\sfrac{1}{2} - \sfrac{x_{30}\cdot x_{31}}{2|x_{30}| |x_{31}|}\right] \dd^3x_3}{|x_{30}|^{2\alpha} |x_{31}|^{2(\beta+S)} |x_{32}|^{2(\gamma+S)}} = \pi^{\frac{3}{2}} S! \frac{\Gamma(3-2\beta)\, \Gamma(3/2-\gamma)}{2^{2\gamma-3} \Gamma(2\alpha)\, \Gamma(\gamma+S)}\\
    \times \frac{\Gamma(\kappa+S+1-\alpha)\, \Gamma(1+\alpha-\kappa)}{\Gamma(5/2-\kappa-\beta)\, \Gamma(\kappa+S+\beta-1/2)\, |x_{10}|^{3-2\gamma} |x_{20}|^{3-2\beta} |x_{21}|^{3+2(S-\alpha)}}\, ,
\end{multline}
which holds when the three points are aligned, $x^\pm_0 = x^\pm_1 =x^\pm_2= 0$, and $x_2$ does not lie between $x_0$ and $x_1$, namely $x_{20}\cdot x_{21} = |x_{20}| |x_{21}|$.

Let us prove these two relations. We define
\begin{equation}
    I_{S,\kappa}(x_0,x_1,x_2) = \frac{\Gamma(\kappa+S+\beta-1/2)}{\Gamma(1+\alpha-\kappa)\, |x_{10}|^{2(\kappa-\alpha)}}\!\! \int\!\! \frac{\left(x_{3}^{-}\right)^S\!\! {}_2F_1\!\left[\genfrac{}{}{0pt}{1}{2\alpha,1-2\kappa}{1+\alpha-\kappa};\sfrac{1}{2} - \sfrac{x_{30}\cdot x_{31}}{2|x_{30}| |x_{31}|}\right] \dd^3x_3}{|x_{30}|^{2\alpha} |x_{31}|^{2(\beta+S)} |x_{32}|^{2\gamma}}\, .
\end{equation}
On the one hand, from $\mathrm{SL}(2,\mathbb{R})\times \mathrm{U}(1)$ covariance, it is clear that
\begin{equation}
    I_{S,\kappa}(x_0,x_1,x_2) = \frac{\left(x_{2}^{-}\right)^S f_{S,\kappa}\left(\sfrac{x_{20}\cdot x_{21}}{|x_{20}| |x_{21}|}\right)}{|x_{10}|^{3+2(\kappa-\alpha-\gamma)} |x_{20}|^{3-2\beta} |x_{21}|^{3+2(S-\alpha)}} 
\end{equation}
for some yet-to-be-determined function $f_{S,\kappa}$. On the other hand, we have
\begin{equation}
    2I_{S,\kappa+1} + \frac{1}{2}\p_{x_0^3}^2 I_{S,\kappa} + \p_{x_2^+} I_{S-1,\kappa+1} = 0\, ,
\end{equation}
for all $S\geqslant 1$, as can be checked by acting directly on the integrand and using integration by parts for the third term (with $\p_{x_2^+}$). Combining the last two equations yields differential equations on the functions $f_{S,\kappa}$. We find that the solution that is regular when $u\to1^-$ is
\begin{equation}
    f_{S,\kappa}(u) = C(\kappa) \frac{\Gamma(\kappa+S+1-\alpha)}{\Gamma(5/2-\kappa-\beta)}{}_2F_1\!\left[\genfrac{}{}{0pt}{1}{3-2\beta,1-2\kappa}{5/2-\beta-\kappa};\sfrac{1-u}{2}\right]\, ,
\end{equation}
where the undetermined function $C(\kappa)$ is independent of $u$ and $S$, but 1-periodic in $\kappa$. The regularity property of the integral was used to  rule out a term proportional to ${}_2F_1[\cdots;\sfrac{1+u}{2}]$. In order to compute the undetermined function, let us focus on the $S=0$ case. We have just shown that there exists $C(\kappa)$ such that
\begin{multline}
    \int \frac{{}_2F_1\!\left[\genfrac{}{}{0pt}{1}{2\alpha,1-2\kappa}{1+\alpha-\kappa};\sfrac{1}{2} - \sfrac{x_{30}\cdot x_{31}}{2|x_{30}| |x_{31}|}\right] \dd^3x_3}{|x_{30}|^{2\alpha} |x_{31}|^{2\beta} |x_{32}|^{2\gamma}}\\
    = C(\kappa) \frac{\Gamma(\kappa+1-\alpha)\, \Gamma(1+\alpha-\kappa)}{\Gamma(5/2-\kappa-\beta)\, \Gamma(\kappa+\beta-1/2)}
    \frac{{}_2F_1\!\left[\genfrac{}{}{0pt}{1}{3-2\beta,1-2\kappa}{5/2-\beta-\kappa};\sfrac{1}{2} - \sfrac{x_{20}\cdot x_{21}}{2|x_{20}| |x_{21}|}\right]}{|x_{10}|^{3-2\gamma} |x_{20}|^{3-2\beta} |x_{21}|^{3-2\alpha}}\, .
\end{multline}
Defining $u=x_{20}\cdot x_{21}/|x_{20}| |x_{21}|$, one can bring the previous equation to the form
\begin{multline}\label{simplified 2F1 int}
    \int_0^{+\infty} \!\!\!\int_{-1}^1 \int_0^{2\pi} \frac{{}_2F_1\!\left[\genfrac{}{}{0pt}{1}{2\alpha,1-2\kappa}{1+\alpha-\kappa};\sfrac{1-v}{2}\right] r^{2-2\beta} \dd \phi\, \dd v\, \dd r}{[r^2+1-2uvr-2r\sin\phi\sqrt{(1-u^2)(1-v^2)}]^{\gamma}}\\
    = C(\kappa) \frac{\Gamma(\kappa+1-\alpha)\, \Gamma(1+\alpha-\kappa)}{\Gamma(5/2-\kappa-\beta)\, \Gamma(\kappa+\beta-1/2)} {}_2F_1\!\left[\genfrac{}{}{0pt}{1}{3-2\beta,1-2\kappa}{5/2-\beta-\kappa};\sfrac{1-u}{2}\right]\, .
\end{multline}
Integrating this relation over $u\in[-1,1]$ yields\footnote{The triple integral over $\phi$, $u$, and $r$ in the left-hand side is nothing else but the $\kappa = 1/2$ version of \eqref{simplified 2F1 int}, which is itself a rewriting of the usual star-triangle relation \eqref{star-triangle}.}
\begin{multline}
    \pi^{\frac{3}{2}} \frac{\Gamma(3/2-\alpha)\, \Gamma(3/2-\beta)\, \Gamma(3/2-\gamma)}{\Gamma(\alpha)\, \Gamma(\beta)\, \Gamma(\gamma)} \int_{-1}^1 {}_2F_1\!\left[\genfrac{}{}{0pt}{1}{2\alpha,1-2\kappa}{1+\alpha-\kappa};\sfrac{1-v}{2}\right] \dd v\\
    = C(\kappa) \frac{\Gamma(\kappa+1-\alpha)\, \Gamma(1+\alpha-\kappa)}{\Gamma(5/2-\kappa-\beta)\, \Gamma(\kappa+\beta-1/2)} \int_{-1}^1 {}_2F_1\!\left[\genfrac{}{}{0pt}{1}{3-2\beta,1-2\kappa}{5/2-\beta-\kappa};\sfrac{1-u}{2}\right]\dd u\, .
\end{multline}
The remaining integrals are straightforward to evaluate and we obtain
\begin{equation}
    C(\kappa) = 2^{3-2\gamma} \pi^{\frac{3}{2}} \frac{\Gamma(3-2\beta)\, \Gamma(3/2-\gamma)}{\Gamma(2\alpha)\, \Gamma(\gamma)}\, ,
\end{equation}
which is 1-periodic in $\kappa$ as expected. This concludes the proof of \eqref{2F1 integral}. It is then easy to obtain \eqref{2F1r integral}: one simply needs to take $S$ derivatives of \eqref{2F1 integral} with respect to $x^-_2$ and evaluate in $x^\pm_2 = 0$.

For the sake of completeness, we conclude this appendix with the generalisation of the integral relation \eqref{2F1 integral} to arbitrary dimension:
\begin{multline}\label{2F1 integral d}
    \int\! \frac{(\zeta\cdot x_3)^S {}_2F_1\!\left[\genfrac{}{}{0pt}{1}{2\alpha,1-2\kappa}{1+\alpha-\kappa};\sfrac{1}{2} \pm \sfrac{x_{30}\cdot x_{31}}{2|x_{30}| |x_{31}|}\right] \dd^dx_3}{|x_{30}|^{2\alpha} |x_{31}|^{2(\beta+S)} |x_{32}|^{2\gamma}} = \pi^{\frac{d}{2}} \frac{\Gamma(d-2\beta)\, \Gamma(\sfrac{d}{2}-\gamma)}{2^{2\gamma-d} \, \Gamma(2\alpha)\, \Gamma(\gamma)}\\
    \times \frac{\Gamma(\kappa+S+(d-1)/2-\alpha)\, \Gamma(1+\alpha-\kappa)}{\Gamma(d/2+1-\kappa-\beta)\, \Gamma(\kappa+S+\beta-1/2)} \frac{(\zeta\cdot x_2)^S {}_2F_1\!\left[\genfrac{}{}{0pt}{1}{d-2\beta,1-2\kappa}{d/2+1-\beta-\kappa};\sfrac{1}{2} \pm \sfrac{x_{20}\cdot x_{21}}{2|x_{20}| |x_{21}|}\right]}{|x_{10}|^{3-2\gamma} |x_{20}|^{3-2\beta} |x_{21}|^{3+2(S-\alpha)}}\, ,
\end{multline}
where $\alpha+\beta+\gamma = d$ and $\zeta\in\mathbb{C}^d$ is such that $\zeta\cdot\zeta =\zeta\cdot x_{10}= 0$. The proof is an obvious generalisation of the one shown for $d=3$.

\subsection{Conformal Changes of Variables}
\label{app:conf int}

We explain here how to make the cross-ratio dependence of conformal integrals explicit. Consider for instance the first integral appearing in the right-hand side of \eqref{RHS equal-equal}. We first make the change of variables $y\to x_0 + y/|y|^2$ to arrive at
\begin{multline}
    \int \frac{|x_{32}| r^2\, \dd^3 y}{|y-x_3|^{2\Delta+1} (|y-x_2||y-x_1|)^{5-2\Delta} |y-x_0|^{2\Delta-1}}\\
    = \frac{|x_{32}|}{|x_{30}|^{2\Delta+1} (|x_{20}||x_{10}|)^{5-2\Delta} } \int \frac{r^2\, \dd^3 y}{|y-\hat{e}_3/x^3_{30}|^{2\Delta+1} (|y-\hat{e}_3/x^3_{20}||y-\hat{e}_3/x^3_{10}|)^{5-2\Delta}}\, .
\end{multline}
We then change variables according to $y\to \hat{e}_3/x^3_{20} + y\, x^3_{21}/(x^3_{10}x^3_{20})$ to obtain
\begin{equation}
    \int \frac{|x_{32}||y-x_0|^{1-2\Delta} r^2\, \dd^3 y}{|y-x_3|^{2\Delta+1} (|y-x_2||y-x_1|)^{5-2\Delta}} = \frac{|x_{20}|\chi}{|x_{30}|^{2\Delta} |x_{21}|^{5-2\Delta} } \! \int \frac{ |y-\hat{e}_3|^{2\Delta-5}\, r^2\, \dd^3 y}{|y+\chi \hat{e}_3|^{2\Delta+1} |y|^{5-2\Delta}}\, ,
\end{equation}
where we recall that $\chi = x^3_{32} x^3_{10}/(x^3_{30} x^3_{21})$.

We perform the same manipulations for more complicated integrals. Take for instance the first integral entering the right-hand side of \eqref{RHS uneq-uneq}. First replacing $y\to x_2 + y/|y|^2$ yields
\begin{multline}
    \frac{1}{|x_{10}|^{2\Delta-1} |x_{32}|^{2\Delta-2}}\int \frac{F_1(u_{32})\, F_2(u_{10})\, \dd^3y}{|y-x_3|^2 |y-x_2|^2 |y-x_1| |y-x_0|} \\
    = \frac{1}{|x_{10}|^{2\Delta-1} |x_{32}|^{2\Delta} |x_{21}| |x_{20}|} \int \frac{F_1\Big(\frac{1/x^3_{32}-y^3}{|y-\hat{e}_3/x^3_{32}|}\Big)\, F_2\Big(\frac{(y-\hat{e}_3/x^3_{12})\cdot (y-\hat{e}_3/x^3_{02})}{|y-\hat{e}_3/x^3_{12}| |y-\hat{e}_3/x^3_{02}|}\Big)\, \dd^3y}{|y-\hat{e}_3/x^3_{32}|^2 |y-\hat{e}_3/x^3_{12}| |y-\hat{e}_3/x^3_{02}|}\, .
\end{multline}
We then make the change of variables $y\to \hat{e}_3/x^3_{02} + y\, x^3_{30}/x^3_{32}x^3_{20}$ to obtain
\begin{multline}
    \frac{1}{|x_{10}|^{2\Delta-1} |x_{32}|^{2\Delta-2}}\int \frac{F_1(u_{32})\, F_2(u_{10})\, \dd^3y}{|y-x_3|^2 |y-x_2|^2 |y-x_1| |y-x_0|} \\
    = \frac{\chi}{|x_{10}|^{2\Delta} |x_{32}|^{2\Delta}} \int \frac{F_1\Big(\frac{1-y^3}{|y-\hat{e}_3|}\Big)\, F_2\Big(\frac{(y+\chi\hat{e}_3)\cdot y}{|y+\chi\hat{e}_3| |y|}\Big)\, \dd^3y}{|y-\hat{e}_3|^2 |y+\chi\hat{e}_3| |y|}\, .
\end{multline}

\subsection{One-Loop Feynman Integrals}
\label{app:3pt int}

We explain here how to compute the cross-ratio functions that appear in Section~\ref{sec:equal-equal}. For $0<\chi<1$, let
\begin{equation}
    J_m (a,b,c;\chi) = \int \frac{r^{2m}\, \dd^3 y}{|y|^{2a} |y-\chi \hat{e}_3|^{2b} |y-\hat{e}_3|^{2c}}\, .
\end{equation}
It can be shown that $\overline{J}_0 = \chi^{a+b-1}(1-\chi)^{b+c-1} J_0$ is a solution to
\begin{multline}\label{eqdiff J0}
    \chi(1-\chi) \overline{J}'''_0 + (2-4\chi) \overline{J}''_0 + \bigg[(a+c)(a+c-3) + \frac{(a+b-1)(2-a-b)}{\chi}\\
    + \frac{(b+c-1)(2-b-c)}{1-\chi}\bigg] \overline{J}'_0 = 0\, .
\end{multline}
When $\chi\to 0$, we also compute using the chain relation \eqref{twopropstar},
\begin{align}
    J_0(a,b,c;\chi) &= \pi^{\frac{3}{2}} \frac{\Gamma(3/2-a-b)\, \Gamma(3/2-c)\, \Gamma(a+b+c-3/2)}{\Gamma(a+b)\, \Gamma(c)\, \Gamma(3-a-b-c)}(1+O(\chi))\\
    &+ \pi^{\frac{3}{2}} \frac{\Gamma(3/2-a)\, \Gamma(3/2-b)\, \Gamma(a+b-3/2)}{\Gamma(a)\, \Gamma(b)\, \Gamma(3-a-b)}\chi^{3-2(a+b)}(1+O(\chi))\, .
\end{align}
Together with the symmetry $J_m(c,b,a;\chi) = J_m(a,b,c; 1-\chi)$, this suffices to determine $J_0$ explicitly.

Equation \eqref{eqdiff J0} can be derived from the Yangian-type differential equations satisfied by the three-point integral in the bulk \cite{Chicherin:2017cns,Chicherin:2017frs,Loebbert:2024qbw}. This was shown to one of the authors (GF) by Florian Loebbert before the authors realised it would be relevant for the present paper. This is not completely immediate since we are only interested in the configuration where all three points are aligned, see \cite{Loebbert:2019vcj} for a similar kinematic restriction, namely $z=\bar z$, in the case of the conformal box integral in four dimensions. This is also the reason why \eqref{eqdiff J0} is a third-order differential equation, when the Yangian equations for generic kinematics are only of second~order.

We also note that
\begin{equation}\label{J0 to J1}
    J_1(a,b,c;\chi) = \frac{2b-3}{2(b-1)} J_0(a,b-1,c;\chi) - \frac{1}{4(b-1)(b-2)} J''_0(a,b-2,c;\chi)\, ,
\end{equation}
and that the functions $\mathcal{G}_{\text{ee}}$ and $\  \mathcal{\widetilde{F}}_{\text{ee}}$ introduced in Subsection \ref{sec:equal-equal} correspond to
\begin{equation}
     \mathcal{\widetilde{F}}_{\text{ee}}(\chi) = \frac{(\Delta-1)\chi}{(1+\chi)^{2\Delta+1}}\, J_1\!\left(\frac{5}{2}-\Delta,\Delta+\frac{1}{2},\Delta;\frac{\chi}{1+\chi}\right)
\end{equation}
and
\begin{equation}
      \mathcal{G}_{\text{ee}}(\chi) = \frac{(1-2\Delta)\chi}{(1+\chi)^{2\Delta+2}}\, J_1\!\left(\frac{5}{2}-\Delta,\Delta+\frac{1}{2},\Delta+\frac{1}{2};\frac{\chi}{1+\chi}\right)\, .
\end{equation}
Using the differential equations \eqref{eqdiff J0} and \eqref{J0 to J1}, we were able to derive the simple result \eqref{g(chi)} for $  \mathcal{G}_{\text{ee}}$. However, we were unable to simplify the result for $  \mathcal{\widetilde{F}}_{\text{ee}}$, so we only give it indirectly as being obtained from \eqref{J0 to J1} using
\begin{equation}
    J_0\!\left(\frac{5}{2}-\Delta,\Delta-\frac{1}{2},\Delta;\chi\right) = \frac{2\pi^2(1-2\Delta)(1-\chi)^{\frac{3}{2}-2\Delta}}{\chi\,\tan(2\pi\Delta)} \int_{1}^\chi {}_2F_1\!\left[\genfrac{}{}{0pt}{1}{2-2\Delta,4-2\Delta}{2};t\right] (1-t)^{\frac{3}{2}-2\Delta} \dd t\, ,
\end{equation}
and
\begin{equation}
    J_0\!\left(\frac{5}{2}-\Delta,\Delta-\frac{3}{2},\Delta;\chi\right) = \frac{4\pi^2(2-\Delta)(1-\chi)^{\frac{5}{2}-2\Delta}}{\tan(2\pi\Delta)} \int_{1}^\chi {}_2F_1\!\left[\genfrac{}{}{0pt}{1}{3-2\Delta,5-2\Delta}{2};t\right] (1-t)^{\frac{5}{2}-2\Delta} \dd t\, .
\end{equation}
As usual, all integrals are understood as analytic continuation whenever $\Delta$ is such that they do not converge. In the first case for instance, the integral defining $J_0\left(\frac{5}{2}-\Delta,\Delta-\frac{1}{2},\Delta;\chi\right)$ converges only if $\Re(\Delta)\in]1;\sfrac{3}{2}[$, and the integral in the right-hand side only if $\Re(\Delta)\in]\frac{3}{4};\sfrac{5}{4}[$. The right-hand side could easily be rewritten in a way that is manifestly meromorphic in $\Delta$ using
\begin{multline}
    \int_{1}^\chi {}_2F_1\!\left[\genfrac{}{}{0pt}{1}{2-2\Delta,4-2\Delta}{2};t\right] (1-t)^{\frac{3}{2}-2\Delta} \dd t = \frac{2\cos^2(\pi\Delta)}{(2\Delta-1)(2\Delta-3)\cos(2\pi\Delta)}\\
    + \int_{0}^\chi {}_2F_1\!\left[\genfrac{}{}{0pt}{1}{2-2\Delta,4-2\Delta}{2};t\right] (1-t)^{\frac{3}{2}-2\Delta} \dd t\, .
\end{multline}

\section{Perturbative Checks in the Quasi-Fermionic Theory}
\label{app:perturbative}

In this appendix, we compare some of our bootstrap results with explicit 1-loop perturbative computations. The computations are performed in the quasi-fermionic theory, which is defined by coupling Chern--Simons theory to a fundamental fermion
\begin{equation}
    S= \frac{\ii k}{4\pi} \varepsilon^{\mu\nu\rho} \int\dd^3x \Tr\big(A_\mu\p_\nu A_\rho-\frac{2\ii}3 A_\mu A_\nu A_\rho\big) + \int\dd^3x\,  \bar\psi\gamma^\mu D_\mu \psi\,,
\end{equation}
where $D_\mu\psi= \p_\mu \psi -\ii A_\mu \psi $, $\bar\psi = \psi^\dagger$ and the gauge group is $\mathrm{SU}(N_c)$. Here, $\psi^i_a$ is a two-component spinor in the defining representation of $\mathrm{SU}(N_c)$, i.e. $a\in\{1,2\}$ and $i\in~\{1,\dots,N_c\}$. The $\gamma$ matrices in Euclidean signature are taken to be the Pauli matrices, $\gamma^\mu=\sigma^\mu$. There are no relevant or marginal operators other than the fermion mass term. Hence, after tuning the mass to zero, the theory is conformal. We consider the CFT in the planar, 't Hooft limit, $N_c\rightarrow\infty$ with $\lambda=N_c/k$ fixed. At large $N_c$, the spectrum of primary operators consists of a tower of conserved twist-1 currents and a single, twist-2 scalar operator $J_0 = \bar\psi \psi$.

The free propagators in Feynman gauge are given by
\begin{equation}
    \< \psi^i_a(x) \bar\psi^j_b(y)\>_0 = \delta^{ij}\frac{(x-y)_\mu (\gamma^\mu)_{ab}}{4\pi |x-y|^3}\,,
\end{equation}
\begin{equation}
    \< A^I_\mu(x)A^J_\nu(y)\>_0 = -\frac{\ii}{k}\delta^{IJ} \varepsilon_{\mu\nu \rho}\frac{(x-y)^\rho}{|x-y|^3}\,,
\end{equation}
where $I,J$ are adjoint, gauge-group indices. The Greek indices run from 1 to 3. Finally, we normalise the adjoint generators $T_I$ of the gauge group such that $\Tr(T_I T_J) = \delta_{IJ}/2$.

The stable conformal line operator in this theory is the standard Wilson line
\begin{equation}\label{wilsonline}
    \mathcal{W}[x(\cdot)] = {\cal P}\,\exp\left[{\ii\int\dd\sigma\,\dot{x}^\mu(\sigma)} A_\mu(x(\sigma)) \right]\,.
\end{equation}
As explained in \cite{Gabai:2022vri,Gabai:2022mya}, the corresponding fundamental and anti-fundamental boundary operators for a line oriented in the $\hat x^3$ direction are
\begin{equation}\label{boundary ferm}
    \overline\cO_{\bar s} = \frac1{\sqrt{N_c}} \begin{cases}
D^{\bar s - \frac12}_+ \bar\psi_2 &\quad \bar s > 0\\
D^{-\bar s-\frac12}_- \bar\psi_1 & \quad \bar s< 0
\end{cases} \qquad\text{and}\qquad \cO_{s} = \frac1{\sqrt{N_c}} \begin{cases}
D^{s - \frac12}_+ \psi_1 &\quad s > 0\\
D^{-s-\frac12}_- \psi_2 & \quad s< 0
\end{cases}\,,
\end{equation}
where $ D_{\pm}\bar\psi = \p_{\pm}\bar\psi + \ii\bar\psi A_{\pm} $. For the reader's convenience, we collect here several one-loop results \cite{Gabai:2022mya,Gabai:2022vri,Ferrando:2025ufj}
\begin{align}
    c_\pm &= -\frac{\lambda\pm2}{8\pi} + O(\lambda^2)\,,\ \, \quad d_0 = \frac{-\sqrt{2}}{(4\pi)^2} +O(\lambda^2)\, , \quad\ \, \mathcal{N}_0=\frac{1}{8\pi^2} +O(\lambda^2)\,, \label{one-loop constants}\\
    \eta_\pm &= \mp 2\sqrt{2}\pi\lambda + O(\lambda^3)\, ,\quad d_1 = \frac{1}{4\sqrt{2}\pi^2} + O(\lambda^2)\, ,\quad \mathcal{N}_1 = \frac{1}{32\pi^2} + O(\lambda^2)\, . \label{one-loop constants 2}
\end{align}
We also recall that \cite{Witten:1988hf}
\begin{equation}
    \langle \Tr(\cW)\rangle = N_c \frac{\sin(\pi\lambda)}{\pi\lambda} + O(1)\, .
\end{equation}
This last equation means that the definition of $N$ we chose in our bootstrap approach does not coincide with the number of colours appearing in this appendix. However, the difference between them is $O(\lambda^2)$ at large $N$, and we will only perform one-loop checks so we abuse notation below and use the same symbol for both.

\begin{figure}
\centering{}\includegraphics[scale=2.5]{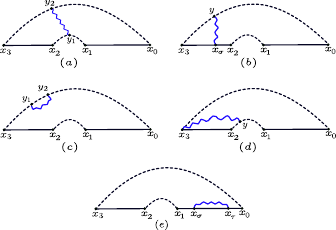}\caption{The dashed line is a fermion propagator, the solid line is a Wilson line, and the blue wiggly line is a gluon propagator. (a) Diagram of an exchange of a gluon between two bulk fermion propagators. (b) Diagram of an exchange of a gluon between a bulk propagator and a Wilson line. (c) Diagram of a self energy  of a bulk fermion propagator. (d) Diagram of an exchange of a gluon between an insertion point with a bulk fermion  propagator. (e) Diagram of a self energy of a Wilson line. }
\label{digrams}
\end{figure}

Lastly, we drew in Figure \ref{digrams} the diagrams that will contribute to the correlators below. We will reference them accordingly. Their actual integral representations depends on the precise configuration of line operators. The diagram of Figure \ref{digrams}(d) contributes only if the operator located at $x_3$ contains some covariant derivatives. The line-to-line diagram of figure \ref{digrams}(e) is a universal contribution to all the cases we study below. Since the line operators are aligned, it does not contribute. Hence we wont consider it further. More generally, it is responsible for the anomalous spin of the boundary operators, see \cite{Gabai:2022mya}.

\subsection{One-Loop Check for $\langle\mathcal{M}^{\left(-\frac{1}{2},-\frac{1}{2}\right)}\mathcal{M}^{\left(\frac{1}{2},\frac{1}{2}\right)}\rangle $}
\label{app:equalpert}

In terms of fields, we have the following two mesonic lines:
\begin{equation}
    \cM_{32}^{(-\frac{1}{2},-\frac{1}{2})} =\frac{1}{N} \bar \psi_1(x_3)\cW\psi_2(x_2)\,,\quad  \cM_{10}^{(\frac{1}{2},\frac{1}{2})} =\frac{1}{N} \bar \psi_2(x_1)\cW\psi_1(x_0)\,.
\end{equation}
At tree level the correlator is a product of two free propagators
\begin{equation}\label{equaltree}
  \langle\mathcal{M}_{32}^{\left(-\frac{1}{2},-\frac{1}{2}\right)}\mathcal{M}_{10}^{\left(\frac{1}{2},\frac{1}{2}\right)}\rangle_{0}=\frac{-1}{N\left(4\pi\right)^{2}\left|x_{30}\right|^{2}\left|x_{12}\right|^{2}}.
\end{equation}

At one-loop we have several diagrams, typical representatives of which can be seen in Figures \ref{digrams}(a), (b), and (c). We go over them separately below and then combine them together with the tree level. Finally, we compare this one-loop result to the non-perturbative result from the Section \ref{sec:equal-equal}.

\paragraph{Bulk-to-bulk diagram} The diagram can be seen in Figure \ref{digrams}(a). Using Feynman rules we find the following integral
\begin{multline}
    I_{1} = \frac{\text{i}\lambda\,\varepsilon_{\mu\nu\rho}}{2\left(4\pi\right)^{4}N} \left(\gamma^{\eta}\gamma^{\nu}\gamma^{\delta}\right)_{11} \left(\gamma^{\beta}\gamma^{\mu}\gamma^{\alpha}\right)_{22}\\
    \times \int \frac{(y_{1}-x_{2})_{\beta} (y_{1}-x_{1})_{\alpha} (y_{2}-x_{3})_{\delta} (x_{0}-y_{2})_{\eta} y_{12}^{\rho}}{|y_{1}-x_{2}|^{3} |x_{1}-y_{1}|^{3} |x_{3}-y_{2}|^{3} |y_{2}-x_{0}|^{3} |y_{12}|^{3}}\, \dd^3 y_1\, \dd^3 y_2\,.
\end{multline}
We are unable to compute it directly but we conjecture that the integral evaluates to 
\begin{equation}
I_1=\frac{-\lambda(|x_{20}|+|x_{31}|)}{2(4\pi)^2N|x_{30}|^2 |x_{20}||x_{31}||x_{21}|}\,  .
\end{equation}
We have evaluated $I_1$ numerically for random values of the positions $x_i^3$, using Mathematica, and found agreement with this conjecture.

\paragraph{Bulk-to-line diagrams}
We have four such diagrams. We drew one of them in Figure \ref{digrams}(b); it corresponds to
\begin{equation}
    I_{2} = \frac{\ii\lambda\,\varepsilon_{\mu\nu 3} |x_{32}|}{2(4\pi)^{3} N |x_{21}|^{2}} \left(\gamma^{\delta}\gamma^{\mu}\gamma^{\alpha}\right)_{11} \int_{0}^{1} \int \frac{(y-x_{3})_{\alpha} (x_{0}-y)_{\delta} (x_{\sigma}-y)^{\nu}}{|y-x_{3}-\epsilon\hat{e}_{3}|^{3} |x_{0}-y|^{3} |x_{\sigma}-y|^{3}}\, \dd^3 y\, \dd\sigma\,,
\end{equation}
with $x_\sigma = x_2+\sigma x_{32}$ and $\epsilon$ a point-splitting cut-off that regularises divergences when $x_\sigma$ approaches $x_3$. Using the integral identity \cite{Guadagnini:1989am}
\begin{equation}\label{identity1}
    \int\frac{y^{\mu}\,\dd^{3}y}{|y|^{3}|y-a||y-b|}=\frac{4\pi}{|a|+|b|+|a-b|}\left(\frac{a^{\mu}}{|a|}+\frac{b^{\mu}}{|b|}\right)\,, 
\end{equation}
and performing the line integration explicitly using  Mathematica we find that
\begin{equation} \label{equalI2}
    I_2 = \frac{ \lambda\ln(\frac{\epsilon |x_{20}|}{|x_{30}| |x_{32}| })    }{2(4\pi)^2 N |x_{30}|^2 |x_{21}|^2}\,.
\end{equation}
Combining with contribution from the other three bulk-to-line diagrams, we find
\begin{equation}
   I_2+I_3+I_4+I_5=\lambda \frac{|x_{21}|(|x_{20}|+|x_{31}|)-2|x_{20}| |x_{31}|+  2|x_{20}|  |x_{31}| \ln (\frac{|x_{21}|}{|x_{30}|}) }{2(4\pi)^2 N|x_{30}|^2 |x_{20}| |x_{31}| |x_{21}|^2 }\,.
\end{equation}

\paragraph{Propagator corrections}
We have two fermion propagator corrections by gluon exchange, see one of them in Figure \ref{digrams}(c).  They have been computed in \cite{Gabai:2022mya}. They combine to 
\begin{equation}
       I_{6}+I_7= \frac{\lambda}{(4\pi)^2 N |x_{30}|^2 |x_{21}|^2}\,.
\end{equation}

\paragraph{Full conformal result}
Through one-loop order, the correlator is thus given by
\begin{align}
    \langle\mathcal{M}_{32}^{\left(-\frac{1}{2},-\frac{1}{2}\right)}\mathcal{M}_{10}^{\left(\frac{1}{2},\frac{1}{2}\right)}\rangle &= \langle\mathcal{M}_{32}^{\left(-\frac{1}{2},-\frac{1}{2}\right)}\mathcal{M}_{10}^{\left(\frac{1}{2},\frac{1}{2}\right)}\rangle_0 + \sum_{k=1}^7 I_k + O(\lambda^2)\\
    &= - \frac{1+\lambda\ln(\sfrac{|x_{30}|}{|x_{21}|})}{(4\pi)^2 N |x_{30}|^2 |x_{21}|^2} + O(\lambda^2)\, .
\end{align}
Let us now compare this to the exact result
\begin{equation}
    \langle\mathcal{M}_{32}^{\left(-\frac{1}{2},-\frac{1}{2}\right)}\mathcal{M}_{10}^{\left(\frac{1}{2},\frac{1}{2}\right)}\rangle = \frac{c_+\,c_-\,\sin^2(\pi\Delta)[\mathcal{F}_{\text{ee}}(\chi) + \mathcal{\widetilde{F}}_{\text{ee}}(\chi)]}{N\pi^3(1-\Delta)|x_{30}|^{4-2\Delta} |x_{21}|^{2\Delta}}\, .
\end{equation}
We need to expand this to first order in $\lambda = 2(\Delta-1)$. For the prefactor, we use the values computed perturbatively in \cite{Gabai:2022mya,Gabai:2022vri,Ferrando:2025ufj} and recalled in \eqref{one-loop constants} and~\eqref{one-loop constants 2}.
For the functions, we first notice that $  \mathcal{\widetilde{F}}_{\text{ee}}= O(\lambda)$ and that, according to \eqref{ode equal-equal} and \eqref{f(0)},
\begin{equation}
    \mathcal{G}_{\text{ee}}(\chi) = \frac{-4\pi}{(1+\chi)^2} + O(\lambda) \Longrightarrow   \mathcal{F}_{\text{ee}}(\chi) =   \mathcal{F}_{\text{ee}}(0) + O(\lambda) = -\pi^2 \cot(\pi\Delta) + O(\lambda)\, .
\end{equation}
Thus \eqref{equal equal} indeed agrees with the perturbative computation.

\subsection{One-Loop Check for $\langle\mathcal{M}^{\left(\frac{3}{2},-\frac{1}{2}\right)}\mathcal{M}^{\left(-\frac{1}{2},-\frac{1}{2}\right)}\rangle $}
\label{app:structper}

In terms of fields, we have the following two mesonic lines:
\begin{equation}
    \cM_{32}^{(\frac{3}{2},-\frac{1}{2})} =\frac{1}{N}  D_+ \bar\psi_2(x_3)\cW\psi_2(x_2)\,,\quad  \cM_{10}^{(-\frac{1}{2},-\frac{1}{2})} =\frac{1}{N} \bar \psi_1(x_1)\cW\psi_2(x_0)\,.
\end{equation}
There is no tree-level contribution and only two types of diagrams contribute at one lopp: bulk-to-bulk ($L_1$) and vertex-to-bulk ($L_2$), see Figure \ref{digrams}(a) and (d) respectively. 

\paragraph{Bulk-to-bulk diagram} The integral takes the form
\begin{multline}
    L_1 = \frac{\ii\lambda\,\varepsilon_{\mu\nu\rho}}{2(4\pi)^4 N} \left(\gamma^{\eta}\gamma^{\mu}\gamma^{\delta}\right)_{22} \left(\gamma^{\beta}\gamma^{\nu}\gamma^{\alpha}\right)_{21} \\
    \times \p_{x_{3}^{+}} \int \frac{(x_{3}-y_2)_{\delta} (y_{2}-x_0)_{\eta} (x_{2}-y_1)_{\beta} (y_{1}-x_1)_{\alpha} y_{12}^{\rho}}{|y_1-x_2|^3 |x_1-y_1|^3 |x_3-y_2|^3 |y_{2}-x_0|^3 |y_{21}|^3}\, \dd^3 y_1\,\dd^3 y_2\, .
\end{multline}
As in the previous cases, we did not compute this double integral but we simply conjecture that it evaluates to 
\begin{equation}
    L_1 = \frac{-\lambda (|x_{30}|+|x_{32}|)}{32\sqrt{2}\pi^{2} |x_{20}| |x_{30}|^2 |x_{31}|^2 |x_{32}|}\,.
\end{equation}
And this was tested against numerical evaluation for random values of the positions $x_i^3$, using Mathematica.

\paragraph{Vertex-to-bulk diagram} This contribution is
\begin{align}
    L_{2} &= \frac{\ii\lambda\,\varepsilon_{+\mu\rho}}{2 (4\pi)^3 N} \left(\gamma^{\delta}\gamma^{\mu}\gamma^{\beta}\right)_{21} \frac{x_{03\alpha}\gamma_{22}^{\alpha}}{|x_{30}|^{3}}\int \frac{(y-x_2)_{\delta} (y-x_1)_{\beta} (x_3-y)^{\rho}}{|y-x_2|^3 |y-x_1|^3 |x_3-y|^3} \dd^3 y \nonumber\\
    &=\frac{-\lambda}{32\sqrt{2}\pi^2 |x_{30}|^2 |x_{31}|^2 |x_{32}|}\, ,
\end{align}
where we used \eqref{identity1}.

\paragraph{Full conformal result} We add the two integrals to get 
\begin{equation} 
    \langle\mathcal{M}_{32}^{\left(\frac{3}{2},-\frac{1}{2}\right)}\mathcal{M}_{10}^{\left(-\frac{1}{2},-\frac{1}{2}\right)}\rangle = \frac{-\lambda}{16\sqrt{2}\pi^{2}N\left|x_{20}\right|\left|x_{30}\right|\left|x_{31}\right|^{2}\left|x_{32}\right|} + O(\lambda^2)\, .
\end{equation}
Let us now compare this to the exact result \eqref{unequal equal}. We use Mathematica to guess the value of $\mathcal{F}_{\text{ue}}^{(1)}$ as $\lambda=2(\Delta-1)\to 0$. We then use \eqref{rel1} and \eqref{rel2} to determine $\mathcal{F}_{\text{ue}}^{(2)}$ and $\mathcal{F}_{\text{ue}}^{(3)}$. Thus, we find
\begin{align}
    \mathcal{F}_{\text{ue}}^{(1)}(\zeta) &= \frac{8\pi}{3} + O(\lambda)\,, \quad 
    \mathcal{F}_{\text{ue}}^{(2)}(\zeta) =\frac{4\pi}{3}(1+\zeta) + O(\lambda)\,,\\
    \mathcal{F}_{\text{ue}}^{(3)}(\zeta)&= \frac{4\pi}{1-\zeta}+O(\lambda)\,,\quad \widetilde{\mathcal{F}}_{\text{ue}}(\zeta) =O(1)\,,
\end{align}
which reproduces the perturbative result.
Furthermore, taking the limit $x_2\rightarrow x_1$ produces the three-point function with an insertion of the displacement operator, provided we also multiply by $\eta_-$. This gives
\begin{equation}\label{threeC12}
    \langle\bcO_{\frac{3}{2}}(x_3) \mathbb{D}_{-}(x_1) \cW\mathcal{O}_{-\frac{1}{2}}(x_0)\rangle = \frac{-\lambda^{2}}{8\pi N |x_{10}| |x_{30}| |x_{31}|^{3}}+O(\lambda^3)\,.
\end{equation}
This agrees with the non-perturbative result \eqref{struc} for $C_{-\frac{1}{2}}$.

\subsection{One-Loop Check for $\langle\mathcal{M}^{\left(\frac{1}{2},-\frac{1}{2}\right)}\mathcal{M}^{\left(\frac{1}{2},-\frac{1}{2}\right)}\rangle_c$} \label{app: pert unequal unequal}

In terms of fields, we have the following two mesonic lines:
\begin{equation}
    \cM_{32}^{(\frac{1}{2},-\frac{1}{2})} =\frac{1}{N} \bar \psi_2(x_3)\cW\psi_2(x_2)\,,\quad  \cM_{10}^{(\frac{1}{2},-\frac{1}{2})} =\frac{1}{N} \bar \psi_2(x_1)\cW\psi_2(x_0)\,.
\end{equation}
At tree level the connected part of the correlator is simply a product of two free propagators
\begin{equation}\label{unequaltree}
  \langle\mathcal{M}_{32}^{\left(\frac{1}{2},-\frac{1}{2}\right)}\mathcal{M}_{10}^{\left(\frac{1}{2},-\frac{1}{2}\right)}\rangle_{c,0} = \frac{1}{N(4\pi)^{2} |x_{30}|^{2} |x_{12}|^{2}}\, .
\end{equation}

Let us now compute the one-loop (connected) correlator and compare to the non-perturbative result from  Section \ref{sec:unequal-unequal}.

\paragraph{Bulk-to-bulk diagram}
This corresponds to
\begin{multline}
    T_{1} = \frac{\ii\lambda\,\varepsilon_{\mu\nu\rho}}{2(4\pi)^{4}N} \left(\gamma^{\eta}\gamma^{\nu}\gamma^{\delta}\right)_{22} \left(\gamma^{\beta}\gamma^{\mu}\gamma^{\alpha}\right)_{22} \\
    \times \int \frac{(y_1-x_2)_{\beta} (y_1-x_1)_{\alpha} (y_2-x_3)_{\delta} (x_0-y_2)_{\eta}\, y_{12}^{\rho}}{|y_1-x_2|^{3} |x_1-y_1|^{3} |x_{3}-y_2|^{3} |y_2-x_0|^{3} |y_{12}|^{3}}\, \dd^3 y_1 \, \dd^3 y_2\,.
\end{multline}
We are not able to compute this integral analytically, but using numerical checks in Mathematica for random values of $x_i$ we conjecture the result to be
\begin{equation}
    T_1 = \frac{-\lambda(|x_{32}|+|x_{10}|)}{2(4\pi)^2\,|x_{20}|\,|x_{21}|\,|x_{30}|^{2}\,|x_{31}|}\, .
\end{equation}

\paragraph{Bulk-to-line diagrams}
The bulk to line diagram can be seen in Figure \ref{digrams}(b). We have 4 such diagrams here, each line ending on either fermion propagator. For instance one of them takes the following integral form 
\begin{equation}
    T_2 = \frac{\ii\lambda\,\varepsilon_{\mu\nu\rho}\, |x_{32}|}{2 (4\pi)^3 N |x_{21}|^2} \left(\gamma^{\delta}\gamma^{\nu}\gamma^{\alpha}\right)_{22} \int_0^1 \int \frac{(y-x_3)_{\alpha} (x_0-y)_{\delta} (x_{\sigma}-y)^{\rho}}{|y-x_{3}+\epsilon\hat{e}_{3}|^3 |x_0-y|^{3} |x_{\sigma}-y|^{3}} \dd^3y\, \dd\sigma \,,
\end{equation}
with $x_\sigma = x_2+\sigma x_{32}$ and $\epsilon$ a point-splitting cut-off that  regularises divergences when $x_\sigma$ approaches $x_3$. Using the integral relation \eqref{identity1} and performing the line integration explicitly using Mathematica we find the same result as in \eqref{equalI2}. Summing with the rest of the diagrams, we find
\begin{multline}
    T_2+T_3+T_4+T_5=\frac{\lambda}{2(4\pi)^2\,
|x_{30}|^{2}\,
|x_{20}|\,
|x_{31}|\,
|x_{21}|^{2}}
\Bigg[
- |x_{20}|^{2}
- |x_{10}|^{2}
+ |x_{30}|\big(|x_{20}|+|x_{10}|\big)
\\+ 2\,|x_{20}|\,|x_{31}|\,
\ln\!\left(
\frac{|x_{20}|\,|x_{31}|\,\epsilon^{2}}
{|x_{30}|\,|x_{32}|\,|x_{21}|\,|x_{10}|}
\right)
\Bigg]\,.
\end{multline}

\paragraph{Propagator corrections}
We have two fermion propagator corrections by gluon exchange, see Figure \ref{digrams}(c). We find that they cancel one another.

\paragraph{Full conformal result}
Summing the previous results, the connected part of the correlator is given by
\begin{equation}
    \langle\cM_{32}^{\left(\frac{1}{2},-\frac{1}{2}\right)}\mathcal{M}_{10}^{\left(\frac{1}{2},-\frac{1}{2}\right)}\rangle_c = \frac{1 + \lambda\left(\frac{\chi}{1+\chi} + \ln\left(\frac{\epsilon^{2} (1+\chi)}{|x_{10}| |x_{32}|}\right)\right)}{(4\pi)^2 N|x_{21}|^2 |x_{30}|^2} + O(\lambda^2)\, .
\end{equation}
Let us now compare this to the result we have obtained in Section \ref{sec:unequal-unequal}, i.e.
\begin{multline}
    \langle\mathcal{M}_{32}^{\left(\frac{1}{2},-\frac{1}{2}\right)}\mathcal{M}_{10}^{\left(\frac{1}{2},-\frac{1}{2}\right)}\rangle_c = \frac{c_+^2\, \sin^3(\pi\Delta) \cos(\pi\Delta)}{2\pi^4(1-\Delta)^2 N\, |x_{32}|^{2\Delta} |x_{10}|^{2\Delta}}\\
    \times\left[\mathcal{F}_{\text{uu}}(\chi)+\mathcal{G}_{\text{uu}}(\chi) + \frac{2(1-\Delta)}{2\Delta-1} \widetilde{\mathcal{F}}_{\text{uu}}(\chi)\right].
\end{multline}
First, we note that the expansion of the constant prefactor is
\begin{equation}
    \frac{c_+^2\, \sin^3(\pi\Delta) \cos(\pi\Delta)}{2\pi^4(1-\Delta)^2} = \frac{\lambda(1+\lambda)}{64\pi^3} + O(\lambda^3)\, .
\end{equation}
Then, we point out that $\widetilde{\mathcal{F}}_{\text{uu}}(\chi) = O(1)$ is not relevant to the order we are working at, and that, using \eqref{identity1}, $\mathcal{G}_{\text{uu}}(\chi) = \frac{4\pi\chi}{1+\chi} + O(\lambda)$. This last point implies, by 
solving \eqref{ode unequal-unequal} perturbatively using the boundary condition $\mathcal{F}_{\text{uu}}(\chi)\sim b\chi^{2\Delta}$ when $\chi\to+\infty$ with $b$ given by \eqref{boundary unequal}, that
\begin{equation}
    \mathcal{F}_{\text{uu}}(\chi) = 4\pi \left[\frac{\chi^2}{\lambda} + \chi^2 \ln(1+\chi) - \chi\right] +O(\lambda)\, .
\end{equation}
This means that
\begin{equation}
    \frac{\mathcal{F}_{\text{uu}}(\chi)+\mathcal{G}_{\text{uu}}(\chi)}{|x_{32}|^{2\Delta} |x_{10}|^{2\Delta}} \epsilon^{2\lambda} = \frac{4\pi}{\lambda  |x_{21}|^2 |x_{30}|^2}\left[ 1 + \lambda\left(\ln\left(\frac{\epsilon^{2}(1+\chi)}{|x_{10}| |x_{32}|}\right) -\frac{1}{1+\chi}\right)\right]\, .
\end{equation}
Combining the previous equations, we recover the result obtained through Feynman integral computations.

\section{More Four-Point Functions}
\label{app:4pt}

In this appendix, we compute more general four-point functions than the ones presented in Section \ref{sec:bootstrap2lines}.

\subsection{General Equal-Equal Case}
\label{app: equal-equal-gen}

It possible to generalise the result of Section \ref{sec:equal-equal} to any configuration of spins such that $s_3 s_2 >0$ and $s_1 s_0 >0$. We derive the explicit result for $s_3,s_2<0$ and $s_1,s_0>0$. The configurations with opposite signs of the spins can simply be reached by a conformal transformation that exchanges $(x_3,x_2)\leftrightarrow (x_1,x_0)$.

Our starting point is the equation
\begin{multline}\label{variation ee}
    q\left(\p_{x_2^3}^{2} - \p_{x_3^3}^{2} + \p_{x_0^3}^{2} - \p_{x_1^3}^{2}\right) \langle\mathcal{M}_{32}^{(s_3,s_2)} \mathcal{M}_{10}^{(s_1,s_0)}\rangle\\
    =\! \frac{5a_1}{N}\!\! \int \!\! \left[\langle\mathcal{M}_{32}^{(s_3,s_2)}\p_3 J_0(y)\rangle \langle\mathcal{M}_{10}^{(s_1,s_0)} J_1^3(y)\rangle\! +\! \langle\mathcal{M}_{32}^{(s_3,s_2)} J_1^3(y)\rangle \langle\mathcal{M}_{10}^{(s_1,s_0)}\p_3 J_0(y)\rangle\right]\!\dd^{3}y\,.
\end{multline}
The absence of double-trace terms and line integrations is explained by the same arguments as in Section \ref{sec:equal-equal}. Inputting the correlators from Appendix \ref{app:corr}, and following similar steps as in Section \ref{sec:bootstrap2lines}, we arrive at
\begin{multline} \label{app: gen equal equal}
   \<\mathcal{M}_{32}^{(s_3,s_2)} \mathcal{M}_{10}^{(s_1,s_0)}\>
    =\frac{ 2^{s_{0}+s_{1}+1}\,\sin^3(\pi \Delta)\cos(\pi \Delta)}{\pi^4\,\eta_{-}\eta_{+}\left|x_{31}\right|^{2\Delta+2s_{1}-1} \left|x_{30}\right|^{4-4\Delta+2(s_0+s_2)} \left|x_{20}\right|^{2\Delta-2s_{2}-1}}\\
    \times \Big[2 (\sfrac{3}{2} -\Delta)_{\frac{1}{2}-s_{3}} (\Delta-\sfrac{1}{2})_{\frac{1}{2}-s_{2}} (\Delta)_{-\frac{1}{2}+s_{1}} (1-\Delta)_{\frac{1}{2}+s_{0}} \mathcal{\widetilde{H}}_{\text{ee}}(\zeta) - \mathcal{G}_{\text{ee}}(\zeta)\Big]\,,
\end{multline}
where $\zeta=x^3_{21} x^3_{30}/(x^3_{31} x^3_{20}) = 1/(1+\chi)$ and
\begin{equation}
\mathcal{\widetilde{H}}_{\text{ee}}(\zeta)=\int\frac{r^{2\left(s_{0}+s_{1}\right)}\dd^{3}y}{\left|y\right|^{2\Delta-2s_{2}}\left|y-\zeta\hat{e}_{3}\right|^{2\Delta+2s_{1}-1}\left|y-\hat{e}_{3}\right|^{3+2s_0-2\Delta}}\,. 
\end{equation}
The explicit form of $\mathcal{G}_{\text{ee}}(\zeta)$ depends on whether $s_2=-s_1$ or $s_2\neq -s_1$. We point out that in the $s_2=-s_1$ case, the identity operator is among the exchanged operator, and that the homogeneous part of \eqref{variation ee}, i.e. with vanishing right-hand side, admits non-trivial solutions.

Let us first look at the case $s_2=-s_1 \Leftrightarrow s_3=-s_0$. We find that $\mathcal{G}_{\text{ee}}(\zeta)$ reads
\begin{equation}
    \mathcal{G}_{\text{ee}}(\zeta) = -\frac{4}{s_0+1-\Delta} \left[(\sfrac{3}{2}-\Delta)_{\frac{1}{2}+s_0} (\Delta-\sfrac{1}{2})_{\frac{1}{2}+s_1} \right]^2 \mathcal{K}_{\text{ee}}(\zeta)\,,
\end{equation}
with $\mathcal{K}_{\text{ee}}$ given by the first order differential equation
\begin{equation}
\mathcal{K}_{\text{ee}}(\zeta)(2-4\Delta-4s_{1})-2\zeta\mathcal{K}_{\text{ee}}'(\zeta)=\mathcal{K}_{\text{aux}}(\zeta)(3-4\Delta-2(s_{1}-s_{0}))-\zeta\mathcal{K}_{\text{aux}}'(\zeta)\,,
\end{equation}
and 
\begin{equation}
\mathcal{K}_{\text{aux}}(\zeta)=\int\frac{r^{2\left(s_{0}+s_{1}\right)}\dd^{3}y}{\left|y\right|^{2\Delta+2s_{1}}\left|y-\zeta\hat{e}_{3}\right|^{2\Delta+2s_{1}}\left|y-\hat{e}_{3}\right|^{2+2s_{0}-2\Delta}}\,.
\end{equation}
The equation for $\mathcal{K}_{\text{ee}}$ has a homogeneous solution with an integration constant which can be fixed by examining the $\zeta\rightarrow1^-$ limit of the full correlator. This is the OPE limit in which $x_3\rightarrow x_2$ (or $x_1\rightarrow x_0$). We show later in this section a consistency check with the main example, namely, $s_0=s_1=1/2$.

When $s_2\neq -s_1$, we find instead 
\begin{multline}
    \mathcal{G}_{\text{ee}}(\zeta) = \frac{1}{(s_{1}+s_{2})\zeta}\times\\\Bigg[(\sfrac{3}{2}-\Delta)_{\frac{1}{2}-s_{3}} (\Delta-\sfrac{1}{2})_{\frac{1}{2}-s_{2}} \bigg((s_{0}-\Delta+1) (\Delta-\sfrac{1}{2})_{\frac{1}{2}+s_{1}} (\sfrac{3}{2}-\Delta)_{-\frac{1}{2}+s_{0}} \zeta\,\mathcal{H}_{\text{ee}}^{(1)}(\zeta)\\
    +(\Delta-\sfrac{1}{2})_{\frac{1}{2}+s_{1}} (\sfrac{3}{2}-\Delta)_{\frac{1}{2}+s_{0}} \mathcal{H}_{\text{ee}}^{(2)}(\zeta)\bigg)\\
    +\frac{(\Delta-1)_{\frac{1}{2}+s_{1}} (1-\Delta)_{\frac{1}{2}+s_{0}}}{\tan^{2}(\pi\Delta)} \bigg(-(2-\Delta)_{\frac{1}{2}-s_{3}} (\Delta)_{-\frac{1}{2}-s_{2}} (s_{2}-\Delta+1) \mathcal{H}_{\text{ee}}^{(3)}(\zeta)\\
    +(1-s_{3}-\Delta)(2-\Delta)_{-\frac{1}{2}-s_{3}} (\Delta)_{\frac{1}{2}-s_{2}} \zeta\,\mathcal{H}_{\text{ee}}^{(4)}(\zeta) \bigg)\Bigg]\,,
\end{multline}
with 
\begin{align}
\mathcal{H}_{\text{ee}}^{(1)}(\zeta)&=\int\frac{r^{2\left(s_{0}+s_{1}\right)}\dd^{3}y}{\left|y\right|^{2\Delta-2s_{2}}\left|y-\zeta\hat{e}_{3}\right|^{2\Delta+2s_{1}}\left|y-\hat{e}_{3}\right|^{2-2\Delta+2s_{0}}}\,,\\\mathcal{H}_{\text{ee}}^{(2)}(\zeta)&=\int\frac{r^{2\left(s_{0}+s_{1}\right)}\dd^{3}y}{\left|y\right|^{2\Delta-2s_{2}}\left|y-\zeta\hat{e}_{3}\right|^{2\Delta+2s_{1}-2}\left|y-\hat{e}_{3}\right|^{4-2\Delta+2s_{0}}}\,,\\\mathcal{H}_{\text{ee}}^{(3)}(\zeta)&=\int\frac{r^{2\left(s_{0}+s_{1}\right)}\dd^{3}y}{\,\left|y\right|^{2\Delta-2s_{2}-1}\left|y-\zeta\hat{e}_{3}\right|^{2\Delta+2s_{1}-1}\left|y-\hat{e}_{3}\right|^{3-2\Delta+2s_{0}}}\,,\\\mathcal{H}_{\text{ee}}^{\left(4\right)}(\zeta)&=\int\frac{r^{2\left(s_{0}+s_{1}\right)}\dd^{3}y}{\left|y\right|^{2\Delta-2s_{2}+1}\left|y-\zeta\hat{e}_{3}\right|^{2\Delta+2s_{1}-1}\left|y-\hat{e}_{3}\right|^{3-2\Delta+2s_{0}}}\,.
\end{align}
However, these functions are not independent. They are related via
\begin{equation}
(2s_{2}-2\Delta+1)\mathcal{H}_{\text{ee}}^{(1)}(\zeta)-\zeta(\mathcal{H}_{\text{ee}}^{(1)})'(\zeta) + K_1 [z(\mathcal{H}_{\text{ee}}^{(4)})'(\zeta)
    +(2\Delta+2s_{1}-1)\mathcal{H}_{\text{ee}}^{(4)}(\zeta)] = 0\,,
\end{equation}
\begin{equation}
    (2\Delta+2s_{1}-2) \mathcal{H}_{\text{ee}}^{(2)}(\zeta) + \zeta(\mathcal{H}_{\text{ee}}^{(2)})'(\zeta) + K_{2}((2-2\Delta+2s_{2})\mathcal{H}_{\text{ee}}^{(3)}(\zeta) - \zeta(\mathcal{H}_{\text{ee}}^{(3)})'(\zeta))=0\,,
\end{equation}
with the auxiliary constants 
\begin{equation}
    K_{1}=\frac{(s_{0}-\Delta+1)_{\frac{1}{2}}(\Delta-s_{2})_{\frac{1}{2}}}{\left(s_{1}+\Delta-\frac{1}{2}\right)_{\frac{1}{2}} \left(\sfrac{3}{2}-s_{3}-\Delta\right)_{\frac{1}{2}}},\quad K_{2}=\frac{(s_{1}+\Delta-1)_{\frac{1}{2}}(2-s_{3}-\Delta)_{\frac{1}{2}}}{\left(s_{0}-\Delta+\frac{3}{2}\right)_{\frac{1}{2}}\left(\Delta-s_{2}-\frac{1}{2}\right)_{\frac{1}{2}}}\,.
   \end{equation}
We have verified this equations numerically We finish this section with several consistency checks.

\paragraph{The case $s_3=s_2=-1/2$ and $s_1=s_0=1/2$}  Plugging the explicit spin values we find obtain
\begin{equation}
  \<\mathcal{M}_{32}^{\left(-\frac{1}{2},-\frac{1}{2}\right)}\mathcal{M}_{10}^{\left(\frac{1}{2},\frac{1}{2}\right)}\rangle=\frac{5a_{1}d_{0}\eta_{-}d_{1}\left(3-2\Delta\right)\left(2\Delta-1\right)}{2qN\eta_{+}\left|x_{31}\right|^{2\Delta}\left|x_{30}\right|^{4-4\Delta}\left|x_{20}\right|^{2\Delta}}\big[(\Delta-1)\widetilde{\mathcal{H}}_{\text{ee}}(\zeta)+(1-2\Delta)\mathcal{K}_{\text{ee}}(\zeta)\big]\,,
\end{equation}
with an equation that determines $\mathcal{K}_{\text{ee}}$ given by
\begin{equation}
4\Delta\mathcal{K}_{\text{ee}}(\zeta)+2\zeta\mathcal{K}'_{\text{ee}}(\zeta)=\zeta\mathcal{K}_{\text{aux}}'(\zeta)+(4\Delta-3)\mathcal{K}_{\text{aux}}(\zeta)\,.
\end{equation}
The general solution has a homogeneous solution  $c\, \zeta^{-2\Delta}$ which by examining the OPE limit $x_3\rightarrow x_2$ is fixed with  $c=\pi^{2}\left(4(\Delta-1)^{2}+1\right)\cot(\pi\Delta)/(3-2\Delta)(2\Delta-1)^{2}$. 
This should coincide with the result from Section \ref{sec:equal-equal}. Comparing the current expression to the result~\eqref{equal equal}, we see that they match because
\begin{equation}
    \mathcal{\widetilde{F}}_{\text{ee}}(\chi) = \frac{\Delta-1}{(1+\chi)^{2\Delta}} \widetilde{\mathcal{H}}_{\text{ee}}\left(\frac{1}{1+\chi}\right) \quad\text{and}\quad
    \mathcal{F}_{\text{ee}}(\chi) = \frac{2\Delta-1}{(1+\chi)^{2\Delta}} \mathcal{K}_{\text{ee}}\left(\frac{1}{1+\chi}\right)\, ,
\end{equation}
as follows from a simple change of variables.

\paragraph{The case $s_3=-1/2, s_2=-3/2$ and $s_1=1/2, s_0=3/2$}

We can perform a simple consistency check in which we check that in the limit $\zeta\rightarrow1^-$ we have agreement with the OPE limit when $x_1\rightarrow x_0$. On the one hand, taking the $\zeta\rightarrow1^-$ limit in \eqref{app: gen equal equal} and using relation \eqref{twopropstar} for the integrals gives
\begin{equation}
    \langle\mathcal{M}_{32}^{\left(-\frac{1}{2},-\frac{3}{2}\right)} \mathcal{M}_{10}^{\left(\frac{1}{2},\frac{3}{2}\right)}\rangle = \frac{-3(\Delta-1)^{2}\,d_{1}^{2}\,\eta_{-}}{8\,\eta_{+}\,\mathcal{N}_{1}\left|x_{30}\right|^{4-2\Delta}\left|x_{20}\right|^{2\Delta+2}}\,,
\end{equation}
On the other hand, from the OPE expansion, we have
\begin{multline}
    \langle\mathcal{M}_{32}^{\left(-\frac{1}{2},-\frac{3}{2}\right)}\mathcal{M}_{10}^{\left(\frac{1}{2},\frac{3}{2}\right)}\rangle\sim\frac{\langle\mathcal{M}^{\left(\frac{1}{2},\frac{3}{2}\right)}\partial_{-}J_{1}^{+}\rangle}{\langle\partial_{-}J_{1}^{+}\partial_{+}J_{1}^{-}\rangle}\langle\mathcal{M}_{32}^{\left(-\frac{1}{2},-\frac{3}{2}\right)}\partial_{+}J_{1}^{-}(x_0)\rangle\\+\frac{\langle\mathcal{M}^{\left(\frac{1}{2},\frac{3}{2}\right)}J_{2}^{++}\rangle}{\langle J_{2}^{++}J_{2}^{--}\rangle}\langle\mathcal{M}_{32}^{\left(-\frac{1}{2},-\frac{3}{2}\right)}J_{2}^{--}(x_0)\rangle\,,
\end{multline}
which by plugging the correlators and OPE data is in agreement. For the reader's convenience we provide the missing data:
\begin{equation}
    \langle\mathcal{M}^{\left(\frac{1}{2},\frac{3}{2}\right)}J_{2}^{++}\rangle=\frac{-3\,d_{2}\,\eta_{-}}{\eta_{+}}\, ,\qquad\langle\mathcal{M}_{32}^{\left(-\frac{1}{2},-\frac{3}{2}\right)}J_{2}^{--}\left(x_{0}\right)\rangle=\frac{3\,d_{2}}{|x_{30}|^{2\Delta+2}|x_{20}|^{4-2\Delta}}\, .
\end{equation}

\subsection{Calculations for the Unequal-Equal Case}
\label{app:unequal-equal}

Elementary manipulations---the only non-trivial step being the use of \eqref{F2eqdiff} ---allow us to first rewrite the right-hand side of \eqref{Quneqeq} as
\begin{align*}
    \int \!&\left[\langle\mathcal{M}_{32}^{\left(\frac{3}{2},-\frac{1}{2}\right)} \p_3 J_0(y)\rangle \langle\mathcal{M}_{10}^{\left(-\frac{1}{2},-\frac{1}{2}\right)} J_1^3(y)\rangle +\langle\mathcal{M}_{32}^{\left(\frac{3}{2},-\frac{1}{2}\right)} J_1^3(y)\rangle \langle\mathcal{M}_{10}^{\left(-\frac{1}{2},-\frac{1}{2}\right)} \p_3 J_0(y)\rangle \right]\! \dd^3 y\\
    &=  d_0\, d_1\, \eta_-\, c_+\Bigg[ \left(\p^2_{x^3_0} - \p^2_{x^3_1}\right) \int \frac{(\Delta+\sfrac{1}{2}) F_1(u_{32})\, x^+_4 x^-_4\, \dd^3 x_4}{|x_{32}|^{2\Delta-2} |x_{43}|^{4} |x_{42}|^{2} |x_{41}|^{4-2\Delta} |x_{40}|^{2\Delta}}\\
    &+ \left(\p^2_{x^3_2} - \p^2_{x^3_3}\right) \int \frac{\Delta(3-2\Delta)\, |x_{10}| \, F_3(u_{32})\, x^+_4 x^-_4\, \dd^3 x_4}{|x_{32}|^{2\Delta-1} |x_{43}|^3 |x_{42}| |x_{41}|^{5-2\Delta} |x_{40}|^{2\Delta+1}}\\
    &+ \frac{(4\Delta^2-1)(3-2\Delta)}{4(\Delta-1)} \left(\p_{x^3_0} + \p_{x^3_1}\right) \Bigg( \int \frac{(x^3_{40} + x^3_{41}) \, F_1(u_{32})\, x^+_4 x^-_4\, \dd^3 x_4}{|x_{32}|^{2\Delta-2} |x_{43}|^4 |x_{42}|^2 |x_{41}|^{5-2\Delta} |x_{40}|^{2\Delta+1}}\\
    &+\left(\p_{x^3_2} - \frac{2\Delta-1}{2\Delta+1} \p_{x^3_3} - \frac{4\Delta}{(2\Delta+1)|x_{32}|}\right) \int \frac{|x_{10}| \, F_2(u_{32})\, x^+_4 x^-_4\, \dd^3 x_4}{|x_{32}|^{2\Delta-1} |x_{43}|^3 |x_{42}| |x_{41}|^{5-2\Delta} |x_{40}|^{2\Delta+1}}\Bigg)\Bigg]\, .
\end{align*}
We now notice that the integrals appearing in the second and third lines are equal, as can be shown using the integral relation \eqref{2F1 integral} and the star-triangle relation. We then perform a conformal change of variables in each of the integrals so that equation \eqref{Quneqeq} now reads
\begin{align}\label{uneq-eq full diffeq}
    q&\left(\p_{x_2^3}^2 - \p_{x_3^3}^2 + \p_{x_0^3}^2 - \p_{x_1^3}^2\right) \langle\cM_{32}^{\left(\frac{3}{2},-\frac{1}{2}\right)} \cM_{10}^{\left(-\frac{1}{2},-\frac{1}{2}\right)}\rangle + \frac{\bar{\rho}^{(2)}_{\frac{3}{2},\frac{1}{2}}}{N} \<\cM_{32}^{\left(\frac{1}{2},-\frac{1}{2}\right)}\> \<\p_3J^-_1(x_3) \cM_{10}^{\left(-\frac{1}{2},-\frac{1}{2}\right)}\> \nonumber\\
    &= \frac{5(2\Delta+1)\,a_1\,d_0\, d_1\, \eta_-\, c_+}{8(\Delta-1)N} \Bigg[\left(\p_{x_2^3}^2 - \p_{x_3^3}^2 + \p_{x_0^3}^2 - \p_{x_1^3}^2\right) \frac{2 (\Delta-1) \widetilde{\mathcal{F}}_{\mathrm{ue}}(\zeta)}{|x_{32}|^{2\Delta-1} |x_{31}|^{4-2\Delta} |x_{30}|^{2\Delta-1} |x_{20}|}\nonumber\\
    &+ \left(\p_{x^3_0} + \p_{x^3_1}\right)\! \Bigg(\!\p_{x^3_0} \frac{(3-2\Delta) \mathcal{F}^{(1)}_{\mathrm{ue}}(\zeta)}{|x_{32}|^{2\Delta-1} |x_{31}|^{5-2\Delta} |x_{30}|^{2\Delta-2} |x_{20}|} + \p_{x^3_1} \frac{(2\Delta-1) \mathcal{F}^{(2)}_{\mathrm{ue}}(\zeta)}{|x_{32}|^{2\Delta-1} |x_{31}|^{3-2\Delta} |x_{30}|^{2\Delta} |x_{20}|}\nonumber\\
    &+\left(\p_{x^3_2} - \frac{2\Delta-1}{2\Delta+1} \p_{x^3_3} - \frac{4\Delta}{(2\Delta+1)|x_{32}|}\right) \frac{(2\Delta-1) (3-2\Delta) |x_{10}| \mathcal{F}^{(3)}_{\mathrm{ue}}(\zeta)}{|x_{32}|^{2\Delta-2} |x_{31}|^{5-2\Delta} |x_{30}|^{2\Delta-1} |x_{20}|^2}\Bigg) \Bigg]\, ,
\end{align}
where $\zeta=x^3_{21} x^3_{30}/x^3_{31} x^3_{20}$. The factorised term is simply
\begin{equation}
    \frac{\bar{\rho}^{(2)}_{\frac{3}{2},\frac{1}{2}}}{N} \<\cM_{32}^{\left(\frac{1}{2},-\frac{1}{2}\right)}\> \<\p_3 J^-_1(x_3) \cM_{10}^{\left(-\frac{1}{2},-\frac{1}{2}\right)}\> = -\frac{q\, d^2_1\, \eta_-\, c_+}{8\,N\,\mathcal{N}_1\, |x_{32}|^{2\Delta}} \p_{x^3_3} \frac{1}{|x_{31}|^{4-2\Delta} |x_{30}|^{2\Delta}}\, ,
\end{equation}
and, from conformal covariance, we may write
\begin{equation}
    \langle\cM_{32}^{\left(\frac{3}{2},-\frac{1}{2}\right)} \cM_{10}^{\left(-\frac{1}{2},-\frac{1}{2}\right)}\rangle = \frac{(2\Delta+1) d^2_1\, \eta_-\, c_+ \tan(\pi\Delta)}{32\pi^2\, N\, \mathcal{N}_1 |x_{32}|^{2\Delta-1} |x_{31}|^{4-2\Delta} |x_{30}|^{2\Delta-1} |x_{20}|} \times \mathcal{G}_{\mathrm{ue}}(\zeta)
\end{equation}
for some function $\mathcal{G}_{\mathrm{ue}}$ that we seek to compute. Plugging this in equation \eqref{uneq-eq full diffeq}, we obtain three independent differential equations. One of them is trivially solved to give 
\begin{equation}
    \mathcal{G}_{\mathrm{ue}}(\zeta) = \frac{b_2}{\zeta}
    - \widetilde{\mathcal{F}}_{\text{ue}}(\zeta) + \frac{(2\Delta-1)(2\Delta-3)}{4(\Delta-1)} \left( \frac{\mathcal{F}^{(1)}_{\text{ue}}(\zeta)}{2\Delta-1} + \frac{\mathcal{F}^{(2)}_{\text{ue}}(\zeta)}{(2\Delta-3)\zeta} + \frac{(1-\zeta)^2}{\zeta} \mathcal{F}^{(3)}_{\text{ue}}(\zeta)\right)\, ,
\end{equation}
where $b_2$ is an integration constant that can be fixed by considering the limit $x_1\to x_0$ or $x_2\to x_3$, i.e. $\zeta\to 1^-$. In such a limit, the functions behave as
\begin{equation}
    \widetilde{\mathcal{F}}_{\text{ue}}(1) = \mathcal{F}^{(1)}_{\text{ue}}(1) = \mathcal{F}^{(2)}_{\text{ue}}(1)\, ,\quad \mathcal{F}^{(3)}_{\text{ue}}(\zeta) = O(\sfrac{1}{1-\zeta})\, ,
\end{equation}
so that $\mathcal{G}_{\mathrm{ue}}(1) = b_2$. On the other hand, the OPE is dominated by the exchange of $J^\pm_1$ and the correlator becomes
\begin{equation}
    \lim_{x_1\to x_0} \langle\cM_{32}^{\left(\frac{3}{2},-\frac{1}{2}\right)} \cM_{10}^{\left(-\frac{1}{2},-\frac{1}{2}\right)}\rangle = \frac{d^2_1\, \eta_-\, c_+}{16(2\Delta-1)\,N\,\mathcal{N}_1\, |x_{32}|^{2\Delta-1} |x_{30}|^3 |x_{20}|}\, .
\end{equation}
Matching the two results fixes
\begin{equation}
    b_2 = \frac{2\pi^2 \cot(\pi\Delta)}{4\Delta^2-1}\, ,
\end{equation}
and we have thus proved \eqref{unequal equal}.

The other two constraints deduced from \eqref{uneq-eq full diffeq} can then be rewritten as non-trivial first-order differential equations satisfied by the functions 
$\mathcal{F}^{(i)}_{\mathrm{ue}}$. We show them here for completeness:
\begin{multline}\label{rel1}
    (1-\zeta)(2\Delta-1) [\zeta(1-\zeta)(\mathcal{F}^{(3)}_{\mathrm{ue}})'(\zeta) + (2+(2\Delta-5)\zeta)\mathcal{F}^{(3)}_{\mathrm{ue}}(\zeta)]\\
    + (2\Delta+1) \zeta \mathcal{F}^{(1)}_{\mathrm{ue}}(\zeta) = \frac{8\pi^2 (\Delta-1)\cot(\pi\Delta)}{(2\Delta-1)(3-2\Delta)}\, ,
\end{multline}
\begin{multline}\label{rel2}
    \zeta(1-\zeta) \left[(3-2 \Delta) \zeta (\mathcal{F}^{(1)}_{\mathrm{ue}})'(\zeta)+(1-2 \Delta) \left((\mathcal{F}^{(2)}_{\mathrm{ue}})'(\zeta)+2 (2 \Delta -3) (1-\zeta) (\mathcal{F}^{(3)}_{\mathrm{ue}})'(\zeta)\right)\right]\\
    +2 (4\Delta^2 -8\Delta +3) (1+2
   (\Delta -2) \zeta) (\zeta-1) \mathcal{F}^{(3)}_{\mathrm{ue}}(z\zeta) +(2 \Delta -3) (\zeta-2 (\Delta +1))  \zeta \mathcal{F}^{(1)}_{\mathrm{ue}}(\zeta)\\
   - (2 \Delta -1) (2+(2 \Delta -3) \zeta) \mathcal{F}^{(2)}_{\mathrm{ue}}(\zeta)  = \frac{16\pi^2 (\Delta-1)\cot(\pi\Delta)}{4\Delta^2-1}\, .
\end{multline}
We have not tried to prove these relations but we tested them numerically.

\subsection{Additional Unequal-Unequal Configurations}
\label{app:unequal-unequal}

We consider here the simple case where $s_3=-s_2=s_1=-s_0=-s>0$. Repeating the derivation of Section \ref{sec:unequal-unequal}, we find that
\begin{equation}
    \frac{\langle\cM_{32}^{(-s,s)} \cM_{10}^{(-s,s)}\rangle_c}{\langle\cM_{32}^{(-s,s)}\rangle \langle\cM_{10}^{(-s,s)}\rangle} = \frac{\sin^3(\pi\Delta) \cos(\pi\Delta)}{2\pi^4 N (1-\Delta_s)} \left(\frac{\mathcal{F}_{\text{uu},s}(\chi) + \mathcal{\mathcal{G}}_{\text{uu},s}(\chi)}{1-\Delta_s} + \frac{2\mathcal{\widetilde{F}}_{\text{uu},s}(\chi)}{2\Delta_{s}-1}\right)\,,
\end{equation}
where  $\Delta_{s}=\Delta-s-1/2$ and 
\begin{equation}
\mathcal{G}_{\text{uu},s}(\chi) = \chi\int\frac{F_{1,s}\left(\frac{1-y^{3}}{\left|y-\hat{e}_{3}\right|}\right)F_{2,s}\left(\frac{y\cdot\left(y+\chi\hat{e}_{3}\right)}{\left|y+\chi\hat{e}_{3}\right|\left|y\right|}\right)}{\left|y+\chi\hat{e}_{3}\right|\left|y\right|\left|y-\hat{e}_{3}\right|^{2}} \dd^{3}y\,,
\end{equation}
\begin{equation}
\mathcal{\widetilde{F}}_{\text{uu},s}(\chi)=\chi\int\frac{F_{1,s}\left(\frac{1-y^{3}}{\left|y-\hat{e}_{3}\right|}\right)F_{3,s}\left(\frac{y\cdot\left(y+\chi\hat{e}_{3}\right)}{\left|y+\chi\hat{e}_{3}\right|\left|y\right|}\right)}{\left|y+\chi\hat{e}_{3}\right|\left|y\right|\left|y-\hat{e}_{3}\right|^{2}} \dd^3 y\,.
\end{equation}
The definitions for $F_{1,s},F_{2,s}$ and $F_{3,s}$ can be found in \eqref{2f1gen1} and \eqref{2f1gen2}. To obtain $\mathcal{F}_{\text{uu}}$ one solves
\begin{equation}
    \chi\mathcal{F}'_{\text{uu},s}(\chi)-2\Delta_s \mathcal{F}_{\text{uu},s}(\chi) = \mathcal{{G}}_{\text{uu},s}(\chi)\, .
\end{equation}
This equation has a homogeneous solution that can be fixed by considering the limit $x_2\rightarrow x_1$.

\paragraph{Generalisation with a double-trace}
An interesting generalisation is one where the double-trace terms in the action of the pseudo-charge on the boundary operators contribute to the four-point function. For simplicity, we focus on particular spin choices, namely $\langle\cM_{32}^{\left(\frac{3}{2},-\frac{1}{2}\right)} \cM_{10}^{\left(\frac{1}{2},-\frac{3}{2}\right)}\rangle$, but the derivation can be extended. The relevant
boundary equations are  
\begin{equation}
\left[Q_{33}^{(3)},\overline{\mathcal{O}}_{\frac{3}{2}}\right]=-q\delta_{3}^{2}\overline{\mathcal{O}}_{\frac{3}{2}}+\frac{\bar{\rho}^{(2)}_{\frac{3}{2},\frac{1}{2}}}{N}:\p_{3}J_{1}^{-}\overline{\mathcal{O}}_{\frac{1}{2}}: + \dots\,,
\end{equation}
and similarly for the fundamental operator
\begin{equation}
\left[Q_{33}^{(3)},\mathcal{O}_{-\frac{3}{2}}\right]=q\delta_{3}^{2}\mathcal{O}_{-\frac{3}{2}}+\frac{\rho^{(2)}_{-\frac{3}{2},-\frac{1}{2}}}{N}:\p_{3}J_{1}^{+}\mathcal{O}_{-\frac{1}{2}}:+\dots\,.
\end{equation}
We did not include the admissible terms $:J^{-3}_2\overline\cO_{\frac12}:$ or $:J^{+3}_2\cO_{-\frac12}:$ as they can be shown to come with a vanishing coefficient, using the techniques of Appendix~\ref{app:coeffs}. The value of $\bar\rho^{(2)}_{\frac{3}{2},\frac{1}{2}}$ is also fixed in Appendix \ref{app:J1dt} and a short side calculation shows that $\bar\rho^{(2)}_{\frac{3}{2},\frac{1}{2}}=-\rho^{(2)}_{-\frac{3}{2},-\frac{1}{2}}$. The equation for the correlator thus takes the form 
\begin{multline}
    q\left(\p_{x_{2}^{3}}^{2}-\p_{x_{3}^{3}}^{2}+\p_{x_{0}^{3}}^{2}-\p_{x_{1}^{3}}^{2}\right)\langle\cM_{32}^{\left(\frac{3}{2},-\frac{1}{2}\right)} \cM_{10}^{\left(\frac{1}{2},-\frac{3}{2}\right)}\rangle\\
    + \frac{\bar{\rho}^{(2)}_{\frac{3}{2},\frac{1}{2}}}{N}\langle\cM_{32}^{\left(\frac{1}{2},-\frac{1}{2}\right)}\rangle\langle\cM_{10}^{\left(\frac{1}{2},-\frac{3}{2}\right)}\p_{3}J_{1}^{-}\left(x_{3}\right)\rangle + \frac{\rho^{(2)}_{-\frac{3}{2},-\frac{1}{2}}}{N}\langle\cM_{32}^{\left(\frac{3}{2},-\frac{1}{2}\right)} \p_{3}J_{1}^{+}(x_0)\rangle\langle\cM_{10}^{\left(\frac{1}{2},-\frac{1}{2}\right)}\rangle\\
    = \frac{5a_{1}}{N}\int\left[\langle\cM_{32}^{\left(\frac{3}{2},-\frac{1}{2}\right)}\p_{3}J_{0}(y)\rangle\langle\cM_{10}^{\left(\frac{1}{2},-\frac{3}{2}\right)}J_{1}^{3}(y)\rangle+\langle\cM_{32}^{\left(\frac{3}{2},-\frac{1}{2}\right)}J_{1}^{3}(y)\rangle\langle\cM_{10}^{\left(\frac{1}{2},-\frac{3}{2}\right)}\p_{3}J_{0}(y)\rangle\right]\dd^3 y\, .
\end{multline}
Using details from
Appendix~\ref{app:corr} and \cite{Ferrando:2025ufj}, one can show that
\begin{multline}
    \frac{\bar{\rho}^{(2)}_{\frac{3}{2},\frac{1}{2}}}{N} \langle\cM_{32}^{\left(\frac{1}{2},-\frac{1}{2}\right)}\rangle \langle\cM_{10}^{\left(\frac{1}{2},-\frac{3}{2}\right)}\p_3 J_1^-(x_3) \rangle + \frac{\rho^{(2)}_{-\frac{3}{2},-\frac{1}{2}}}{N}\langle\cM_{32}^{\left(\frac{3}{2},-\frac{1}{2}\right)}\p_3 J_1^+(x_0) \rangle \langle\cM_{10}^{\left(\frac{1}{2},-\frac{1}{2}\right)}\rangle\\
    = \frac{c_+^2 \eta_{-}\, d_1 \bar{\rho}^{(2)}_{\frac{3}{2},\frac{1}{2}}}{2(2\Delta-1)N} \left(\partial_{x_{2}^{3}}^{2}-\p_{x_3^3}^{2} + \p_{x_0^3}^{2}-\p_{x_1^3}^{2}\right)\frac{\chi\,{}_2F_{1}\!\left[\genfrac{}{}{0pt}{1}{1,1-2\Delta}{2-2\Delta};-\chi\right]}{(2-4\Delta) |x_{32}|^{2\Delta} |x_{10}|^{2\Delta} |x_{30}|^{2}}\,.
\end{multline}
Combining together with
the right-hand side contributions, we find 
\begin{multline}\label{unequal-unequal 31}
    \langle\cM_{32}^{\left(\frac{3}{2},-\frac{1}{2}\right)}\cM_{10}^{\left(\frac{1}{2},-\frac{3}{2}\right)}\rangle=\frac{5a_{1}d_{1}d_{0}\eta_{-}^{2}c_{+}^{2}\left(2\Delta+1\right)}{4qN\left|x_{32}\right|^{2\Delta}\left|x_{10}\right|^{2\Delta}\left|x_{30}\right|^{2}\left(\Delta-1\right)}\\
    \times \Bigg[\frac{\Delta\widetilde{\mathcal{V}}_{\text{uu}}(\chi)}{2\Delta-1} + \frac{\mathcal{V}_{\text{uu}}(\chi)}{4(\Delta-1)}-\frac{2\pi^{2} (\Delta-1)\chi\, {}_2F_{1}\!\left[\genfrac{}{}{0pt}{1}{1,1-2\Delta}{2-2\Delta};-\chi\right]}{(2\Delta+1) (2\Delta-1)^{2} \tan(\Delta\pi)}\Bigg]\,,
\end{multline}
where we used $\bar{\rho}^{(2)}_{\frac{3}{2},\frac{1}{2}}=-10\pi^{2}a_{1}d_{0}\eta_{-}\cot\left(\Delta\pi\right)$
which was derived in Appendix \ref{app:J1dt} and the cross-ratio functions are given by the following integrals
\begin{align}
\widetilde{\mathcal{V}}_{\text{uu}}(\chi)&=\chi\int\frac{F_{1}\left(\frac{1-y^{3}}{\left|y-\hat{e}_{3}\right|}\right)F_{3}\left(\frac{y\cdot\left(y+\chi\hat{e}_{3}\right)}{\left|y+\chi\hat{e}_{3}\right|\left|y\right|}\right)r^{2}\dd^{3}y}{\left|y-\hat{e}_{3}\right|^{4}\left|y+\chi\hat{e}_{3}\right|\left|y\right|^{3}}\,,\\\mathcal{T}_{\text{uu}}(\chi)&=\chi\int\frac{F_{1}\left(\frac{1-y^{3}}{\left|y-\hat{e}_{3}\right|}\right)F_{2}\left(\frac{y\cdot\left(y+\chi\hat{e}_{3}\right)}{\left|y+\chi\hat{e}_{3}\right|\left|y\right|}\right)r^{2}\dd^{3}y}{\left|y-\hat{e}_{3}\right|^{4}\left|y+\chi\hat{e}_{3}\right|\left|y\right|^{3}}\,,
\end{align}
with 
\begin{equation}
    \chi\mathcal{V}'_{\text{uu}}(\chi)-2\Delta\mathcal{V}_{\text{uu}}(\chi) = 2\Delta[(2\Delta-1)\mathcal{T}_{\text{uu}}(\chi)-\chi\mathcal{T}'_{\text{uu}}(\chi)]
\end{equation}
which determines $\mathcal{V}_{\text{uu}}(\chi)$ up to a constant of integration which can be taken to determine the behaviour at infinity according to $\mathcal{V}_{\text{uu}}(\chi)\sim b_3 \,\chi^{2\Delta}$. To fix the value of $b_3$, we consider the correlator in the limit $x_2\rightarrow x_1$, which indeed corresponds to $\chi\rightarrow+\infty$. The leading exchanged operator, as in the previous example, is the identity, and we find that
\begin{equation}
    b_3 = \frac{\left(2\pi\Delta(4\Delta^{2}-1) + \sin(2\pi\Delta)\right)b}{(1-2\Delta)(2\Delta+1)^{2}\pi}\,,
\end{equation}
where $b$ is defined in \eqref{boundary unequal}. We also remark that the integrals no longer have a logarithmic divergence at large $\chi$. This is consistent with the absence of disconnected piece to this correlator.

We end with a simple consistency check of the result in the $x_1\rightarrow x_0$ limit, where
\begin{equation}
    \widetilde{\mathcal{V}}_{\text{uu}}(\chi)\sim \mathcal{T}_{\text{uu}}(\chi) \sim \alpha\chi\,,\quad \mathcal{V}_{\text{uu}}(\chi) \sim \frac{4(\Delta-1)\Delta\alpha\chi}{1-2\Delta}\,,
\end{equation}
and $\alpha$ is some constant that cancels out in the final result so that only the last term in~\eqref{unequal-unequal 31} contributes to
\begin{equation}
   \langle\cM_{32}^{\left(\frac{3}{2},-\frac{1}{2}\right)}\cM_{10}^{\left(\frac{1}{2},-\frac{3}{2}\right)}\rangle \sim \frac{\chi\left(c_{+}d_{1}\eta_{-}\right)^{2}}{16(2\Delta-1)^{2}N\mathcal{N}_{1}\left|x_{10}\right|^{2\Delta}\left|x_{32}\right|^{2\Delta}\left|x_{30}\right|^{2}}\,.
\end{equation}
We compare this to the contribution of the exchange of $J^\pm_1$, which is the lightest operator exchanged in this OPE channel, 
\begin{equation}
    \langle\cM_{32}^{\left(\frac{3}{2},-\frac{1}{2}\right)}\cM_{10}^{\left(\frac{1}{2},-\frac{3}{2}\right)}\rangle\sim\frac{\langle\cM^{\left(\frac{1}{2},-\frac{3}{2}\right)}J_{1}^{-}\rangle}{\langle J_{1}^{+}J_{1}^{-}\rangle}\frac{\langle\cM_{32}^{\left(\frac{3}{2},-\frac{1}{2}\right)}J_{1}^{+}(x_0)\rangle}{\left|x_{10}\right|^{2\Delta-1}}\,,
\end{equation}
and we find agreement after inserting the OPE data and the correlator.

\end{appendix}

\bibliography{Refs.bib}

@article{Aharony:2018npf,
    author = "Aharony, Ofer and Alday, Luis F. and Bissi, Agnese and Yacoby, Ran",
    title = "{The Analytic Bootstrap for Large $N$ Chern-Simons Vector Models}",
    eprint = "1805.04377",
    archivePrefix = "arXiv",
    primaryClass = "hep-th",
    doi = "10.1007/JHEP08(2018)166",
    journal = "JHEP",
    volume = "08",
    pages = "166",
    year = "2018"
}

@article{Aharony:2011jz,
    author = "Aharony, Ofer and Gur-Ari, Guy and Yacoby, Ran",
    title = "{d=3 Bosonic Vector Models Coupled to Chern-Simons Gauge Theories}",
    eprint = "1110.4382",
    archivePrefix = "arXiv",
    primaryClass = "hep-th",
    doi = "10.1007/JHEP03(2012)037",
    journal = "JHEP",
    volume = "03",
    pages = "037",
    year = "2012"
}

@article{Bedhotiya:2015uga,
    author = "Bedhotiya, Akshay and Prakash, Shiroman",
    title = "{A test of bosonization at the level of four-point functions in Chern-Simons vector models}",
    eprint = "1506.05412",
    archivePrefix = "arXiv",
    primaryClass = "hep-th",
    doi = "10.1007/JHEP12(2015)032",
    journal = "JHEP",
    volume = "12",
    pages = "032",
    year = "2015"
}

@article{Belitsky:2019fan,
    author = "Belitsky, A. V. and Korchemsky, G. P.",
    title = "{Exact null octagon}",
    eprint = "1907.13131",
    archivePrefix = "arXiv",
    primaryClass = "hep-th",
    reportNumber = "IPhT-T19/097",
    doi = "10.1007/JHEP05(2020)070",
    journal = "JHEP",
    volume = "05",
    pages = "070",
    year = "2020"
}

@article{Brezin:1972se,
    author = "Brézin, E. and Wallace, D. J.",
    title = "{Critical Behavior of a Classical Heisenberg Ferromagnet with Many Degrees of Freedom}",
    reportNumber = "COO-3072-12",
    doi = "10.1103/PhysRevB.7.1967",
    journal = "Phys. Rev. B",
    volume = "7",
    number = "5",
    pages = "1967",
    year = "1973"
}

@article{Caron-Huot:2017vep,
    author = "Caron-Huot, Simon",
    title = "{Analyticity in Spin in Conformal Theories}",
    eprint = "1703.00278",
    archivePrefix = "arXiv",
    primaryClass = "hep-th",
    doi = "10.1007/JHEP09(2017)078",
    journal = "JHEP",
    volume = "09",
    pages = "078",
    year = "2017"
}

@article{Chicherin:2017cns,
    author = {Chicherin, Dmitry and Kazakov, Vladimir and Loebbert, Florian and M{\"u}ller, Dennis and Zhong, De-liang},
    title = "{Yangian Symmetry for Bi-Scalar Loop Amplitudes}",
    eprint = "1704.01967",
    archivePrefix = "arXiv",
    primaryClass = "hep-th",
    reportNumber = "HU-EP-17-09, MITP-17-022, LPTENS-17-07",
    doi = "10.1007/JHEP05(2018)003",
    journal = "JHEP",
    volume = "05",
    pages = "003",
    year = "2018"
}

@article{Chicherin:2017frs,
    author = {Chicherin, Dmitry and Kazakov, Vladimir and Loebbert, Florian and M{\"u}ller, Dennis and Zhong, De-liang},
    title = "{Yangian Symmetry for Fishnet Feynman Graphs}",
    eprint = "1708.00007",
    archivePrefix = "arXiv",
    primaryClass = "hep-th",
    reportNumber = "MITP-17-049, HU-EP-17-20, LPTENS-17-32",
    doi = "10.1103/PhysRevD.96.121901",
    journal = "Phys. Rev. D",
    volume = "96",
    number = "12",
    pages = "121901",
    year = "2017"
}

@article{Coronado:2018cxj,
    author = "Coronado, Frank",
    title = "{Bootstrapping the Simplest Correlator in Planar $\mathcal N = 4$ Supersymmetric Yang-Mills Theory to All Loops}",
    eprint = "1811.03282",
    archivePrefix = "arXiv",
    primaryClass = "hep-th",
    doi = "10.1103/PhysRevLett.124.171601",
    journal = "Phys. Rev. Lett.",
    volume = "124",
    number = "17",
    pages = "171601",
    year = "2020"
}

@article{Ferrando:2025ufj,
    author = {Ferrando, Gwena{\"e}l and Sever, Amit and Urisman, Elior},
    title = "{Correlators of line defect and local operator in conformal field theories with a slightly broken higher-spin symmetry}",
    eprint = "2505.10232",
    archivePrefix = "arXiv",
    primaryClass = "hep-th",
    reportNumber = "BONN-TH-2025-19",
    doi = "10.1007/JHEP10(2025)204",
    journal = "JHEP",
    volume = "10",
    pages = "204",
    year = "2025"
}

@article{Gabai:2022vri,
    author = "Gabai, Barak and Sever, Amit and Zhong, De-liang",
    title = "{Line Operators in Chern-Simons\textendash{}Matter Theories and Bosonization in Three Dimensions}",
    eprint = "2204.05262",
    archivePrefix = "arXiv",
    primaryClass = "hep-th",
    doi = "10.1103/PhysRevLett.129.121604",
    journal = "Phys. Rev. Lett.",
    volume = "129",
    number = "12",
    pages = "121604",
    year = "2022"
}

@article{Gabai:2022mya,
    author = "Gabai, Barak and Sever, Amit and Zhong, De-liang",
    title = "{Line operators in Chern-Simons-Matter theories and Bosonization in Three Dimensions II: Perturbative analysis and all-loop resummation}",
    eprint = "2212.02518",
    archivePrefix = "arXiv",
    primaryClass = "hep-th",
    doi = "10.1007/JHEP04(2023)070",
    journal = "JHEP",
    volume = "04",
    pages = "070",
    year = "2023"
}

@article{Gabai:2023lax,
    author = "Gabai, Barak and Sever, Amit and Zhong, De-liang",
    title = "{Bootstrapping smooth conformal defects in Chern-Simons-matter theories}",
    eprint = "2312.17132",
    archivePrefix = "arXiv",
    primaryClass = "hep-th",
    doi = "10.1007/JHEP03(2024)055",
    journal = "JHEP",
    volume = "03",
    pages = "055",
    year = "2024",
    note = "[Erratum: JHEP 12, 083 (2024)]"
}

@article{Giombi:2011kc,
    author = "Giombi, Simone and Minwalla, Shiraz and Prakash, Shiroman and Trivedi, Sandip P. and Wadia, Spenta R. and Yin, Xi",
    title = "{Chern-Simons Theory with Vector Fermion Matter}",
    eprint = "1110.4386",
    archivePrefix = "arXiv",
    primaryClass = "hep-th",
    doi = "10.1140/epjc/s10052-012-2112-0",
    journal = "Eur. Phys. J. C",
    volume = "72",
    pages = "2112",
    year = "2012"
}

@article{Giombi:2016zwa,
    author = "Giombi, S. and Gurucharan, V. and Kirilin, V. and Prakash, S. and Skvortsov, E.",
    title = "{On the Higher-Spin Spectrum in Large N Chern-Simons Vector Models}",
    eprint = "1610.08472",
    archivePrefix = "arXiv",
    primaryClass = "hep-th",
    reportNumber = "PUPT-2512, LMU-ASC-52-16",
    doi = "10.1007/JHEP01(2017)058",
    journal = "JHEP",
    volume = "01",
    pages = "058",
    year = "2017"
}

@article{Grabner:2017pgm,
    author = "Grabner, David and Gromov, Nikolay and Kazakov, Vladimir and Korchemsky, Gregory",
    title = "{Strongly $\gamma$-Deformed $\mathcal{N}=4$ Supersymmetric Yang-Mills Theory as an Integrable Conformal Field Theory}",
    eprint = "1711.04786",
    archivePrefix = "arXiv",
    primaryClass = "hep-th",
    reportNumber = "KCL-MTH-17-04, LPTENS-17-31, IPHT-T17-171, LPTENS--17-31, IPHT--T17-171",
    doi = "10.1103/PhysRevLett.120.111601",
    journal = "Phys. Rev. Lett.",
    volume = "120",
    number = "11",
    pages = "111601",
    year = "2018"
}

@article{Gromov:2018hut,
    author = "Gromov, Nikolay and Kazakov, Vladimir and Korchemsky, Gregory",
    title = "{Exact Correlation Functions in Conformal Fishnet Theory}",
    eprint = "1808.02688",
    archivePrefix = "arXiv",
    primaryClass = "hep-th",
    doi = "10.1007/JHEP08(2019)123",
    journal = "JHEP",
    volume = "08",
    pages = "123",
    year = "2019"
}

@article{Guadagnini:1989am,
    author = "Guadagnini, E. and Martellini, M. and Mintchev, M.",
    title = "{Wilson Lines in Chern-Simons Theory and Link Invariants}",
    reportNumber = "CERN-TH-5420/89, IFUP-TH-24/89",
    doi = "10.1016/0550-3213(90)90124-V",
    journal = "Nucl. Phys. B",
    volume = "330",
    pages = "575--607",
    year = "1990"
}

@article{Jain:2022ajd,
    author = "Jain, Prabhav and Jain, Sachin and Sahoo, Bibhut and Dhruva, K. S. and Zade, Aashna",
    title = "{Mapping Large N Slightly Broken Higher Spin (SBHS) theory correlators to free theory correlators}",
    eprint = "2207.05101",
    archivePrefix = "arXiv",
    primaryClass = "hep-th",
    doi = "10.1007/JHEP12(2023)173",
    journal = "JHEP",
    volume = "12",
    pages = "173",
    year = "2023"
}

@article{Jain:2020puw,
    author = "Jain, Sachin and John, Renjan Rajan and Malvimat, Vinay",
    title = "{Constraining momentum space correlators using slightly broken higher spin symmetry}",
    eprint = "2008.08610",
    archivePrefix = "arXiv",
    primaryClass = "hep-th",
    doi = "10.1007/JHEP04(2021)231",
    journal = "JHEP",
    volume = "04",
    pages = "231",
    year = "2021"
}

@article{Kalloor:2019xjb,
    author = "Kalloor, Rohit R.",
    title = "{Four-point functions in large $N$ Chern-Simons fermionic theories}",
    eprint = "1910.14617",
    archivePrefix = "arXiv",
    primaryClass = "hep-th",
    doi = "10.1007/JHEP10(2020)028",
    journal = "JHEP",
    volume = "10",
    pages = "028",
    year = "2020"
}

@article{Kostov:2019stn,
    author = "Kostov, Ivan and Petkova, Valentina B. and Serban, Didina",
    title = "{Determinant Formula for the Octagon Form Factor in $N$=4 Supersymmetric Yang-Mills Theory}",
    eprint = "1903.05038",
    archivePrefix = "arXiv",
    primaryClass = "hep-th",
    doi = "10.1103/PhysRevLett.122.231601",
    journal = "Phys. Rev. Lett.",
    volume = "122",
    number = "23",
    pages = "231601",
    year = "2019"
}

@article{Kukolj:2024yyo,
    author = "Kukolj, Trivko",
    title = "{Four-point functions and contact terms from higher-spin Ward identities of Chern-Simons-matter theory}",
    eprint = "2406.17011",
    archivePrefix = "arXiv",
    primaryClass = "hep-th",
    doi = "10.1007/JHEP11(2024)147",
    journal = "JHEP",
    volume = "11",
    pages = "147",
    year = "2024"
}

@article{Lang:1992pp,
    author = "Lang, K. and Rühl, W.",
    title = "{The critical O(N) sigma model at dimensions 2 {\ensuremath{<}} d {\ensuremath{<}} 4: a list of quasiprimary fields}",
    reportNumber = "KL-TH-92-7",
    doi = "10.1016/0550-3213(93)90119-A",
    journal = "Nucl. Phys. B",
    volume = "402",
    pages = "573--603",
    year = "1993"
}

@article{Lemos:2017vnx,
    author = "Lemos, Madalena and Liendo, Pedro and Meineri, Marco and Sarkar, Sourav",
    title = "{Universality at large transverse spin in defect CFT}",
    eprint = "1712.08185",
    archivePrefix = "arXiv",
    primaryClass = "hep-th",
    reportNumber = "DESY 17-239, HU-EP-17/31, DESY-17-239, HU-EP-17-31",
    doi = "10.1007/JHEP09(2018)091",
    journal = "JHEP",
    volume = "09",
    pages = "091",
    year = "2018"
}

@article{Li:2019twz,
    author = "Li, Zhijin",
    title = "{Bootstrapping conformal four-point correlators with slightly broken higher spin symmetry and $3D$ bosonization}",
    eprint = "1906.05834",
    archivePrefix = "arXiv",
    primaryClass = "hep-th",
    doi = "10.1007/JHEP10(2020)007",
    journal = "JHEP",
    volume = "10",
    pages = "007",
    year = "2020"
}

@article{Loebbert:2024qbw,
    author = "Loebbert, Florian and Mathur, Harshad",
    title = "{The Feyn-structure of Yangian symmetry}",
    eprint = "2410.11936",
    archivePrefix = "arXiv",
    primaryClass = "hep-th",
    reportNumber = "BONN-TH-2024-14",
    doi = "10.1007/JHEP01(2025)112",
    journal = "JHEP",
    volume = "01",
    pages = "112",
    year = "2025"
}

@article{Loebbert:2019vcj,
    author = {Loebbert, Florian and M{\"u}ller, Dennis and M{\"u}nkler, Hagen},
    title = "{Yangian Bootstrap for Conformal Feynman Integrals}",
    eprint = "1912.05561",
    archivePrefix = "arXiv",
    primaryClass = "hep-th",
    reportNumber = "HU-EP-19/39",
    doi = "10.1103/PhysRevD.101.066006",
    journal = "Phys. Rev. D",
    volume = "101",
    number = "6",
    pages = "066006",
    year = "2020"
}

@article{Maldacena:2012sf,
    author = "Maldacena, Juan and Zhiboedov, Alexander",
    title = "{Constraining conformal field theories with a slightly broken higher spin symmetry}",
    eprint = "1204.3882",
    archivePrefix = "arXiv",
    primaryClass = "hep-th",
    reportNumber = "PUPT-2410",
    doi = "10.1088/0264-9381/30/10/104003",
    journal = "Class. Quant. Grav.",
    volume = "30",
    pages = "104003",
    year = "2013"
}

@article{Ribault:2024rvk,
    author = "Ribault, Sylvain",
    title = "{Exactly solvable conformal field theories}",
    eprint = "2411.17262",
    archivePrefix = "arXiv",
    primaryClass = "hep-th",
    month = "11",
    year = "2024"
}

@article{Silva:2021ece,
    author = "Silva, Joao A.",
    title = "{Four point functions in CFT{\textquoteright}s with slightly broken higher spin symmetry}",
    eprint = "2103.00275",
    archivePrefix = "arXiv",
    primaryClass = "hep-th",
    doi = "10.1007/JHEP05(2021)097",
    journal = "JHEP",
    volume = "05",
    pages = "097",
    year = "2021"
}

@article{Teschner:1995yf,
    author = "Teschner, Jorg",
    title = "{On the Liouville three point function}",
    eprint = "hep-th/9507109",
    archivePrefix = "arXiv",
    doi = "10.1016/0370-2693(95)01200-A",
    journal = "Phys. Lett. B",
    volume = "363",
    pages = "65--70",
    year = "1995"
}

@article{Trevisani:2024djr,
    author = "Trevisani, Emilio",
    title = "{The Parisi-Sourlas uplift and infinitely many solvable 4d CFTs}",
    eprint = "2405.00771",
    archivePrefix = "arXiv",
    primaryClass = "hep-th",
    doi = "10.21468/SciPostPhys.18.2.056",
    journal = "SciPost Phys.",
    volume = "18",
    number = "2",
    pages = "056",
    year = "2025"
}

@article{Turiaci:2018nua,
    author = "Turiaci, Gustavo J. and Zhiboedov, Alexander",
    title = "{Veneziano Amplitude of Vasiliev Theory}",
    eprint = "1802.04390",
    archivePrefix = "arXiv",
    primaryClass = "hep-th",
    doi = "10.1007/JHEP10(2018)034",
    journal = "JHEP",
    volume = "10",
    pages = "034",
    year = "2018"
}

@article{Yacoby:2018yvy,
    author = "Yacoby, Ran",
    title = "{Scalar Correlators in Bosonic Chern-Simons Vector Models}",
    eprint = "1805.11627",
    archivePrefix = "arXiv",
    primaryClass = "hep-th",
    month = "5",
    year = "2018"
}

@article{Witten:1988hf,
    author = "Witten, Edward",
    editor = "Mitra, Asoke N.",
    title = "{Quantum Field Theory and the Jones Polynomial}",
    reportNumber = "IASSNS-HEP-88-33",
    doi = "10.1007/BF01217730",
    journal = "Commun. Math. Phys.",
    volume = "121",
    pages = "351--399",
    year = "1989"
}

@article{Andrei:2018die,
    author = "Andrei, N. and others",
    title = "{Boundary and Defect CFT: Open Problems and Applications}",
    eprint = "1810.05697",
    archivePrefix = "arXiv",
    primaryClass = "hep-th",
    doi = "10.1088/1751-8121/abb0fe",
    journal = "J. Phys. A",
    volume = "53",
    number = "45",
    pages = "453002",
    year = "2020"
}

@article{Komargodski:2025jbu,
    author = "Komargodski, Zohar and Popov, Fedor K. and Rayhaun, Brandon C.",
    title = "{Defect anomalies, a spin-flux duality, and Boson-Kondo problems}",
    eprint = "2508.14963",
    archivePrefix = "arXiv",
    primaryClass = "hep-th",
    doi = "10.1007/JHEP04(2026)071",
    journal = "JHEP",
    volume = "04",
    pages = "071",
    year = "2026"
}

@article{Giombi:2018qox,
    author = "Giombi, Simone and Komatsu, Shota",
    title = "{Exact Correlators on the Wilson Loop in $\mathcal{N}=4$ SYM: Localization, Defect CFT, and Integrability}",
    eprint = "1802.05201",
    archivePrefix = "arXiv",
    primaryClass = "hep-th",
    reportNumber = "PUTP-2549",
    doi = "10.1007/JHEP05(2018)109",
    journal = "JHEP",
    volume = "05",
    pages = "109",
    year = "2018",
    note = "[Erratum: JHEP 11, 123 (2018)]"
}

@article{Pestun:2007rz,
    author = "Pestun, Vasily",
    title = "{Localization of gauge theory on a four-sphere and supersymmetric Wilson loops}",
    eprint = "0712.2824",
    archivePrefix = "arXiv",
    primaryClass = "hep-th",
    reportNumber = "ITEP-TH-41-07, PUTP-2248",
    doi = "10.1007/s00220-012-1485-0",
    journal = "Commun. Math. Phys.",
    volume = "313",
    pages = "71--129",
    year = "2012"
}

@article{Correa:2012at,
    author = "Correa, Diego and Henn, Johannes and Maldacena, Juan and Sever, Amit",
    title = "{An exact formula for the radiation of a moving quark in N=4 super Yang Mills}",
    eprint = "1202.4455",
    archivePrefix = "arXiv",
    primaryClass = "hep-th",
    doi = "10.1007/JHEP06(2012)048",
    journal = "JHEP",
    volume = "06",
    pages = "048",
    year = "2012"
}

@article{Giombi:2018hsx,
    author = "Giombi, Simone and Komatsu, Shota",
    title = "{More Exact Results in the Wilson Loop Defect CFT: Bulk-Defect OPE, Nonplanar Corrections and Quantum Spectral Curve}",
    eprint = "1811.02369",
    archivePrefix = "arXiv",
    primaryClass = "hep-th",
    reportNumber = "PUTP-2572",
    doi = "10.1088/1751-8121/ab046c",
    journal = "J. Phys. A",
    volume = "52",
    number = "12",
    pages = "125401",
    year = "2019"
}

@article{Mazac:2016qev,
    author = "Mazac, Dalimil",
    title = "{Analytic bounds and emergence of AdS$_{2}$ physics from the conformal bootstrap}",
    eprint = "1611.10060",
    archivePrefix = "arXiv",
    primaryClass = "hep-th",
    doi = "10.1007/JHEP04(2017)146",
    journal = "JHEP",
    volume = "04",
    pages = "146",
    year = "2017"
}

@article{Lanzetta:2025xfw,
    author = "Lanzetta, Ryan A. and Liu, Shang and Metlitski, Max A.",
    title = "{The beginning of the endpoint bootstrap for conformal line defects}",
    eprint = "2508.14964",
    archivePrefix = "arXiv",
    primaryClass = "cond-mat.str-el",
    month = "8",
    year = "2025"
}
\bibliographystyle{JHEP.bst}

\end{document}